\documentclass[11pt]{article}
\usepackage[utf8]{inputenc}
\usepackage{amsfonts,algorithmic}
\usepackage[para]{footmisc}
\usepackage{xcolor}
\usepackage{qcircuit}
\usepackage{bm}
\usepackage{amsmath}
\usepackage{enumerate}
\usepackage{appendix}
\usepackage{amsthm,mathrsfs}
\usepackage{multirow}
\usepackage{enumitem}
\usepackage{tikz}
\usepackage{multicol}
\usepackage[pagebackref]{hyperref}
\hypersetup{
    colorlinks,
    linkcolor={red!100!black},
    citecolor={blue!100!black},
}
\usepackage{color, colortbl}
\definecolor{Gray}{gray}{0.89}
\usetikzlibrary{matrix,decorations.pathreplacing}
\usepackage{mleftright}
\usepackage{amssymb}
\usepackage[linesnumbered,ruled,vlined]{algorithm2e}
\usepackage{mathabx}
\usepackage{geometry}
\usepackage{comment}
\usepackage{subcaption}
\usepackage{soul}
\usepackage{bbm}

\usepackage[capitalize,nameinlink,noabbrev]{cleveref}

\usepackage{authblk}
\usepackage[normalem]{ulem}
\usepackage{euscript}
\usepackage{scalerel}
\usepackage{makecell}

\newcommand\independent{\protect\mathpalette{\protect\independenT}{\perp}}
\def\independenT#1#2{\mathrel{\rlap{$#1#2$}\mkern2mu{#1#2}}}

\newtheorem{theorem}{Theorem}
\newtheorem{lemma}[theorem]{Lemma}
\newtheorem{remark}[theorem]{Remark}

\newtheorem{definition}[theorem]{Definition}

\newtheorem{fact}[theorem]{Fact}

\newtheorem{corollary}[theorem]{Corollary}
\newtheorem{result}{Result}

\crefname{lemma}{Lemma}{lemma}
\crefname{definition}{Definition}{definition}
\crefname{corollary}{Corollary}{corollary}
\crefname{result}{Result}{result}

\usepackage{mathtools}

\DeclareMathOperator*{\argmax}{arg\,max}

\usepackage{geometry}
\def\be{\begin{eqnarray}}
\def\ee{\end{eqnarray}}

\newcommand{\e}{\mathrm{e}}
\newcommand{\ri}{\mathrm{i}}

\title{Reconquering Bell sampling on qudits: stabilizer learning and testing, quantum pseudorandomness bounds, and more}

\author[1]{Jonathan Allcock\thanks{\scriptsize jonallcock@tencent.com}}
\author[2]{Joao F. Doriguello\thanks{\scriptsize doriguello@renyi.hu}}
\author[3]{G\'abor Ivanyos\thanks{\scriptsize gabor.ivanyos@sztaki.hu}}
\author[4]{Miklos Santha\thanks{\scriptsize miklos.santha@gmail.com}}
\affil[1]{Tencent Quantum Laboratory, Hong Kong, China}
\affil[2]{HUN-REN Alfr\'ed R\'enyi Institute of Mathematics, Budapest, Hungary}
\affil[3]{HUN-REN Institute for Computer Science and Control, Budapest, Hungary}
\affil[4]{CNRS, IRIF, Universit\'e Paris Cit\'e}

\date{\today}

\begin{document}
\maketitle

\begin{abstract}        
    Bell sampling is a simple yet powerful tool based on measuring two copies of a quantum state in the Bell basis, and has found applications in a plethora of problems related to stabiliser states and measures of magic. However, it was not known how to generalise the procedure from qubits to $d$-level systems -- qudits -- for all dimensions $d > 2$ in a useful way. Indeed, a prior work of the authors (arXiv'24) showed that the natural extension of Bell sampling to arbitrary dimensions fails to provide meaningful information about the quantum states being measured. In this paper, we  overcome the difficulties encountered in previous works and develop a useful generalisation of Bell sampling to qudits of all dimensions $d\geq 2$. At the heart of our primitive is a new unitary, based on Lagrange's four-square theorem, that maps four copies of any stabiliser state $|\mathcal{S}\rangle$ to four copies of its complex conjugate $|\mathcal{S}^\ast\rangle$  (up to some Pauli operator), which may be of independent interest. We then demonstrate the utility of our new Bell sampling technique by lifting several known results from qubits to qudits for any $d\geq 2$ (which involves working with submodules instead of subspaces):
    \begin{itemize}
        \item Learning an unknown stabiliser state $|\mathcal{S}\rangle\in(\mathbb{C}^d)^{\otimes n}$ in $O(n^3)$ time with $O(n)$ samples;
        \item Solving the Hidden Stabiliser Group Problem (a stabiliser version of the State Hidden Subgroup Problem) in $\widetilde{O}(n^3/\varepsilon)$ time with $\widetilde{O}(n/\varepsilon)$ samples;
        \item Testing whether $|\psi\rangle\in(\mathbb{C}^d)^{\otimes n}$ has stabiliser size (a generalisation of stabiliser dimension for submodules) at least $d^t$ or is $\varepsilon$-far from all such states in $\widetilde{O}(n^3/\varepsilon)$ time with $\widetilde{O}(n/\varepsilon)$ samples if $\varepsilon = O(d^{-2})$;
        \item Testing whether $|\psi\rangle\in(\mathbb{C}^d)^{\otimes n}$ is Haar-random or the output of a Clifford circuit augmented with less than $n/2$ single-qudit non-Clifford gates in $O(n^3)$ time using $O(n)$ samples. As a corollary, we show that Clifford circuits with at most $n/2$ single-qudit non-Clifford gates cannot prepare pseudorandom states, an exponential improvement over previous works;
        \item Testing whether $|\psi\rangle\in(\mathbb{C}^d)^{\otimes n}$ has stabiliser fidelity at least $1-\varepsilon_1$ or at most $1-\varepsilon_2$ with $O(d^2\varepsilon_2/(\varepsilon_2-\varepsilon_1)^2)$ samples if $\varepsilon_1 = O(\varepsilon_2/d^{2})$.
    \end{itemize}
\end{abstract}

\newpage

\tableofcontents

\newpage

\section{Introduction}

Stabiliser states are central to quantum computation, and find applications in quantum error correction~\cite{shor1995scheme,calderbank1996good,gottesman1996class,gottesman1997stabilizer}, efficient classical simulations of quantum circuits~\cite{aaronson2004improved,bravyi2016trading,bravyi2019simulation}, low-rank recovery~\cite{kueng2016low}, randomized benchmarking~\cite{knill2008randomized,magesan2011scalable,helsen2019multiqubit}, measurement-based quantum computation~\cite{raussendorf2000quantum}, tensor networks for holography codes~\cite{hayden2016holographic,nezami2020multipartite}, and quantum learning algorithms~\cite{huang2020predicting}. A major tool in exploring stabiliser states is \emph{Bell sampling}~\cite{montanaro2017learning}, which involves measuring two copies of a given $n$-qubit state $|\psi\rangle\in(\mathbb{C}^2)^{\otimes n}$ in the Bell basis. In spite of its simplicity, Bell sampling -- along with a variant called Bell difference sampling~\cite{gross2021schur} -- has proved to be a powerful primitive, and has found widespread applications.

The first application of Bell sampling was due to Montanaro~\cite{montanaro2017learning} to identify an unknown stabiliser state, which simplified or improved upon previous works on the same task~\cite{aaronson2008stabiliser,rotteler2009quantum,zhao2016fast}. Later, \cite{gross2021schur} built on the same sampling procedure to decide whether a given unknown state is either a stabiliser state or is far from one. Bell sampling has since been employed to infer several properties associated with stabiliser states~\cite{gs007,Anshu2024}, e.g., to detect and learn circuit errors and to extract information such as the depth and number of T-gates in a circuit~\cite{hangleiter2023bell}; to measure stabiliser entropies~\cite{haug2023efficient,Haug2023scalable}; to obtain the classical description of the output of a Clifford circuit augmented with a few non-Clifford single-qubit gates (called \emph{doped Clifford circuits})~\cite{leone2023learning,salvatore2024unscrambling,leone2024learning,grewal2023efficient}; to test whether a state is Haar-random or the output of a doped Clifford circuit~\cite{grewal2022low,grewal2023improved}; perform agnostic tomography on stabilizer product states~\cite{grewal2024agnostic}.


While the aforementioned works focused on qubits, the scenario regarding \emph{qudits} is still mostly unexplored, the main reason being that it is not known how to extend Bell (difference) sampling to qudits in a meaningful and useful way that retains most of the compelling properties from the qubit case. The limitations of Bell sampling on qudits was recently explored by the authors~\cite{allcock2024beyond}, who showed that a natural extension based on measuring two copies $|\psi\rangle^{\otimes 2}$ of a given $n$-qudit state $|\psi\rangle\in(\mathbb{C}^d)^{\otimes n}$ in the generalised Bell basis is doomed to fail: even in the simplest case when $|\psi\rangle$ is a stabiliser state, Bell difference sampling can return a completely random output. This is in stark contrast to the qubit case where the output is a generator of the stabiliser group. At a high level, this stems from the well-known fact that the transpose map $\psi\mapsto \bar{\psi} = \psi^\top$ is not completely positive.

In this work, we successfully extend Bell sampling to qudits of all dimensions $d$, such that the main and useful properties from the qubit case are preserved. Our generalisation is non-trivial yet simple, and at its heart it is a novel unitary that maps four copies $|\mathcal{S}\rangle^{\otimes 4}$ of a stabiliser state into their complex conjugate $|\mathcal{S}^\ast\rangle^{\otimes 4}$ with respect to the computational basis  (up to some qudit Pauli operator), which we believe might be of independent interest. Given our new Bell sampling primitive on qudits, we then manage to lift several known results from qubits to qudits.

\subsection{Bell sampling on qudits}

To explain Bell (difference) sampling and our generalisation in more detail, let us introduce some notation and recall various concepts relating to the Pauli group on qudits. Let $\omega \triangleq \e^{\ri 2\pi/d}$ be the standard primitive $d$-th root of unity and $\tau \triangleq (-1)^{d}\e^{\ri \pi/d} = \e^{\ri \pi(d^2+1)/d}$ such that $\tau^2 = \omega$. Let $D=d$ if $d$ is odd and $D=2d$ if $d$ is even, so that $D$ is the order of $\tau$. The generalised Pauli operators are defined using the shift and clock operators $\mathsf{X}$ and $\mathsf{Z}$, respectively, as
\begin{align*}
    \mathsf{X}|q\rangle = |q+1\rangle, \quad \mathsf{Z}|q\rangle = \omega^q |q\rangle, \quad \forall q\in\mathbb{Z}_d.
\end{align*}
A useful way of writing Pauli operators on $n$ qudits is via the so-called \emph{Weyl operators}
\begin{align*}
    \mathcal{W}_{\mathbf{x}} = \mathcal{W}_{\mathbf{v},\mathbf{w}} \triangleq \tau^{\langle \mathbf{v},\mathbf{w}\rangle} (\mathsf{X}^{w_1}\mathsf{Z}^{v_1})\otimes \cdots \otimes (\mathsf{X}^{w_n}\mathsf{Z}^{v_n}) \qquad \forall\mathbf{x} = (\mathbf{v},\mathbf{w})\in\mathbb{Z}_d^{2n},
\end{align*}
where $\langle \mathbf{v},\mathbf{w}\rangle = \sum_{i=1}^n v_i w_i$ is the usual scalar product, considering $\mathbf{v},\mathbf{w}$ as embedded into~$\mathbb{Z}^{n}$. One can think of Weyl operators as Pauli operators labeled by a $2n$-bit string. There is then a nice connection between operators in the Hilbert space $(\mathbb{C}^d)^{\otimes n}$ and strings in $\mathbb{Z}_d^{2n}$. The generalised Bell basis is defined as
\begin{align*}
	|\mathcal{W}_{\mathbf{x}}\rangle \triangleq (\mathcal{W}_{\mathbf{x}}\otimes \mathbb{I})|\Phi^+\rangle, \qquad\text{where}\quad |\Phi^+\rangle \triangleq d^{-\frac{n}{2}}\sum_{\mathbf{q}\in\mathbb{Z}_d^{n}}|\mathbf{q},\mathbf{q}\rangle \quad\text{is a maximally entangled state}.
\end{align*}
The $n$-qudit Pauli group $\mathscr{P}_d^n \triangleq \{\tau^s \mathcal{W}_{\mathbf{x}}:\mathbf{x}\in\mathbb{Z}_d^{2n},s\in\mathbb{Z}\}$ is the group of all Weyl operators up to a power of $\tau$, while the $n$-qudit Clifford group $\mathscr{C}^n_{d} \triangleq \{\mathcal{U}\in\mathbb{C}^{d^n\times d^n}: \mathcal{U}^\dagger\mathcal{U} = \mathbf{I},\, \mathcal{U}\mathscr{P}^n_{d}\mathcal{U}^\dagger = \mathscr{P}^n_{d}\}$ is the group of $d^n\times d^n$ unitary operators that normalise the Pauli group. A stabilizer group $\mathcal{S}\subset \mathscr{P}_d^n$ is an abelian subgroup of the $n$-qudit Pauli group of size $d^n$.

Bell sampling on qubits~\cite{montanaro2017learning} measures two copies of an $n$-qubit state $|\psi\rangle = \sum_{\mathbf{q}\in\mathbb{F}_2^{2n}}\alpha_{\mathbf{q}}|\mathbf{q}\rangle$ in the Bell basis $|\mathcal{W}_{\mathbf{
x}}\rangle$ to obtain a string $\mathbf{x}\in\mathbb{F}_2^{2n}$ from the distribution $2^{-n}|\langle\psi|\mathcal{W}_{\mathbf{x}}|\psi^\ast\rangle|^2$, where $|\psi^\ast\rangle = \sum_{\mathbf{q}\in\mathbb{F}_2^{2n}}\alpha_{\mathbf{q}}^\ast|\mathbf{q}\rangle$ is the complex conjugate of $|\psi\rangle$ with respect to the computational basis. Bell difference sampling~\cite{montanaro2017learning,gross2021schur} performs Bell sampling twice and computes the difference of the two outcome strings. Gross, Nezami, and Walter~\cite{gross2021schur} proved that Bell difference sampling on $|\psi\rangle^{\otimes 4}$ returns a string $\mathbf{x}\in\mathbb{F}_2^{2n}$ with probability
\begin{align*}
    q_\psi(\mathbf{x}) \triangleq \sum_{\mathbf{y}\in\mathbb{F}_2^{2n}} p_\psi(\mathbf{y}) p_\psi(\mathbf{x}-\mathbf{y}), \qquad\text{where}\quad p_\psi(\mathbf{x}) \triangleq 2^{-n}|\langle\psi|\mathcal{W}_{\mathbf{x}}|\psi\rangle|^2
\end{align*}
is called the \emph{characteristic distribution}. If $|\psi\rangle$ is a stabiliser state $|\mathcal{S}\rangle$ of some stabiliser group $\mathcal{S}$, then $p_{\mathcal{S}}(\mathbf{x}) = 2^{-n}$ for all strings $\mathbf{x}\in\mathbb{F}_2^{2n}$ such that $\omega^s\mathcal{W}_{\mathbf{x}}\in\mathcal{S}$ for some $s\in\mathbb{Z}_d$ and $0$ otherwise. Therefore $q_{\mathcal{S}}(\mathbf{x}) = 2^{-n}$ if $\omega^s\mathcal{W}_{\mathbf{x}} \in \mathcal{S}$ and $0$ otherwise. Bell difference sampling on copies of a stabiliser state thus returns an element of its stabiliser group, and that is where its power lies.

As mentioned, the generalisation of Bell (difference) sampling to $d$-level systems has proved challenging.  Indeed,~\cite{allcock2024beyond} recently showed that the natural extension of the qubit scheme, based on measuring two copies of a given $n$-qudit state $|\psi\rangle\in(\mathbb{C}^d)^{\otimes n}$ in the generalised Bell basis, can return a uniformly random string in the worst case, and thus may provide no information about the corresponding stabilizer group at all. 
It is clear from the results of~\cite{allcock2024beyond} that a useful generalisation of Bell difference sampling, one which carried over important properties like sampling a generator of the stabiliser group given copies of its stabiliser state, is non-trivial.

Here we successfully propose a useful Bell sampling routine for qudits of all dimensions $d$. The starting motivation behind our approach is the observation that, if one is given the complex conjugate $|\psi^\ast\rangle$ of some state $|\psi\rangle\in(\mathbb{C}^d)^{\otimes n}$, then the standard Bell sampling on $|\psi\rangle|\psi^\ast\rangle$ returns a string $\mathbf{x}\in\mathbb{Z}_d^{2n}$ with probability $p_\psi(\mathbf{x}) = d^{-n}|\langle\psi|\mathcal{W}_{\mathbf{x}}|\psi\rangle|^2$. In other words, we get a sample from the characteristic distribution, which is exactly what we want since $p_{\mathcal{S}}(\mathbf{x}) = d^{-n}$ if $\omega^s\mathcal{W}_{\mathbf{x}} \in \mathcal{S}$ and $0$ otherwise for a stabiliser state $|\mathcal{S}\rangle$. The crucial fact for qubits is that $|\mathcal{S}^\ast\rangle = \mathcal{W}_{\mathbf{z}}|\mathcal{S}\rangle$ for some Weyl operator $\mathcal{W}_{\mathbf{z}}$, so Bell sampling on two copies of $|\mathcal{S}\rangle\in(\mathbb{C}^2)^{\otimes n}$ is equivalent to sampling from a shifted version of the characteristic distribution $p_{\mathcal{S}}$. Bell difference sampling then removes this shift by $\mathbf{z}$. The same cannot be said for qudits in general, i.e., $|\mathcal{S}\rangle$ and $|\mathcal{S}^\ast\rangle$ are not necessarily related by a Pauli operator. If it were possible to transform $|\mathcal{S}\rangle$ into $|\mathcal{S}^\ast\rangle$, then Bell difference sample would work on qudits as well.

Our generalisation of Bell (difference) sampling is based exactly on this idea: transforming enough copies of a stabiliser state $|\mathcal{S}\rangle$ into its complex conjugate $|\mathcal{S}^\ast\rangle$. More specifically, we propose a unitary transformation that maps at most four copies of any stabiliser state, $|\mathcal{S}\rangle^{\otimes 4}$, into $|\mathcal{S}^\ast\rangle^{\otimes 4}$ up to a Pauli operator. Interestingly enough, our unitary is based on Lagrange's four-square theorem, which states that every positive integer can be written as the sum of four integer squares!
\begin{result}\label{res:res1}
    For every integer $d\geq 2$, let $a_1,a_2,a_3,a_4$ be non-negative integers such that $a_1^2 + a_2^2 + a_3^2 + a_4^2 = d-1$ coming from Lagrange's four-square theorem. Define the unitary $\mathcal{B}_{\mathbf{R}}\in((\mathbb{C}^d)^{\otimes n})^{\otimes 4}$
    \begin{align*}
         \mathcal{B}_{\mathbf{R}}: |\mathbf{Q}\rangle \mapsto |\mathbf{Q}\mathbf{R} \rangle, \quad\text{where} \quad \mathbf{Q} = [\mathbf{q}_1,\mathbf{q}_2,\mathbf{q}_3,\mathbf{q}_4]\in\mathbb{Z}_d^{n\times 4}
    \end{align*}
    and $\mathbf{R}\in\mathbb{Z}_d^{4\times 4}$ is the matrix
    \begin{align*}
        \mathbf{R} = \begin{pmatrix}
            a_1 & a_2 & a_3 & a_4 \\
            a_2 & -a_1 & a_4 & -a_3 \\
            a_3 & -a_4 & -a_1 & a_2\\
            a_4 & a_3 & -a_2 & -a_1
        \end{pmatrix} \operatorname{mod}d.
    \end{align*}
    Given any stabiliser state $|\mathcal{S}\rangle\in(\mathbb{C}^d)^{\otimes n}$, then $\mathcal{B}_{\mathbf{R}}|\mathcal{S}\rangle^{\otimes 4} = \mathcal{P}(\mathcal{S})|\mathcal{S}^\ast\rangle^{\otimes 4}$ for some Pauli operator $\mathcal{P}(\mathcal{S})$ that depends on $|\mathcal{S}\rangle$.
\end{result}

The notation $|\mathbf{Q}\rangle$ is short for $|\mathbf{q}_1,\mathbf{q}_2,\mathbf{q}_3,\mathbf{q}_4\rangle$, while $|(\mathbf{Q}\mathbf{R})_1,(\mathbf{Q}\mathbf{R})_2,(\mathbf{Q}\mathbf{R})_3,(\mathbf{Q}\mathbf{R})_4\rangle$ is more compactly represented as $|\mathbf{Q}\mathbf{R}\rangle$, where $(\mathbf{Q}\mathbf{R})_i$ is the $i$-th column of $\mathbf{Q}\mathbf{R}$. More explicitly, the first $n$-qudit register $|\mathbf{q}_1\rangle$ is transformed into $|a_1\mathbf{q}_1 + a_2\mathbf{q}_2 + a_3\mathbf{q}_3 + a_4\mathbf{q}_4\rangle$ and similarly for the other three registers. The unitary $\mathcal{B}_{\mathbf{R}}$ thus mixes four states in a simple but non-trivial way.

The crucial property behind the matrix $\mathbf{R}$ that guarantees the transformation of $|\mathcal{S}\rangle^{\otimes 4}$ into $|\mathcal{S}^\ast\rangle^{\otimes 4}$ is that $\mathbf{R}^\top\mathbf{R} =  -\mathbf{I}\operatorname{mod}d$. It is well known that stabiliser states are quadratic phase states~\cite{hostens2005stabilizer}, meaning they are of the form $|\mathscr{A}|^{-\frac{1}{2}}\sum_{\mathbf{q}\in \mathscr{A}} \omega^{L(\mathbf{q})}\tau^{Q(\mathbf{q})}|\mathbf{q}\rangle$ where $\mathscr{A}$ is some affine submodule of $\mathbb{Z}_d^{n}$ and $L,Q:\mathbb{Z}^n\to\mathbb{Z}$ are linear and quadratic polynomials, respectively. The unitary $\mathcal{B}_{\mathbf{R}}:|\mathbf{Q}\rangle \mapsto |\mathbf{Q}\mathbf{R}\rangle$ basically introduces, by a change of variables, the matrix $\mathbf{R}$ into the joint phases of $|\mathcal{S}\rangle^{\otimes 4}$, and the factor $\mathbf{R}^\top\mathbf{R} = -\mathbf{I}\operatorname{mod}d$ appears in the quadratic polynomial $Q(\mathbf{Q})$, mapping $\tau^{Q(\mathbf{Q})} \mapsto \tau^{-Q(\mathbf{Q})}$. The complex conjugate of the linear part can be obtained by some Pauli operator $\mathcal{P}(\mathcal{S})$. Together, $\mathcal{P}(\mathcal{S})\mathcal{B}_{\mathbf{R}}|\mathcal{S}\rangle^{\otimes 4} = |\mathcal{S}^\ast\rangle^{\otimes 4}$.

The number of required stabiliser copies is dictated by the number of quadratic terms needed to decompose $d-1$, so in special cases fewer than four copies can suffice. For example, in prime dimensions $d$ there always exist $a_1,a_2\in\mathbb{F}_d$ such that $a_1^2 + a^2_2 \equiv -1~(\operatorname{mod}d)$, so only two copies of $|\mathcal{S}\rangle$ suffice and we employ $\mathbf{R} = \bigl(\begin{smallmatrix}
    a_1 & a_2\\
    a_2 & -a_1
\end{smallmatrix}\bigr)\operatorname{mod}d$. If moreover $d\equiv 1~(\operatorname{mod}4)$, $-1$ is a quadratic residue modulo $d$, meaning that only one copy of $|\mathcal{S}\rangle$ suffices. Furthermore, by picking integers $a_1,\dots,a_4$ whose squares sum to any positive integer $m$ (not necessarily $d-1$), the unitary $\mathcal{B}_{\mathbf{R}}$ can more generally be seen as transforming the quadratic part $Q(\mathbf{q})$ of any quantum phase state into $mQ(\mathbf{q})$, a fact that might be of independent interest.

Our new Bell sampling method then proceeds in a very simple way: take $8$ copies of a stabiliser state $|\mathcal{S}\rangle$, apply $\mathcal{B}_{\mathbf{R}}$ onto $4$ out the $8$ copies to obtain $|\mathcal{S}\rangle^{\otimes 4}|\mathcal{S}^\ast\rangle^{\otimes 4}$ up to some Pauli operators, and carry on with the standard procedure of measuring each pair $|\mathcal{S}\rangle|\mathcal{S}^\ast\rangle$ in the Bell basis.  The result is four strings in $\mathbb{Z}_d^{2n}$. See \Cref{algo:bell_sampling} for a summary of the procedure. If we repeat this procedure again and take the difference between the outcomes, the four resulting strings $\mathbf{x}_1,\dots,\mathbf{x}_4\in\mathbb{Z}_d^{2n}$ are samples from the characteristic distribution $p_{\mathcal{S}}(\mathbf{x})$ which, as mentioned above, is uniformly distributed over the strings corresponding to the stabiliser group $\mathcal{S}$. We thus successfully sample four generators of the stabiliser group through this procedure. We formalise this in the next definition.
\begin{definition}[Skewed Bell difference sampling]
    Let $\mathcal{B}_{\mathbf{R}}\in((\mathbb{C}^d)^{\otimes n})^{\otimes 4}$ be unitary from {\rm \Cref{res:res1}} and let $\mathcal{U}_{\mathbf{R}}\in((\mathbb{C}^d)^{\otimes n})^{\otimes 8}$ be the unitary acting on $\bigotimes_{i\in[8]}|\mathbf{q}_i\rangle$ that applies $\mathcal{B}_{\mathbf{R}}$ to $|\mathbf{q}_2,\mathbf{q}_4,\mathbf{q}_6,\mathbf{q}_8\rangle$ and the identity to the remaining registers. The \emph{skewed Bell difference sampling} on $16$ quantum states in $(\mathbb{C}^d)^{\otimes n}$ is the projective measurement given by 
    \begin{align*}
        &\mathcal{U}_{\mathbf{R}}\otimes\mathcal{U}_{\mathbf{R}}\left(\bigotimes_{i\in[4]}\Pi_{\mathbf{x}_i}\right) \mathcal{U}_{\mathbf{R}}^\dagger\otimes\mathcal{U}_{\mathbf{R}}^\dagger\\
        \text{where}\qquad 
        \Pi_{\mathbf{x}_i} &= \sum_{\mathbf{y} \in\mathbb{Z}_d^{2n}}\ |\mathcal{W}_{\mathbf{y}}\rangle\langle\mathcal{W}_{\mathbf{y}}| \otimes |\mathcal{W}_{\mathbf{x}_i+\mathbf{y}}\rangle\langle\mathcal{W}_{\mathbf{x}_i+\mathbf{y}}|, \qquad\forall \mathbf{x}_1,\dots,\mathbf{x}_4\in\mathbb{Z}_d^{2n}.
    \end{align*}
\end{definition}
\begin{algorithm}[t]
\caption{Skewed Bell sampling}
\DontPrintSemicolon
\label{algo:bell_sampling}
\SetKwInput{KwPromise}{Promise}

    \KwIn{$8$ copies of $|\psi\rangle\in(\mathbb{C}^d)^{\otimes n}$.}
    Let $\mathcal{B}_{\mathbf{R}}:|\mathbf{Q}\rangle\mapsto |\mathbf{Q}\mathbf{R}\rangle$ where $\mathbf{Q} = [\mathbf{q}_1,\mathbf{q}_2,\mathbf{q}_3,\mathbf{q}_4]\in\mathbb{Z}_d^{n\times 4}$ and $\mathbf{R}\in\mathbb{Z}_d^{2n\times 4}$ is such that $\mathbf{R}^\top\mathbf{R} = -\mathbf{I}\operatorname{mod}d$, e.g., as in \Cref{res:res1}

    Measure $|\psi\rangle^{\otimes 4}\mathcal{B}_{\mathbf{R}}|\psi\rangle^{\otimes 4}$ in the Bell basis $\{\bigotimes_{i=1}^4 |\mathcal{W}_{\mathbf{x}_i}\rangle\langle \mathcal{W}_{\mathbf{x}_i}|\}_{\mathbf{x}_1,\dots,\mathbf{x}_4\in\mathbb{Z}_d^{2n}}$ (matching the $i$-th and $(i+4)$-th states) to obtain $\mathbf{x}_1,\mathbf{x}_2,\mathbf{x}_3,\mathbf{x}_4\in\mathbb{Z}_d^{2n}$
\end{algorithm}

More generally, our Bell difference sampling also has a useful interpretation if we are given copies of an arbitrary (not necessarily stabiliser) state $|\psi\rangle$ (see \Cref{algo:bell_sampling}).
\begin{result}
    Given $|\psi\rangle\in(\mathbb{C}^d)^{\otimes n}$, skewed Bell difference sampling on $|\psi\rangle^{\otimes 16}$ corresponds to sampling from the distribution
    \begin{align*}
        b_\psi(\mathbf{x}_1,\mathbf{x}_2,\mathbf{x}_3,\mathbf{x}_4) 
        = \sum_{\mathbf{Y}\in\mathbb{Z}_d^{2n\times 4}} \prod_{i=1}^4 p_\psi(\mathbf{x}_i + \mathbf{Y}_i) p_\psi((\mathbf{Y}\mathbf{R})_i), \qquad\text{where}\quad p_\psi(\mathbf{x}) \triangleq d^{-n}|\langle\psi|\mathcal{W}_{\mathbf{x}}|\psi\rangle|^2
    \end{align*}
    and $\mathbf{Y}_i$ and $(\mathbf{Y}\mathbf{R})_i$ denote the $i$-th column of $\mathbf{Y}$ and $\mathbf{Y}\mathbf{R}$.
    If $|\psi\rangle = |\mathcal{S}\rangle$ is a stabiliser state,
    \begin{align*}
        b_{\mathcal{S}}(\mathbf{x}_1,\mathbf{x}_2,\mathbf{x}_3,\mathbf{x}_4) = \begin{cases}
            d^{-4n} &\text{if}~\omega^{s_1}\mathcal{W}_{\mathbf{x}_1},\omega^{s_2}\mathcal{W}_{\mathbf{x}_2},\omega^{s_3}\mathcal{W}_{\mathbf{x}_3},\omega^{s_4}\mathcal{W}_{\mathbf{x}_4}\in\mathcal{S},\\
            0 &\text{otherwise}.
        \end{cases}
    \end{align*}
\end{result}
The above result is a direct generalisation of the distribution $q_\psi(\mathbf{x}) = \sum_{\mathbf{y}\in\mathbb{F}_2^{2n}} p_\psi(\mathbf{x}+\mathbf{y}) p_\psi(\mathbf{y})$ for qubits, the main difference being the presence of more factors involving the characteristic distribution and the matrix $\mathbf{R}$ which connects these factors in a non-trivial way. Notice that sampling four times from the distribution $q_\psi(\mathbf{x})$ on qubits can alternatively be written as $\sum_{\mathbf{Y}\in\mathbb{F}_2^{2n\times 4}}\prod_{i=1}^4 p_\psi(\mathbf{x}_i+\mathbf{Y}_i) p_\psi(\mathbf{Y}_i)$.


\paragraph*{Related work.} Recently, Hinsche, Eisert, and Carrasco~\cite{hinsche2025abelianstatehiddensubgroup} introduced a general measurement primitive called \emph{Fourier sampling}. They showed that, when applied to the stabiliser formalism, Fourier sampling reduces to the standard Bell (difference) sampling on qubits ($d=2$) and yields new measurement primitives for dimensions $d>2$ (their procedure is defined for prime dimensions but appears to be valid for all $d\geq 2$). More precisely, they introduced the POVM defined by the projectors $\Pi_{\mathbf{x}}^{\rm HEC} \triangleq d^{-2n}\sum_{\mathbf{y}\in\mathbb{Z}_d^{2n}} \omega^{[\mathbf{y},\mathbf{x}]} \mathcal{W}_{\mathbf{y}}^{\otimes D}$ acting on $((\mathbb{C}^d)^{\otimes n})^{\otimes D}$ and showed that the probability distribution behind this POVM has interesting and desirable properties like returning generators of the stabiliser group $\mathcal{S}$ when measuring copies of its stabiliser state $|\mathcal{S}\rangle$. Compared to our results, the POVM of~\cite{hinsche2025abelianstatehiddensubgroup} requires $D$ copies of a state $|\psi\rangle$, while our new Bell difference sampling subroutine only needs at most $8$ copies at a time ($4$ copies for prime dimensions). The two measurement primitives yield different distributions, so one might be preferable over the other depending on the application.

\subsection{Applications of Bell sampling on qudits}

Bell (difference) sampling has been applied to a plethora of problems~\cite{montanaro2017learning,gross2021schur,hangleiter2023bell,haug2023efficient,Haug2023scalable,grewal2023efficient,grewal2023improved,grewal2024pseudoentanglement,grewal2024agnostic}, almost all on qubits. By employing our generalised Bell (difference) sampling primitive, we manage to lift several of these results to qudits for all $d\geq 2$. These generalisations are often non-trivial since the distribution $b_\psi$ is slightly more involved than the corresponding distribution $q_\psi$ for qubits, given the presence of the matrix $\mathbf{R}$. Moreover, for general dimensions $d$ one loses several properties pertaining to subspaces that are normally taken for granted, and submodules must be analysed instead. See \Cref{tab:results} for a summary of our results and a comparison to previous works. In what follows, we assume a computational model in which classical operations over the ring $\mathbb{Z}_d$, measurements in the computational basis, and single and two-qudit gates from a universal gate set~\cite{Muthukrishnan2000multivalued,brylinski2002universal,Brennen2005criteria,brennen2006efficient} all require $O(1)$ time to be performed. 

\subsubsection{Stabiliser State Learning and Hidden Stabiliser Group Problem}

Stabiliser states are among the classes of states that can be efficiently identified~\cite{aaronson2008stabiliser,rotteler2009quantum,zhao2016fast,montanaro2017learning}, in contrast to arbitrary $n$-qubit states which require exponentially many copies (in $n$) to determine. 
The simplest and most efficient quantum algorithm for learning an unknown stabiliser state $|\mathcal{S}\rangle\in(\mathbb{C}^2)^{\otimes n}$ on qubits is due to~\cite{montanaro2017learning}, who employed Bell (difference) sampling to identity $|\mathcal{S}\rangle$ in $O(n^3)$ time using $O(n)$ copies of $|\mathcal{S}\rangle$ and making collective measurements across at most two copies of $|\mathcal{S}\rangle$ at a time. For qudits, we are only aware of the works~\cite{allcock2024beyond,hinsche2025abelianstatehiddensubgroup}. The former focuses on prime dimensions and proposes an algorithm based on techniques from hidden polynomial problems~\cite{decker2013polynomial,IVANYOS201773} to learn $n$-qudit stabiliser states $|\mathcal{S}\rangle\in(\mathbb{C}^d)^{\otimes n}$ in $O(n^3)$ time using $O(n)$ copies of $|\mathcal{S}\rangle$. On the other hand, the algorithm of \cite{hinsche2025abelianstatehiddensubgroup} applies to all dimensions and requires $O(nd)$ copies of $|\mathcal{S}\rangle$ and $O(n^3+nd)$ time.

The problem of identifying an unknown stabiliser state can be generalised via the \emph{Hidden Stabiliser Group Problem} recently introduced by  Hinsche, Eisert, and Carrasco~\cite{hinsche2025abelianstatehiddensubgroup}, who were inspired by the work of~\cite{bouland2025state}. Here one is given copies of a state $|\psi\rangle\in(\mathbb{C}^d)^{\otimes n}$ promised to satisfy $|\langle\psi|\mathcal{P}|\psi\rangle| \leq 1-\varepsilon$ for any Pauli $\mathcal{P}\in \mathscr{P}_d^n$ that does not leave $|\psi\rangle$ invariant up a power of $\omega$. The problem is then to identify the hidden group of Pauli operators such that $\mathcal{P}|\psi\rangle = \omega^s|\psi\rangle$ for some $s\in\mathbb{Z}_d$, which is called the \emph{unsigned stabiliser group} of $|\psi\rangle$, ${\operatorname{Weyl}}(|\psi\rangle) \triangleq \{\mathbf{x}\in\mathbb{Z}_d^{2n}: \mathcal{W}_{\mathbf{x}}|\psi\rangle = \omega^s|\psi\rangle~\text{for some}~s\in\mathbb{Z}_d\}$. When the hidden group ${\operatorname{Weyl}}(|\psi\rangle)$ is a stabiliser group, $\varepsilon = 1$ and the problem basically to identifying an unknown stabiliser state. By employing their new Fourier sampling procedure, \cite{hinsche2025abelianstatehiddensubgroup} gave a quantum algorithm for the Hidden Stabiliser Group Problem that uses $O(n\log d\max\{d,\varepsilon^{-1}\})$ copies of $|\psi\rangle$ and runs in polynomial time.

By employing our new Bell difference sampling procedure, we improve the results of~\cite{hinsche2025abelianstatehiddensubgroup}. As a simple corollary, we also obtain a quantum algorithm for identifying an unknown stabiliser state $|\mathcal{S}\rangle\in(\mathbb{C}^d)^{\otimes n}$ for all dimensions $d\geq 2$.
\begin{result}\label{res:hidden_stabiliser_group}
    For all $d\geq 2$ with prime decomposition $d = \prod_{i=1}^\ell p_i^{k_i}$ and largest prime factor $p_\ell$, there is a quantum algorithm that solves the Hidden Stabiliser Group Problem with high probability by identifying ${\operatorname{Weyl}}(|\psi\rangle)$ using $O\big(\frac{n}{\varepsilon}\min\{\sum_{i=1}^\ell k_i, \log{p_\ell}\}\big)$ copies of $|\psi\rangle\in(\mathbb{C}^d)^{\otimes n}$ in $O\big(\frac{n^3}{\varepsilon}\min\{\sum_{i=1}^\ell k_i, \log{p_\ell}\}\big)$ time. The quantum algorithm acts on $8$ copies of $|\psi\rangle$ at a time.
\end{result}
\begin{result}\label{res:res3}
    For $d\geq 2$, there is a quantum algorithm that identifies an unknown stabiliser state $|\mathcal{S}\rangle\in(\mathbb{C}^d)^{\otimes n}$ in $O(n^3)$ time using $O(n)$ copies of $|\mathcal{S}\rangle$ and acting on $\leq 8$ copies of $|\psi\rangle$ at a time.
\end{result}

\Cref{res:hidden_stabiliser_group} improves upon the sample complexity $O(n\log{d}\max\{d,\varepsilon^{-1}\})$ of Hinsche, Eisert, and Carrasco when $\varepsilon = \Omega(d^{-1})$. Moreover, our algorithm only acts on at most $8$ copies of $|\psi\rangle$ at a time, whereas~\cite{hinsche2025abelianstatehiddensubgroup} requires $D$ copies, which can be quite demanding for large dimensions. Our algorithm is similar to the one from~\cite{hinsche2025abelianstatehiddensubgroup} (and~\cite{montanaro2017learning} for that matter): we gather enough samples from the skewed Bell distribution $b_\psi$ which we can show is supported on the orthogonal complement of ${\operatorname{Weyl}}(|\psi\rangle)$ (a rigorous definition is left to \Cref{sec:scalar_symplectic_products}). Therefore, with high probability their span can be used to obtain ${\operatorname{Weyl}}(|\psi\rangle)$ by casting the matrix of samples into its Smith normal form in $O(n^3)$ time~\cite{storjohann1996near}. Finally, note that any algorithm for learning a stabiliser state requires $\Omega(n^2)$ time just to write the output.



\subsubsection{Stabiliser size property testing}

For qubits (or qudits over prime dimensions), the dimension of ${\operatorname{Weyl}}(|\psi\rangle)$ is called the \emph{stabiliser dimension}. Grewal, Iyer, Kretschmer, and Liang~\cite{grewal2023efficient} studied the problem of testing whether a given $n$-qubit state $|\psi\rangle\in(\mathbb{C}^2)^{\otimes n}$ either has stabiliser dimension at least $t$,  or else has fidelity less than $1-\varepsilon$ with all states with stabiliser dimension at least $t$, promised that one of these cases hold. Using Bell sampling, they proposed a quantum algorithm that tests stabiliser dimension with high probability when $\varepsilon < \frac{3}{8}$ in $O\big(\frac{n^3}{\varepsilon}\big)$ time and with $O\big(\frac{n}{\varepsilon}\big)$ copies of $|\psi\rangle$.

The concept of dimension of a submodule is not well defined in general, with concepts like rank and length serving as possible generalisations. We find that working with such concepts for ${\operatorname{Weyl}}(|\psi\rangle)$ over a general $d$ can be too cumbersome and instead simply define the \emph{stabiliser size} of $|\psi\rangle\in(\mathbb{C}^d)^{\otimes n}$ as the cardinality of ${\operatorname{Weyl}}(|\psi\rangle)$. We then propose a quantum algorithm for the problem of testing whether a given state has stabiliser size at least $d^t$ or is $\varepsilon$-far in fidelity from any state with stabiliser size at least $d^t$, promised that one of these two cases hold.
\begin{result}
    Let $d=\prod_{i=1}^\ell p_i^{k_i}$ be the prime decomposition of $d$ and $\varepsilon = O(d^{-2})$. Let $|\psi\rangle\in(\mathbb{C}^d)^{\otimes n}$ be a state promised to either have stabiliser size at least $d^t$ or be $\varepsilon$-far in fidelity from all such states. There is a quantum algorithm that distinguishes the two cases with high probability in $O\big(\frac{n^3}{\varepsilon}\sum_{i=1}^\ell k_i\big)$ time using $O\big(\frac{n}{\varepsilon}\sum_{i=1}^\ell k_i\big)$ copies of $|\psi\rangle$.
\end{result}
Our quantum algorithm is similar to that of~\cite{grewal2023efficient}, although its analysis is more involved since we are dealing with submodules. The main idea is to gather $O\big(\frac{n}{\varepsilon}\sum_{i=1}^\ell k_i\big)$ samples from the distribution $b_\psi$ through skewed Bell difference sampling and analyse the cardinality of the submodule generated by them. All the samples are within the orthogonal complement of ${\operatorname{Weyl}}(|\psi\rangle)$, which has size at most $d^{2n - t}$ if we are in the case when $|{\operatorname{Weyl}}(|\psi\rangle)| \geq d^t$. If, on the other hand, $|\psi\rangle$ is $\varepsilon$-far away from states with stabiliser size at least $d^t$, we then prove that, with high probability, the cardinality of the submodule generated by the samples is greater than $d^{2n-t}$. Therefore, we need only to check whether the samples generate a submodule of size smaller or greater than $d^{2n-t}$, which can be done by putting the matrix of all samples into its Smith normal form in $O\big(\frac{n^3}{\varepsilon}\sum_{i=1}^\ell k_i\big)$ time. We note that, just like in~\cite{grewal2023efficient}, our result is valid for \emph{all} choices of $t$ in $d^t$. Increasing the range of $\varepsilon = O(d^{-2})$ is left as an open problem.

\subsubsection{Testing doped Clifford circuits and pseudorandomness lower bounds}

Pseudorandom states were introduced by Ji, Liu, and Song~\cite{ji2018pseudorandom} and have attracted a lot of attention due to their applicability in quantum cryptography and complexity theory~\cite{ji2018pseudorandom,kretschmer2021quantum,ananth2022cryptography,morimae2022quantum,hhan2023hardness,kretschmer2023quantum}. At a high level, a set of quantum states is pseudorandom if they mimic the Haar measure over $n$-qudit states and can be prepared by some efficient, $\operatorname{poly}(n)$-time, quantum algorithm (a formal definition is left to \Cref{sec:pseudorandomness}).

A few works have explored the resources required for constructing pseudorandom states, e.g.,~\cite{aaronson2022quantum} explored possible constructions of pseudorandom states using limited entanglement, while~\cite{haug2023pseudorandom} defined and studied several other pseudorandom resources. Grewal, Iyer, Kretschmer, and Liang~\cite{grewal2022low,grewal2023improved}, on the other hand, analysed pseudorandomness from the perspective of stabiliser complexity. More specifically, they proposed a quantum algorithm to test whether a quantum state is Haar-random or has non-negligible stabiliser fidelity. As a consequence, they proved that any $t$-doped Clifford circuit (a circuit with at most $t$ non-Clifford single-qubit gates) with $t=O(\log{n})$  cannot generate a set of $n$-qubit pseudorandom quantum states~\cite{grewal2022low}. This bound was later improved to $t = \frac{n}{2}$~\cite{grewal2023improved} by employing Bell difference sampling to distinguish Haar-random states from doped Clifford circuit outputs. 

Regarding qudits, \cite{allcock2024beyond} generalised the work of~\cite{grewal2022low} by employing the stabiliser testing POVM of~\cite{gross2021schur}  to efficiently distinguish $t$-doped Clifford circuits from Haar-random states when $t = O(\log_d{n})$. As a consequence, they proved that $O(\log_d{n})$-doped Clifford circuits cannot generate pseudorandom quantum states. This bound was later re-obtained by~\cite{hinsche2025abelianstatehiddensubgroup} as a consequence of learning (not testing) the output state of a $O(\log_d{n})$-doped Clifford circuit.

Here, by employing our skewed Bell difference sampling, we exponentially improve the pseudorandomness bounds on qudits of~\cite{allcock2024beyond,hinsche2025abelianstatehiddensubgroup} from $O(\log_d{n})$ to $\frac{n}{2}$, thus matching the bound of~\cite{grewal2023improved} on qubits.
\begin{result}
    Let $|\psi\rangle\in(\mathbb{C}^d)^{\otimes n}$ be a state promised to either be Haar-random or the output of a $t$-doped Clifford circuit for $t < \frac{n}{2}$. There is a quantum algorithm that distinguishes between the two cases with high probability in $O(n^3)$ time and using $O(n)$ copies of $|\psi\rangle$. As a consequence, any doped Clifford circuit in $(\mathbb{C}^d)^{\otimes n}$ that uses less than $\frac{n}{2}$ non-Clifford single-qudit gates cannot produce an ensemble of pseudorandom quantum states.
\end{result}

Our algorithm works by sampling $O(n)$ strings from the distribution $b_\psi$ using skewed Bell difference sampling and once again studying the cardinality of their generated submodule. If $|\psi\rangle$ is Haar-random, we prove that the $O(n)$ samples generate the whole space $\mathbb{Z}_d^{2n}$ with very high probability, which cannot happen if $|\psi\rangle$ is the output of a $t$-doped Clifford circuit with $t < \frac{n}{2}$, since the samples coming from $b_\psi$ are supported on a submodule or size at most $d^{n+2t}$.


\subsubsection{Tolerant Stabiliser Testing}

Tolerant stabiliser testing is the problem of deciding whether a given quantum state $|\psi\rangle\in(\mathbb{C}^d)^{\otimes n}$ has \emph{stabiliser fidelity}~\cite{bravyi2019simulation}, defined as the maximum overlap between $|\psi\rangle$ and any stabiliser state $|\mathcal{S}\rangle$ according to $F_{\mathcal{S}}(|\psi\rangle) \triangleq \max_{|\mathcal{S}\rangle}|\langle\mathcal{S}|\psi\rangle|^2$, at least $1-\varepsilon_1$ or at most $1-\varepsilon_2$ for $\varepsilon_1 < \varepsilon_2$.

The non-tolerant version of stabiliser testing with $\varepsilon_1=0$ has been successfully addressed by~\cite{gross2021schur}, who proposed efficient testing algorithms for all dimensions $d\geq 2$. Their testing algorithm for \emph{qubits} is very simple: one obtains $\mathbf{x}\in\mathbb{F}_2^{2n}$ via Bell difference sampling, measures the corresponding Weyl operator $\mathcal{W}_{\mathbf{x}}$ twice on two independent copies of $|\psi\rangle$, and checks whether the outcomes are equal or not. Both outcomes will always be equal if $|\psi\rangle$ is a stabiliser state, but might differ with some probability is $|\psi\rangle$ has stabiliser fidelity at most $1-\varepsilon_2$. By repeating this procedure $O(\frac{1}{\varepsilon_2})$ times, one can distinguish both cases with high probability. Their algorithm for higher dimensions $d>2$, however, does not have a simple interpretation. By introducing an ad hoc binary POVM with the desired property of always returning $+1$ if and only if $|\psi\rangle$ is a stabiliser state, the authors showed that $O\big(\frac{d^2}{\varepsilon_2}\big)$ measurements are enough to distinguish between the two cases with high probability.

The tolerant version of stabiliser testing, in contrast with the $\varepsilon_1=0$ case, is still poorly understood and has received a lot of attention recently, all focused on qubits~\cite{grewal2023improved,arunachalam2024notepolynomial,hinsche2025single,iyer2024tolerant,bao2025tolerant,arunachalam2025polynomial,mehraban2025improved,chen2025stabilizer}. Grewal, Iyer, Kretschmer, and Liang~\cite{grewal2023improved} analysed the testing protocol of~\cite{gross2021schur} under the tolerant setting and proved that $O\big(\frac{1}{(\varepsilon_2-\varepsilon_1)^2}\big)$ copies of $|\psi\rangle\in(\mathbb{C}^2)^{\otimes n}$ suffices to determine whether $F_{\mathcal{S}}(|\psi\rangle) \geq 1-\varepsilon_1$ or $F_{\mathcal{S}}(|\psi\rangle) \leq 1-\varepsilon_2$. Their algorithm works as long as $\varepsilon_1 \leq 1 - \big(1-\frac{3\varepsilon_2}{4}\big)^{1/6} \leq 1 - (\frac{1}{2})^{1/3}$. Follow-up works~\cite{arunachalam2024notepolynomial,bao2025tolerant,arunachalam2025polynomial,mehraban2025improved,chen2025stabilizer} extended the range of $\varepsilon_1$ to $1 - \varepsilon_2 \leq (1-\varepsilon_1)^C$ for some universal constant $C>1$ while still requiring $\operatorname{poly}((1-\varepsilon_1)^{-1})$ samples (although the current polynomial dependence can be quite bad, sometimes to the power of hundreds). Finally,~\cite{iyer2024tolerant} considered tolerant testing algorithms that accept mixed state inputs.

Here we initiate the study of tolerant testing algorithms for qudits of all dimensions $d\geq 2$. For such, we propose two different algorithms to decide whether $|\psi\rangle\in(\mathbb{C}^d)^{\otimes n}$ has stabiliser fidelity at least $1-\varepsilon_1$ or at most $1-\varepsilon_2$. Our first algorithm is a simple extension of the stabiliser testing protocol of~\cite{gross2021schur}, while our second algorithm works similarly to the qubit case of~\cite{gross2021schur}: one obtains a string $\mathbf{x}\in\mathbb{Z}_d^{2n}$ via skewed Bell difference sampling, measures the corresponding operators $\mathcal{W}_{\mathbf{x}}\otimes\mathcal{W}_{\mathbf{x}}^\dagger + \mathcal{W}_{\mathbf{x}}^\dagger\otimes\mathcal{W}_{\mathbf{x}}$ on two independent copies of $|\psi\rangle$, and checks whether the outcome is $1$ or not. For both of our algorithms, repeated measurements allow us to estimate whether $F_{\mathcal{S}}(|\psi\rangle)$ is greater than $1-\varepsilon_1$ or smaller than $1-\varepsilon_2$.
%
%
\begin{result}
    Let $|\psi\rangle \in (\mathbb{C}^d)^{\otimes n}$ and $\varepsilon_2 > \varepsilon_1 \geq 0$ such that $\varepsilon_1 = O\big(\frac{\varepsilon_2}{d^{2}}\big)$. There is a quantum algorithm that decides whether $F_{\mathcal{S}}(|\psi\rangle) \geq 1 - \varepsilon_1$ or $F_{\mathcal{S}}(|\psi\rangle) \leq 1 - \varepsilon_2$ with high probability using $O\big(\frac{d^2\varepsilon_2}{(\varepsilon_2-\varepsilon_1)^2}\big)$ copies of $|\psi\rangle$.
\end{result}

The range of parameters $\varepsilon_1,\varepsilon_2$ and sample complexities of both of our algorithms are asymptotically similar. As numerically shown in \Cref{sec:stabiliser_testing}, however, overall the POVM-based algorithm presents better range of parameters, while the Bell-sampling-based one has a better sample complexity for smaller values of $\varepsilon_1$. Finally, we notice that for $d=2$, the sample complexity of our algorithms is $O\big(\frac{\varepsilon_2}{(\varepsilon_2-\varepsilon_1)^2}\big)$, which is better than~\cite{grewal2023improved} by a factor of $O(\varepsilon_2)$.


\begin{table}[t!]
    \def\arraystretch{1.85}
    \centering
    \resizebox{\linewidth}{!}{
    \begin{tabular}{|c|c|c|c|c|c|}
        \hline
         Problem & Dim.\ $d$ & Work & Sample Complexity & Time complexity & Remarks  \\ \hline\hline
         \multirow{6}{*}{\makecell{Stabiliser \\ learning}} & \multirow{3}{*}{$d=2$} & \cite{aaronson2008stabiliser} & $O(n^2)$ & $O(n^3)$ & Operations over $1$ copy \\ \cline{3-6}
         & & \cite{zhao2016fast} & $O(n)$ & $O(n^3)$ & \makecell{For graph states \\ Operations over $2$ copies} \\ \cline{3-6}
         & & \cite{montanaro2017learning} & $O(n)$ & $O(n^3)$ & Operations over $2$ copies \\ \cline{2-6}
         & \makecell{$d\geq 2$ \\ prime} & \cite{allcock2024beyond} & $O(n)$ & $O(n^3)$ & Operations over $3$ copies \\ \cline{2-6}
         & \multirow{2}{*}{$d\geq 2$} & \cite{hinsche2025abelianstatehiddensubgroup} & $O(nd\log{d})$ & $O(n^3\log{d} + nd\log{d})$ & Operations over $D$ copies \\ \cline{3-6}
         &  & \cellcolor{Gray} \Cref{thr:algorithm_bell_sampling} & \cellcolor{Gray} $O(n)$ & \cellcolor{Gray} $O(n^3)$ & \cellcolor{Gray} Operations over $8$ copies \\ \hline

         \multirow{2}{*}{\makecell{Hidden \\ Stabiliser \\ Group}} & \multirow{2}{*}{$d\geq 2$} & \cite{hinsche2025abelianstatehiddensubgroup} & $O(n\max\{d,\varepsilon^{-1}\}\log{d})$ & $\!{O}\big((n+\frac{n^3}{d}) \max\{d,\frac{1}{\varepsilon}\}\log{d}\big)\!$ & Operations over $D$ copies \\ \cline{3-6}
         & & \cellcolor{Gray} \Cref{thr:hidden_stabiliser_group} & \cellcolor{Gray} $O((n/\varepsilon)\min\{\sum_{i=1}^\ell k_i, \log{p_\ell}\})$ & \cellcolor{Gray} $O((n^3/\varepsilon)\min\{\sum_{i=1}^\ell k_i, \log{p_\ell}\})$ & \cellcolor{Gray} Operations over $8$ copies \\ \hline
         
         \multirow{2}{*}{\makecell{Stabiliser \\ size \\ testing}} & $d=2$ & \cite{grewal2023efficient} & $O(n/\varepsilon)$ & $O(n^3/\varepsilon)$ & $\varepsilon < 3/8$ \\ \cline{2-6}
         & $d\geq 2$ & \cellcolor{Gray} \Cref{thr:property_testing_stabiliser_size} & \cellcolor{Gray} $O\big((n/\varepsilon)\sum_{i=1}^\ell k_i\big)$ & \cellcolor{Gray} $O\big((n^3/\varepsilon)\sum_{i=1}^\ell k_i\big)$ & \cellcolor{Gray} $\varepsilon = O(d^{-2})$ \\ \hline 
         
         \multirow{6}{*}{\makecell{$t$-doped \\ Clifford \\ testing}} & \multirow{3}{*}{$d=2$} & \cite{grewal2022low} & $\operatorname{poly}(n)$ if $t=O(\log{n})$ & $\operatorname{poly}(n)$ if $t=O(\log{n})$ & \makecell{$t$-doped Cliffords for $t = O(\log{n})$ \\ cannot generate pseudorandom states; \\ Operations over $2$ copies} \\ \cline{3-6}
         & & \cite{grewal2023improved} & $O(n)$ & $O(n^3)$ & \makecell{$t$-doped Cliffords for $t < n/2$ \\ cannot generate pseudorandom states; \\ Operations over $2$ copies} \\ \cline{2-6}
         & \makecell{$d\geq 2$ \\ prime} & \cite{allcock2024beyond} & $O(d^{4t})$ & $O(nd^{4t})$ & \makecell{$t$-doped Cliffords for $t = O(\log_d{n})$ \\ cannot generate pseudorandom states; \\ Operations over $2$ copies} \\ \cline{2-6}
         & $d\geq 2$ & \cellcolor{Gray} \Cref{thr:algorithm_pseudorandomness} & \cellcolor{Gray} $O(n)$ & \cellcolor{Gray} $O(n^3)$ & \cellcolor{Gray} \Gape[0pt][2pt]{\makecell{$t$-doped Cliffords for $t < n/2$ \\ cannot generate pseudorandom states; \\ Operations over $8$ copies}} \\ \hline
         
         \multirow{6}{*}{\makecell{Tolerant \\ stabiliser \\ testing}} & \multirow{3}{*}{$d=2$} & \cite{grewal2023improved} & $O(1/(\varepsilon_2-\varepsilon_1)^2)$ & $O(n/(\varepsilon_2-\varepsilon_1)^2)$ & $\varepsilon_1 \leq 1 - (1-3\varepsilon_2/4)^{1/6}$ \\ \cline{3-6}
         & & \cite{arunachalam2025polynomial,bao2025tolerant,mehraban2025improved} & $1 / \operatorname{poly}(1-\varepsilon_1)$ & $n / \operatorname{poly}(1-\varepsilon_1)$ & $1-\varepsilon_2 \leq (1-\varepsilon_1)^{O(1)}$ \\ \cline{3-6}
         & & \cite{iyer2024tolerant} & $1 / \operatorname{poly}(1-\varepsilon_1)$ & $n / \operatorname{poly}(1-\varepsilon_1)$ & \makecell{$1-\varepsilon_2 \leq (1-\varepsilon_1)^{O(1)}$ \\ Accepts mixed state inputs} \\ \cline{2-6}
         & \multirow{4}{*}{$d\geq 2$} & \cite{gross2021schur} & $O(d^2/\varepsilon_2)$ & $O(n d^2/\varepsilon_2)$ & \makecell{$\varepsilon_1 = 0$ \\ Operations over $2r$ copies, $\operatorname{gcd}(d,r) = 1$} \\ \cline{3-6}
         & & \cellcolor{Gray} \Cref{thr:tolerant_stabiliser_testing_1} & \cellcolor{Gray} \Gape[0pt][2pt]{$O(d^2\varepsilon_2/(\varepsilon_2-\varepsilon_1)^2)$} & \cellcolor{Gray} $O(n d^2\varepsilon_2/(\varepsilon_2-\varepsilon_1)^2)$ & \cellcolor{Gray} \makecell{$\varepsilon_1 = O(\varepsilon_2/d^{2})$ \\ Operations over $2r$ copies, $\operatorname{gcd}(d,r)=1$} \\ \cline{3-6}
         & & \cellcolor{Gray} \Cref{thr:tolerant_stabiliser_testing_2} & \cellcolor{Gray} $O(d^2\log\log(d)\varepsilon_2/(\varepsilon_2-\varepsilon_1)^2)$ & \cellcolor{Gray} $O(n d^2\log\log(d)\varepsilon_2/(\varepsilon_2-\varepsilon_1)^2)$ & \cellcolor{Gray} \Gape[0pt][2pt]{\makecell{$\varepsilon_1 = \widetilde{O}(\varepsilon_2/d^{2})$ \\ Operations over $8$ copies}} \\ \hline
    \end{tabular}
    }
    \caption{A comparison of our results with several known results in the literature for applications of Bell sampling. All results assume a constant success probability. In the remarks column, \emph{operations over $m$ copies} indicates that quantum operations are performed on $m$-fold tensor products of the state under consideration. The prime decomposition of $d$ is $d=\prod_{i=1}^\ell p_i^{k_i}$ with $p_1<p_2<\cdots<p_\ell$.}
    \label{tab:results}
\end{table}

\section{Preliminaries}

Given $n\in\mathbb{N} \triangleq \{1,2,\dots\}$, let $[n] \triangleq \{1,\dots, n\}$. Throughout this paper, let $d\in\mathbb{N}$ be $d\geq 2$. Let $\omega \triangleq \e^{\ri 2\pi /d}$ be the standard primitive $d$-th root of unity and $\tau \triangleq (-1)^{d}\e^{\ri \pi/d} = \e^{\ri \pi(d^2+1)/d}$. Note that $\tau^2 = \omega$. Let $D$ be the order of $\tau$. Then $D = d$ if $d$ is odd and $D = 2d$ is $d$ is even. Given a group $\mathcal{G}$ and $g_1,\dots,g_k\in\mathcal{G}$, let $\langle g_1,\dots,g_k\rangle$ be the subgroup of $\mathcal{G}$ generated by $g_1,\dots,g_k$. In this paper, we shall be concerned with modules over ring $\mathbb{Z}_d \triangleq \mathbb{Z}/d\mathbb{Z} \cong \{0,1,\dots,d-1\}$ of integers modulo $d$. We refer the reader to \Cref{app:modules} for more information. 

For any state $|\psi\rangle = \sum_{\mathbf{w}\in\mathbb{Z}_d^n}a_{\mathbf{w}}|\mathbf{w}\rangle$ in $(\mathbb{C}^{d})^{\otimes n}$, let $|\psi^\ast\rangle = \sum_{\mathbf{w}\in\mathbb{Z}_d^n}\overline{a_{\mathbf{w}}}|\mathbf{w}\rangle$ denote its complex conjugate with respect to the computational basis. We shall often write $\psi$ to denote $|\psi\rangle\langle\psi|$. Let $\operatorname{vec}$ be the vectorisation operator defined by $\operatorname{vec}(|\mathbf{v}\rangle\langle \mathbf{w}|) = |\mathbf{v}\rangle|\mathbf{w}\rangle$ for computational basis states $\mathbf{v},\mathbf{w}\in\mathbb{Z}_d^n$. Note that $\operatorname{vec}$ preserves inner products, i.e., $\langle \operatorname{vec}(\mathbf{A})|\operatorname{vec}(\mathbf{B})\rangle = \operatorname{Tr}[\mathbf{A}^\dagger \mathbf{B}]$ for all $\mathbf{A},\mathbf{B}\in\mathbb{Z}_d^{m\times n}$. Finally, $\mathbf{1}[\cdot]$ is the indicator function such that $\mathbf{1}[\text{statement}]$ equals $1$ if the statement is true and $0$ if it is false.

\subsection{Scalar product modules and symplectic modules}
\label{sec:scalar_symplectic_products}

For any $L\in\mathbb{N}\setminus\{1\}$, define the \emph{scalar product} $\langle \cdot ,\cdot\rangle_L:\mathbb{Z}_d^n \times \mathbb{Z}_d^n\to \mathbb{Z}$, $(\mathbf{v},\mathbf{w})\mapsto \sum_i v_i w_i \bmod L$ and the \emph{symplectic product} $[\cdot,\cdot]_L : \mathbb{Z}^{2n}_d \times \mathbb{Z}^{2n}_d \to \mathbb{Z}$, $(\mathbf{x},\mathbf{y}) = \sum_{i=1}^n (x_i y_{n+i} - x_{n+i}y_i) \bmod L$. Similarly, $\langle \cdot ,\cdot\rangle:\mathbb{Z}_d^n \times \mathbb{Z}_d^n\to \mathbb{Z}$, $(\mathbf{v},\mathbf{w})\mapsto \sum_i v_i w_i$ and $[\cdot,\cdot] : \mathbb{Z}^{2n}_d \times \mathbb{Z}^{2n}_d \to \mathbb{Z}$, $(\mathbf{x},\mathbf{y}) = \sum_{i=1}^n (x_i y_{n+i} - x_{n+i}y_i)$. We consider the module $\mathbb{Z}_d^n$ equipped with $\langle \cdot ,\cdot \rangle_d$ and the module $\mathbb{Z}_d^{2n}$ equipped with $[\cdot,\cdot]_d$. We shall also consider the module $\mathbb{Z}_D^{2n}$ equipped with $[\cdot, \cdot]_D$. We shall often implicitly embed $\mathbb{Z}_d^{2n}$ into $\mathbb{Z}_D^{2n}$ when operating with elements from both modules together. The products $\langle \cdot, \cdot \rangle_L$ and $[\cdot,\cdot]_L$ are both \emph{non-degenerate}, meaning that the zero module element $\mathbf{0}$ is the only element orthogonal to every element in the module.

\begin{definition}[Involution]
    The involution $J:\mathbb{Z}_d^{2n}\to\mathbb{Z}_d^{2n}$ on the symplectic module $\mathbb{Z}_d^{2n}$ is defined as $J(\mathbf{x}) = J((\mathbf{v},\mathbf{w})) = (-\mathbf{v},\mathbf{w})$ where $\mathbf{x} = (\mathbf{v},\mathbf{w})$ and $\mathbf{v},\mathbf{w}\in\mathbb{Z}_d^n$. 
\end{definition}

Let $\mathbf{e}_i\in\mathbb{Z}^n_d$ be the module element whose $i$-th component is $1$ and the remaining components are $0$. Let $\operatorname{vec}$ be the vectorisation operator defined by $\operatorname{vec}(|\mathbf{v}\rangle\langle \mathbf{w}|) = |\mathbf{v}\rangle|\mathbf{w}\rangle$ for computational basis states $\mathbf{v},\mathbf{w}\in\mathbb{Z}_d^n$. Note that $\operatorname{vec}$ preserves inner products, i.e., $\langle \operatorname{vec}(\mathbf{A})|\operatorname{vec}(\mathbf{B})\rangle = \operatorname{Tr}[\mathbf{A}^\dagger \mathbf{B}]$ for all $\mathbf{A},\mathbf{B}\in\mathbb{Z}_d^{m\times n}$. Given a matrix $\mathbf{M}\in\mathbb{Z}_d^{m\times n}$, let 
\begin{align*}
    \operatorname{col}(\mathbf{M}) &\triangleq \{\mathbf{M}\mathbf{v}\in\mathbb{Z}_d^m : \mathbf{v}\in\mathbb{Z}_d^n \} &&\text{be its column submodule}, \\
    \operatorname{row}(\mathbf{M}) &\triangleq \operatorname{col}(\mathbf{M}^\top) &&\text{be its row submodule},\\
    \operatorname{null}(\mathbf{M}) &\triangleq \{\mathbf{v}\in\mathbb{Z}_d^n: \mathbf{M}\mathbf{v} = \mathbf{0}\} &&\text{be its null submodule}.
\end{align*}

Given a submodule $\mathscr{V}\subseteq\mathbb{Z}_d^{n}$ of the scalar product module $\mathbb{Z}_d^n$, its \emph{orthogonal complement} is $\mathscr{V}^\perp \triangleq \{\mathbf{w}\in\mathbb{Z}_d^n: \langle\mathbf{v}, \mathbf{w}\rangle = 0, \forall \mathbf{v}\in\mathscr{V}\}$. Given a submodule $\mathscr{X}\subseteq\mathbb{Z}_d^{2n}$ of the symplectic module $\mathbb{Z}_d^{2n}$, its \emph{symplectic complement} is $\mathscr{X}^{\independent} \triangleq \{\mathbf{y}\in\mathbb{Z}_d^{2n}: [\mathbf{x}, \mathbf{y}] = 0, \forall \mathbf{x}\in\mathscr{X}\}$. We shall use the following facts about orthogonal and symplectic complements throughout.
\begin{fact}
    Let $\mathscr{V},\mathscr{W}\subseteq \mathbb{Z}_d^n$ and $\mathscr{X},\mathscr{Y}\subseteq \mathbb{Z}_d^{2n}$ be submodules of the scalar product and symplectic modules $\mathbb{Z}_d^n$ and $\mathbb{Z}_d^{2n}$, respectively. Then
    \begin{multicols}{2}
    \begin{enumerate}[label=\alph*.]
        \item $\mathscr{V}^\perp$ is a submodule;
        \item $(\mathscr{V}^\perp)^\perp = \mathscr{V}$;
        \item $|\mathscr{V}||\mathscr{V}^\perp| = d^n$;
        \item $\mathscr{V}\subseteq \mathscr{W} \iff \mathscr{W}^\perp \subseteq \mathscr{V}^\perp$;
        \item $(\mathscr{V} + \mathscr{W})^\perp = \mathscr{V}^\perp \cap \mathscr{W}^\perp$;
        \item $\mathscr{X}^{\independent}$ is a submodule;
        \item $(\mathscr{X}^{\independent})^{\independent} = \mathscr{X}$;
        \item $|\mathscr{X}||\mathscr{X}^{\independent}| = d^{2n}$;
        \item $\mathscr{X}\subseteq \mathscr{Y} \iff \mathscr{Y}^{\independent} \subseteq \mathscr{X}^{\independent}$;
        \item $(\mathscr{X} + \mathscr{Y})^{\independent} = \mathscr{X}^{\independent} \cap \mathscr{Y}^{\independent}$.
    \end{enumerate}
    \end{multicols}
\end{fact}

The concept of isotropic and Lagrangian submodules will be fundamental to our analysis.
\begin{definition}
    A submodule $\mathscr{X}\subset \mathbb{Z}_d^{2n}$ of a symplectic module $\mathbb{Z}_d^{2n}$ is \emph{isotropic} if $[\mathbf{x},\mathbf{y}] = 0$ for all $\mathbf{x},\mathbf{y}\in\mathscr{X}$, and is \emph{Lagrangian} if $\mathscr{X}^{\independent} = \mathscr{X}$. 
\end{definition}
\begin{fact}
    Every Lagrangian submodule of $\mathbb{Z}_d^{2n}$ has size $d^n$. Every isotropic submodule of a symplectic module can be extended to a Lagrangian one.
\end{fact}
\begin{fact}
    For any isotropic submodule $\mathscr{X}\subset\mathbb{Z}_d^{2n}$, $\mathscr{X}\subseteq\mathscr{X}^{\independent}$.
\end{fact}

The following fact will be useful.

\begin{lemma}\label{lem:sum}
    Let $\mathscr{V}\subseteq\mathbb{Z}_d^n$ and $\mathscr{X}\subseteq\mathbb{Z}_d^{2n}$ be submodules and $\mathbf{w} \in\mathbb{Z}_d^n$ and $\mathbf{y} \in\mathbb{Z}_d^{2n}$. Then
    \begin{align*}
        \sum_{\mathbf{v}\in\mathscr{V}} \omega^{\langle\mathbf{v}, \mathbf{w}\rangle} = \begin{cases}
            |\mathscr{V}| &\text{if}~\mathbf{w} \in \mathscr{V}^\perp,\\
            0 &\text{if}~ \mathbf{w}\notin \mathscr{V}^\perp,
        \end{cases}\qquad\qquad
        \sum_{\mathbf{x}\in\mathscr{X}} \omega^{[\mathbf{x}, \mathbf{y}]} = \begin{cases}
            |\mathscr{X}| &\text{if}~\mathbf{y} \in \mathscr{X}^{\independent},\\
            0 &\text{if}~ \mathbf{y}\notin \mathscr{X}^{\independent}.
        \end{cases}
    \end{align*}
\end{lemma}

\subsection{Symplectic Fourier analysis}

Boolean Fourier analysis~\cite{de2008brief,o2014analysis} is a well-established field wherein a Boolean function is studied through its Fourier expansion using the characters $\omega^{\langle \mathbf{v}, \cdot \rangle}$ defined with respect to the scalar product. A character of $\mathbb{Z}_d^{2n}$ is a group homomorphism $\chi:\mathbb{Z}_d^{2n} \to \mathbb{C}$ such that $\chi(\mathbf{x} + \mathbf{y}) = \chi(\mathbf{x})\chi(\mathbf{y})$ for all $\mathbf{x},\mathbf{y}\in\mathbb{Z}_d^{2n}$. It is a standard result that every character of the symplectic module $\mathbb{Z}_d^{2n}$ can be written as $\omega^{[\mathbf{x},\cdot]}$ for some $\mathbf{x}\in\mathbb{Z}_d^{2n}$ (see~\cite[Appendix~IX.C]
{gross2006hudson}). In this paper, we shall work with \emph{symplectic} Fourier analysis, which is similar to the usual Boolean Fourier analysis but the Fourier characters are instead defined with respect to the symplectic product as $\omega^{[\mathbf{x}, \cdot]}$. 
\begin{definition}
    The symplectic Fourier transform $\widehat{f}:\mathbb{Z}_d^{2n}\to\mathbb{C}$ of $f:\mathbb{Z}_d^{2n}\to\mathbb{C}$ is
    \begin{align*}
        \widehat{f}(\mathbf{y}) \triangleq \frac{1}{d^{2n}}\sum_{\mathbf{x}\in\mathbb{Z}_d^{2n}}\omega^{[\mathbf{y},\mathbf{x}]}f(\mathbf{x}).
    \end{align*}
\end{definition}
Using \cref{lem:sum}, one can expand any function $f:\mathbb{Z}_d^{2n}\to\mathbb{C}$ using its symplectic Fourier transform as
\begin{align*}
    f(\mathbf{x}) = \sum_{\mathbf{y}\in\mathbb{Z}_d^{2n}}\omega^{[\mathbf{x},\mathbf{y}]}\widehat{f}(\mathbf{y}). 
\end{align*}
The transformation $f\mapsto \widehat{f}$ is unitary, meaning that (a generalised) Parseval's identity holds.
\begin{lemma}[Parseval's identity]\label{lem:parseval}
    Given $f,g:\mathbb{Z}_d^{2n}\to\mathbb{C}$ and $\mathbf{t}\in\mathbb{Z}_d^{2n}$, then
    \begin{align*}
        \frac{1}{d^{2n}}\sum_{\mathbf{x}\in\mathbb{Z}_d^{2n}} \omega^{[\mathbf{t},\mathbf{x}]} f(\mathbf{x})g(\mathbf{x}) &= \sum_{\mathbf{y}\in\mathbb{Z}_d^{2n}}\widehat{f}(\mathbf{y})\widehat{g}(\mathbf{t}-\mathbf{y}).
    \end{align*}
\end{lemma}
\begin{proof}
    $\begin{aligned}[t]
    \sum_{\mathbf{x}\in\mathbb{Z}_d^{2n}} \omega^{[\mathbf{t},\mathbf{x}]}f(\mathbf{x}) g(\mathbf{x}) &= \sum_{\mathbf{x},\mathbf{y},\mathbf{z}\in\mathbb{Z}_d^{2n}}\omega^{[\mathbf{x},\mathbf{y}+\mathbf{z}-\mathbf{t}]}\widehat{f}(\mathbf{y})\widehat{g}(\mathbf{z}) \\
    &= d^{2n}\sum_{\mathbf{y},\mathbf{z}\in\mathbb{Z}_d^{2n}}\widehat{f}(\mathbf{y})\widehat{g}(\mathbf{z})\cdot\mathbf{1}[\mathbf{z} = \mathbf{t}-\mathbf{y}] = d^{2n}\sum_{\mathbf{y}\in\mathbb{Z}_d^{2n}}\widehat{f}(\mathbf{y})\widehat{g}(\mathbf{t}-\mathbf{y}).
    \hspace{0.52cm}\qedhere
    \end{aligned}$
\end{proof}
The usual Parseval's identity follows from the above lemma by taking $\mathbf{t} = \mathbf{0}$ and $g(\mathbf{x}) = \overline{f(\mathbf{x})}$. We also define the convolution operation.
\begin{definition}[Convolution]
    Let $f,g:\mathbb{Z}_d^{2n}\to\mathbb{C}$. The convolution between $f$ and $g$ is the function $f\ast g:\mathbb{Z}_d^{2n}\to\mathbb{C}$ defined by
    \begin{align*}
        (f\ast g)(\mathbf{x}) \triangleq \frac{1}{d^{2n}}\sum_{\mathbf{y}\in\mathbb{Z}_d^{2n}}f(\mathbf{y})g(\mathbf{x} - \mathbf{y}) = \frac{1}{d^{2n}}\sum_{\mathbf{y}\in\mathbb{Z}_d^{2n}}f(\mathbf{x}-\mathbf{y})g(\mathbf{y}).
    \end{align*}
\end{definition}
\begin{lemma}\label{lem:convolution_Fourier}
    Let $f,g:\mathbb{Z}_d^{2n}\to\mathbb{C}$. Then, for all $\mathbf{x}\in\mathbb{Z}_d^{2n}$, $\widehat{f\ast g}(\mathbf{x}) = \widehat{f}(\mathbf{x})\widehat{g}(\mathbf{x})$.
\end{lemma}
\begin{proof}
    $\begin{aligned}[t]
        \widehat{f\ast g}(\mathbf{x}) &= \frac{1}{d^{2n}} \sum_{\mathbf{y}\in\mathbb{Z}_d^{2n}}\omega^{[\mathbf{x},\mathbf{y}]}(f\ast g)(\mathbf{y}) = \frac{1}{d^{4n}}\sum_{\mathbf{y},\mathbf{z}\in\mathbb{Z}_d^{2n}}\omega^{[\mathbf{x},\mathbf{y}]} f(\mathbf{z})g(\mathbf{y} - \mathbf{z}) \\
        &= \frac{1}{d^{4n}}\sum_{\mathbf{z}\in\mathbb{Z}_d^{2n}} \omega^{[\mathbf{x},\mathbf{z}]}f(\mathbf{z})\sum_{\mathbf{y}\in\mathbb{Z}_d^{2n}}\omega^{[\mathbf{x},\mathbf{y}-\mathbf{z}]}g(\mathbf{y}-\mathbf{z}) = \widehat{f}(\mathbf{x})\widehat{g}(\mathbf{x}). \hspace{3.3cm}\qedhere
    \end{aligned}$
\end{proof}

\subsection{Generalised Pauli group and Weyl operators}
\label{sec:Weyl_operators}

Define the unitary shift and clock operators $\mathsf{X}$ and $\mathsf{Z}$, respectively, as
\begin{align*}
    \mathsf{X}|q\rangle \triangleq |q+1\rangle, \quad \mathsf{Z}|q\rangle \triangleq \omega^q |q\rangle, \quad \forall q\in\mathbb{Z}_d.
\end{align*}
The $n$-qudit Pauli group over $\mathbb{Z}_d$ (also known as Weyl-Heisenberg group) is the group generated by tensor products of clock and shift operators, together with powers of $\tau$,
\begin{align*}
    \mathscr{P}^n_{d} \triangleq \langle \tau\mathbf{I}, \mathsf{X}, \mathsf{Z}\rangle^{\otimes n}. 
\end{align*}
The corresponding $n$-qudit Clifford group $\mathscr{C}^n_{d}$ is defined as the normaliser of the Pauli group in the unitary group,
\begin{align*}
    \mathscr{C}^n_{d} \triangleq \{\mathcal{U}\in\mathbb{C}^{d^n\times d^n}: \mathcal{U}^\dagger\mathcal{U} = \mathbf{I},~ \mathcal{U}\mathscr{P}^n_{d}\mathcal{U}^\dagger = \mathscr{P}^n_{d}\}.
\end{align*}
A Pauli gate is any element from $\mathscr{P}^n_d$ and a Clifford gate is any element from $\mathscr{C}^n_d$. 
More generally, an $n$-qudit quantum gate is any unitary in $\mathbb{C}^{d^n \times d^n}$. We assume that measuring a qudit on the computational basis and single and two-qudit quantum gates from a universal gate set~\cite{Muthukrishnan2000multivalued,brylinski2002universal,Brennen2005criteria,brennen2006efficient} all require constant time to be performed. An example of a universal gate set is the set of all single-qudit gates plus the controlled shift operator $\mathbf{I}_{d(d-1)}\otimes \mathsf{X}$~\cite{brennen2006efficient}.

Another way to characterise the Pauli group $\mathscr{P}^n_{d}$ is through the phase space picture of finite-dimensional quantum mechanics~\cite{wootters1987wigner,appleby2005symmetric,gross2006hudson,de2011linearized}. The main objects here are the \emph{Weyl operators} (also known as the generalised Pauli operators) defined as
\begin{align*}
    \mathcal{W}_{\mathbf{x}} = \mathcal{W}_{\mathbf{v},\mathbf{w}} = \tau^{\langle \mathbf{v},\mathbf{w}\rangle}(\mathsf{X}^{w_1}\mathsf{Z}^{v_1})\otimes \cdots \otimes (\mathsf{X}^{w_n}\mathsf{Z}^{v_n}), \qquad \forall\mathbf{x} = (\mathbf{v},\mathbf{w})\in\mathbb{Z}_D^{2n}.
\end{align*}
Every Weyl operator is an element of $\mathscr{P}_d^n$ and  conversely, every Pauli operator is a Weyl operator up to a phase that is a power of $\tau$. Thus $ \mathscr{P}_d^n = \{\tau^s \mathcal{W}_{\mathbf{x}}: \mathbf{x}\in\mathbb{Z}_D^{2n}, ~s\in\mathbb{Z}_D\}$.
\begin{fact}\label{fact:weyl_properties}
    For any $\mathbf{x},\mathbf{y}\in\mathbb{Z}_D^{2n}$,
    \begin{enumerate}[label=\alph*.]
        \item $\mathcal{W}_{\mathbf{x}}\mathcal{W}_{\mathbf{y}} = \tau^{[ \mathbf{x},\mathbf{y}]}\mathcal{W}_{\mathbf{x}+\mathbf{y}} = \omega^{[\mathbf{x},\mathbf{y}]}\mathcal{W}_{\mathbf{y}}\mathcal{W}_{\mathbf{x}}$;
        \item $[\mathcal{W}_{\mathbf{x}},\mathcal{W}_{\mathbf{y}}] = \mathbf{0}$ if and only if $[\mathbf{x},\mathbf{y}]_d = 0$; 
        \item ${\operatorname{Tr}}[\mathcal{W}_{\mathbf{y}}^\dagger\mathcal{W}_{\mathbf{x}}] = \operatorname{Tr}[\mathcal{W}_{\mathbf{y}}\mathcal{W}_{\mathbf{x}}^\dagger] = \pm d^{n}\cdot \mathbf{1}[\mathbf{x} \equiv \mathbf{y}~(\operatorname{mod}d)]$;
        \item $\mathcal{W}_{\mathbf{x} + d\mathbf{y}} = (-1)^{(d+1)[\mathbf{x},\mathbf{y}]}\mathcal{W}_{\mathbf{x}}$; 
        \item $\{d^{-\frac{n}{2}}\mathcal{W}_{\mathbf{x}}:\mathbf{x}\in\mathbb{Z}_d^{2n}\}$ forms an orthonormal basis with respect to $\langle \mathcal{A},\mathcal{B}\rangle = \operatorname{Tr}[\mathcal{A}^\dagger \mathcal{B}]$.
    \end{enumerate}
\end{fact}
Unless explicitly mentioned otherwise, the label $\mathbf{x}$ of a Weyl operator $\mathcal{W}_{\mathbf{x}}$ is always in $\mathbb{Z}_D^{2n}$. If $\mathbf{x}$ is expressed as a function of other vectors, any underlying operation is performed over $\mathbb{Z}_D$. As an example, we shall often write $\mathbf{x} = \sum_{i=1}^k a_i \mathbf{y}_i$ for $a_1,\dots,a_k\in\mathbb{Z}_d$ and $\mathbf{y}_1,\dots,\mathbf{y}_k\in\mathbb{Z}_d^{2n}$; addition and multiplication should always be performed over $\mathbb{Z}_D$ in this case. 

Let $|\Phi^+\rangle \triangleq d^{-n/2}\sum_{\mathbf{q}\in\mathbb{Z}_d^n} |\mathbf{q}\rangle^{\otimes 2}$ be a maximally entangled state. The generalised Bell states over $\mathbb{Z}_d^n$ are defined as
\begin{align*}
    |\mathcal{W}_{\mathbf{x}}\rangle \triangleq (\mathcal{W}_{\mathbf{x}}\otimes\mathbf{I}) |\Phi^+\rangle = \frac{1}{\sqrt{d^n}}\sum_{\mathbf{q}\in\mathbb{Z}_d^n} \tau^{\langle \mathbf{v},\mathbf{w}\rangle}\cdot \omega^{\langle \mathbf{q}, \mathbf{v}\rangle}|\mathbf{q}+\mathbf{w} ~(\operatorname{mod}d)\rangle|\mathbf{q}\rangle, \quad\forall\mathbf{x}=(\mathbf{v},\mathbf{w})\in\mathbb{Z}_D^{2n}.
\end{align*}
Note that, for all $\mathbf{x}=(\mathbf{v},\mathbf{w})\in\mathbb{Z}_D^{2n}$,
\begin{align*}
    (\mathbf{I}\otimes\mathcal{W}_{\mathbf{x}}^\ast) |\Phi^+\rangle &= \frac{1}{\sqrt{d^n}}\sum_{\mathbf{q}\in\mathbb{Z}_d^n}\tau^{- \langle \mathbf{v},\mathbf{w}\rangle}\cdot\omega^{-\langle \mathbf{q},\mathbf{v}\rangle} |\mathbf{q}\rangle|\mathbf{q}+\mathbf{w} ~(\operatorname{mod}d)\rangle \\
    &= \frac{1}{\sqrt{d^n}}\sum_{\mathbf{q}\in\mathbb{Z}_d^n}\tau^{-\langle \mathbf{v},\mathbf{w}\rangle}\cdot\omega^{-\langle \mathbf{q}-\mathbf{w},\mathbf{v}\rangle} |\mathbf{q}-\mathbf{w}~(\operatorname{mod}d)\rangle|\mathbf{q}\rangle
    = (\mathcal{W}_{\mathbf{x}}^\dagger\otimes \mathbf{I})|\Phi^{+}\rangle.
\end{align*}
Thus $(\mathcal{W}_{\mathbf{x}}\otimes\mathcal{W}_{\mathbf{x}}^\ast) |\Phi^+\rangle = |\Phi^+\rangle$. 
%
The following representation of $|\mathcal{W}_{\mathbf{x}}\rangle\langle \mathcal{W}_{\mathbf{x}}|$ will be useful.
\begin{lemma}\label{lem:bell_basis}
    For all $\mathbf{x}\in\mathbb{Z}_D^{2n}$,
    \begin{align*}
        |\mathcal{W}_{\mathbf{x}}\rangle\langle \mathcal{W}_{\mathbf{x}}| = \frac{1}{d^{2n}}\sum_{\mathbf{y}\in\mathbb{Z}_d^{2n}} \omega^{[\mathbf{x},\mathbf{y}]} \mathcal{W}_{\mathbf{y}}\otimes \mathcal{W}_{\mathbf{y}}^\ast = \frac{1}{D^{2n}}\sum_{\mathbf{y}\in\mathbb{Z}_D^{2n}} \omega^{[\mathbf{x},\mathbf{y}]} \mathcal{W}_{\mathbf{y}}\otimes \mathcal{W}_{\mathbf{y}}^\ast.
    \end{align*}
\end{lemma}
\begin{proof}
    Without loss of generality, we can restrict ourselves to $\mathbf{x}\in\mathbb{Z}_d^{2n}$. For all $\mathbf{z}, \mathbf{z'} \in\mathbb{Z}_d^{2n}$,
    \begin{align*}
        \sum_{\mathbf{y}\in\mathbb{Z}_d^{2n}} \omega^{[\mathbf{x},\mathbf{y}]}\langle\mathcal{W}_{\mathbf{z}}|\mathcal{W}_{\mathbf{y}}\otimes \mathcal{W}_{\mathbf{y}}^\ast|\mathcal{W}_{\mathbf{z}'}\rangle &= \sum_{\mathbf{y}\in\mathbb{Z}_d^{2n}} \omega^{[\mathbf{x},\mathbf{y}]}\langle\Phi^+|(\mathcal{W}_{\mathbf{z}}^\dagger\otimes\mathbf{I})(\mathcal{W}_{\mathbf{y}}\otimes \mathcal{W}_{\mathbf{y}}^\ast)(\mathcal{W}_{\mathbf{z}'}\otimes\mathbf{I})|\Phi^+\rangle\\
        &= \sum_{\mathbf{y}\in\mathbb{Z}_d^{2n}} \omega^{[\mathbf{y},\mathbf{z}'-\mathbf{x}]} \langle\Phi^+|(\mathcal{W}_{\mathbf{z}}^\dagger \mathcal{W}_{\mathbf{z}'}\otimes\mathbf{I})(\mathcal{W}_{\mathbf{y}}\otimes \mathcal{W}_{\mathbf{y}}^\ast)|\Phi^+\rangle\\
        &= \sum_{\mathbf{y}\in\mathbb{Z}_d^{2n}} \omega^{[\mathbf{y},\mathbf{z}'-\mathbf{x}]} \langle\Phi^+|(\mathcal{W}_{\mathbf{z}}^\dagger \mathcal{W}_{\mathbf{z}'}\otimes\mathbf{I})|\Phi^+\rangle\\
        &= \mathbf{1}[\mathbf{z} = \mathbf{z}']\sum_{\mathbf{y}\in\mathbb{Z}_d^{2n}} \omega^{[\mathbf{y},\mathbf{z}'-\mathbf{x}]}\\
        &= d^{2n}\mathbf{1}[\mathbf{z} = \mathbf{z}' = \mathbf{x}]. \tag*{(by \cref{lem:sum} with $(\mathbb{Z}_d^{2n})^{\independent} = \{\mathbf{0}\}$) \quad\qedhere}
    \end{align*}
    %
\end{proof}

Any operator $\mathcal{A}$ on $(\mathbb{C}^d)^{\otimes n}$ can be expanded as
\begin{align*}
    \mathcal{A} = d^{-\frac{n}{2}}\sum_{\mathbf{x}\in\mathbb{Z}_d^{2n}} c_{\mathcal{A}}(\mathbf{x})\mathcal{W}_{\mathbf{x}} \quad \text{where}~ c_{\mathcal{A}}:\mathbb{Z}_d^{2n}\to \mathbb{C} ~\text{given by}~c_{\mathcal{A}}(\mathbf{x}) = d^{-\frac{n}{2}}\operatorname{Tr}[\mathcal{W}_{\mathbf{x}}^\dagger \mathcal{A}]
\end{align*}
is the \emph{characteristic function} of the operator $\mathcal{A}$. We note that $\operatorname{Tr}[\mathcal{A}^\dagger \mathcal{B}] = \sum_{\mathbf{x}\in\mathbb{Z}_d^{2n}}\overline{c_{\mathcal{A}}(\mathbf{x})}c_{\mathcal{B}}(\mathbf{x})$. In particular, for any quantum state $|\psi\rangle\in(\mathbb{C}^d)^{\otimes n}$,
\begin{align*}
    \psi \triangleq |\psi\rangle\langle\psi| = d^{-\frac{n}{2}}\sum_{\mathbf{x}\in\mathbb{Z}_d^{2n}} c_{\psi}(\mathbf{x})\mathcal{W}_{\mathbf{x}} \quad\text{where}~c_\psi(\mathbf{x}) = d^{-\frac{n}{2}}\operatorname{Tr}[\mathcal{W}_{\mathbf{x}}^\dagger \psi] = d^{-\frac{n}{2}} \langle \psi|\mathcal{W}_{\mathbf{x}}^\dagger|\psi\rangle.
\end{align*}
Note that $\overline{c_\psi(\mathbf{x})} = c_\psi(-\mathbf{x})$ since $\mathcal{W}_{\mathbf{x}}^\dagger = \mathcal{W}_{-\mathbf{x}}$. 

From the characteristic function of $\psi = |\psi\rangle\langle\psi|$ we define its \emph{characteristic distribution}
\begin{align*}
    p_{\psi}:\mathbb{Z}_d^{2n}\to[0,d^{-n}], \quad p_{\psi}(\mathbf{x}) = |c_{\psi}(\mathbf{x})|^2 = d^{-n} |\langle \psi|\mathcal{W}_{\mathbf{x}}|\psi\rangle|^2 = d^{-n}\operatorname{Tr}[\psi\mathcal{W}_{\mathbf{x}}\psi\mathcal{W}_{\mathbf{x}}^\dagger].
\end{align*}
Note that $p_\psi(\mathbf{x} \operatorname{mod}d) = p_\psi(\mathbf{x})$ is invariant modulo $d$, hence we can restrict its domain to $\mathbb{Z}_d^{2n}$. Furthermore, $p_\psi(\mathbf{x}) \leq d^{-n}$ since $|\langle\psi|\mathcal{W}_{\mathbf{x}}|\psi\rangle| \leq 1$ for all $\mathbf{x}\in\mathbb{Z}_d^{2n}$. To check that $p_\psi$ is indeed a probability distribution, note that $\sum_{\mathbf{x}\in\mathbb{Z}_d^{2n}}p_{\psi}(\mathbf{x}) = \sum_{\mathbf{x}\in\mathbb{Z}_d^{2n}}c_{\psi}(\mathbf{x}) c_{\psi}(\mathbf{x})^\ast = \operatorname{Tr}[|\psi\rangle\langle\psi|\psi\rangle\langle\psi|] = 1$. The next result, which is a slight generalisation of~\cite[Eq.~(3.5)]{gross2021schur} for $d>2$, shows that $p_\psi$ is invariant (up to renormalisation) under the symplectic Fourier transform.

\begin{lemma}\label{lem:invariant}
    For any $|\psi\rangle\in(\mathbb{C}^d)^{\otimes n}$ and $\mathbf{x}\in\mathbb{Z}_d^{2n}$, $\widehat{p_\psi}(\mathbf{x)} = d^{-n}p_\psi(\mathbf{x})$.
\end{lemma}
\begin{proof}
    $\begin{aligned}[t]
        \widehat{p_\psi}(\mathbf{x)} &= \frac{1}{d^{2n}}\sum_{\mathbf{y}\in\mathbb{Z}_d^{2n}}\omega^{[\mathbf{x},\mathbf{y}]}p_\psi(\mathbf{y}) = \frac{1}{d^{3n}}\sum_{\mathbf{y}\in\mathbb{Z}_d^{2n}}\omega^{[\mathbf{x},\mathbf{y}]}\langle \psi|\mathcal{W}_{\mathbf{y}}|\psi\rangle \langle \psi|\mathcal{W}_{\mathbf{y}}^\dagger|\psi\rangle\\
        &= \frac{1}{d^{3n}}\sum_{\mathbf{y}\in\mathbb{Z}_d^{2n}} \langle \psi|\mathcal{W}_{\mathbf{y}}|\psi\rangle \langle \psi|\mathcal{W}_{\mathbf{x}}^\dagger\mathcal{W}_{\mathbf{y}}^\dagger\mathcal{W}_{\mathbf{x}}|\psi\rangle = \frac{1}{d^{2n}}\sum_{\mathbf{y}\in\mathbb{Z}_d^{2n}} \overline{c_\psi(\mathbf{y})} c_{\mathcal{W}_{\mathbf{x}} \psi \mathcal{W}_{\mathbf{x}}^\dagger}(\mathbf{y})\\
        &= \frac{1}{d^{2n}}\operatorname{Tr}[\psi\mathcal{W}_{\mathbf{x}} \psi \mathcal{W}_{\mathbf{x}}^\dagger] = d^{-n}p_\psi(\mathbf{x}). \hspace{3.3cm}\text{(by Parseval's identity) \quad \qedhere}
    \end{aligned}$
\end{proof}
Moreover, the next identity regarding the mass on a submodule $\mathscr{X}\subseteq\mathbb{Z}_d^{2n}$ under $p_\psi$ is true.
\begin{lemma}\label{lem:properties_chateristic_Weyl_1}
    Let $\mathscr{X}\subseteq\mathbb{Z}_d^{2n}$ be a submodule. Then $\sum_{\mathbf{x}\in\mathscr{X}}p_\psi(\mathbf{x}) = \frac{|\mathscr{X}|}{d^n}\sum_{\mathbf{x}\in\mathscr{X}^{\independent}} p_\psi(\mathbf{x})$.
\end{lemma}
\begin{proof}
    Using \Cref{lem:sum,lem:invariant},
    \begin{align*}
        \sum_{\mathbf{x}\in\mathscr{X}}p_\psi(\mathbf{x}) &= \sum_{\mathbf{x}\in\mathscr{X}}\sum_{\mathbf{y}\in\mathbb{Z}_d^{2n}} \omega^{[\mathbf{x},\mathbf{y}]} \widehat{p_\psi}(\mathbf{y}) 
        = \frac{1}{d^n}\sum_{\mathbf{x}\in\mathscr{X}}\sum_{\mathbf{y}\in\mathbb{Z}_d^{2n}} \omega^{[\mathbf{x},\mathbf{y}]} p_\psi(\mathbf{y}) 
        = \frac{|\mathscr{X}|}{d^n}\sum_{\mathbf{y}\in\mathscr{X}^{\independent}} p_\psi(\mathbf{y}). \qedhere
    \end{align*}
\end{proof}

\subsection{Stabiliser groups}
\label{sec:stabiliser_groups}

We now review \emph{stabiliser groups}, which are among the most important subgroups of $\mathscr{P}_d^n$. For completeness and the reader's convenience, we provide proofs for a number of important claims in \Cref{app:stabiliser_groups}.
See~\cite{appleby2005symmetric,gross2006hudson,appleby2009properties,de2011linearized,gross2021schur} for more information. 
\begin{definition}[Partial stabiliser group]
    A partial stabiliser group $\mathcal{X}$ is a subgroup of $\mathscr{P}^n_d$ which contains only the identity from the center of $\mathscr{P}^n_d$, i.e., $\mathcal{X}\cap\mathrm{Z}(\mathscr{P}_d^n) = \{\mathbf{I}\}$. 
    A stabiliser group $\mathcal{S}$ is a partial stabiliser group which is also maximal.
\end{definition}

\begin{lemma}\label{lem:stabiliser_group_form}
    Any partial stabiliser group $\mathcal{X}\subset\mathscr{P}^n_d$ can be written as $\mathcal{X} = \{\omega^{s(\mathbf{x})} \mathcal{W}_{\mathbf{x}} :  \mathbf{x}\in \mathscr{X}\}$, where $\mathscr{X}\subset\mathbb{Z}_d^{2n}$ is an isotropic submodule and $s:\mathbb{Z}_d^{2n}\to\mathbb{Z}_d$. If $d$ is odd, then $s$ is linear, i.e., $s(\mathbf{x} + \mathbf{y}) = s(\mathbf{x}) + s(\mathbf{y})$. 
    If $\mathcal{X}$ is a stabiliser group, then $\mathscr{X}$ is also Lagrangian and $|\mathscr{X}| = d^n$.
\end{lemma}
\begin{proof}
    See \Cref{app:stabiliser_groups}.
\end{proof}

\begin{corollary}
     Any stabiliser group 
    $\mathcal{S}$ of $\mathscr{P}_d^n$ is commutative (abelian) and has size $|\mathcal{S}| = d^n$.
\end{corollary}
\begin{remark}
    We note that, when operating with elements of a partial stabiliser group like $\omega^{s(\mathbf{x}) + s(\mathbf{y})}\mathcal{W}_{\mathbf{x}}\mathcal{W}_{\mathbf{y}} = \omega^{s(\mathbf{x} + \mathbf{y})}\mathcal{W}_{\mathbf{x}+\mathbf{y}}$ for $\mathbf{x},\mathbf{y}\in\mathscr{X}$, the resulting label $\mathbf{x}+\mathbf{y}$ of $\mathcal{W}_{\mathbf{x}+\mathbf{y}}$ is in $\mathbb{Z}_D^{2n}$ according to our convention, even though $\mathscr{X}$ is a submodule. In order to return to $\mathscr{X}$, we employ the rule $\omega^{s(\mathbf{x})}\mathcal{W}_{\mathbf{x}} = \omega^{s(\mathbf{x}\operatorname{mod}d)}\mathcal{W}_{\mathbf{x}\operatorname{mod}d}$. We shall use the notation $\mathbf{x}+\mathbf{y}\operatorname{mod}d$ within $\mathcal{W}_{\mathbf{x}+\mathbf{y}\operatorname{mod}d}$ or $s(\mathbf{x}+\mathbf{y}\operatorname{mod}d)$ to stress that when $\mathbf{x}+\mathbf{y}$ should be interpreted as being in $\mathscr{X}$.
\end{remark}
Given a generating set $\{(\mathbf{v}_i,\mathbf{w}_i)\}_{i\in[\ell]}$ for $\mathscr{M}$, let the matrices $\mathbf{V},\mathbf{W}\in\mathbb{Z}_d^{n\times \ell}$ with columns $\mathbf{v}_1,\dots,\mathbf{v}_\ell$ and $\mathbf{w}_1,\dots,\mathbf{w}_\ell$, respectively. Then $\mathscr{M} = {\operatorname{col}}\big(\big[\begin{smallmatrix} \mathbf{V} \\ \mathbf{W} \end{smallmatrix}\big]\big)$. Note that $\mathbf{V}^\top \mathbf{W} = \mathbf{W}^\top \mathbf{V}$ since $\mathscr{M}$ is isotropic.

Together with the concept of stabiliser group is the concept of stabiliser state, which is the joint $+1$-eigenstate of all elements of a stabiliser group.
\begin{definition}[Partial stabiliser state]\label{def:partial_stabiliser_state}
    A non-zero state $|\psi\rangle\in(\mathbb{C}^{d})^{\otimes n}$ is \emph{stabilised} by $\mathcal{P}\in\mathscr{P}^n_d$ if $\mathcal{P} |\psi\rangle = |\psi\rangle$. A non-zero state $|\Psi\rangle\in(\mathbb{C}^{d})^{\otimes n}$ is a \emph{partial stabiliser state} of a partial stabiliser group $\mathcal{X}$ if $\mathcal{P} |\Psi\rangle = |\Psi\rangle$ for all $\mathcal{P}\in\mathcal{X}$. Let $\EuScript{X}_d^n(K)$ denote the set of all partial stabiliser states in $(\mathbb{C}^d)^{\otimes n}$ of partial stabiliser groups $\mathcal{X}$ of size at least $K$.
\end{definition}
\begin{definition}[Stabiliser state]\label{def:stabiliser_state}
    A non-zero state $|\Psi\rangle\in(\mathbb{C}^{d})^{\otimes n}$ is a \emph{stabiliser state} of a stabiliser group $\mathcal{S}$ if $\mathcal{P} |\Psi\rangle = |\Psi\rangle$ for all $\mathcal{P}\in\mathcal{S}$. Let $\EuScript{S}_d^n$ be the set of all stabiliser states in $(\mathbb{C}^d)^{\otimes n}$. Clearly $\EuScript{S}_d^n = \EuScript{X}_d^n(d^n)$.
\end{definition}

Consider the stabiliser group $\mathcal{S} = \{\omega^{s(\mathbf{x})}\mathcal{W}_{\mathbf{x}}:\mathbf{x}\in\mathscr{M}\}$ and let $V_{\mathcal{S}} \triangleq \{|\Psi\rangle\in(\mathbb{C}^{d})^{\otimes n}: \omega^{s(\mathbf{x})}\mathcal{W}_{\mathbf{x}}|\Psi\rangle = |\Psi\rangle, \forall \mathbf{x}\in\mathscr{M}\}$ be the set of stabiliser states of $\mathcal{S}$. The next lemma constructs an projector onto $V_{\mathcal{S}}$ and shows that any stabiliser group has a unique stabiliser state.
\begin{lemma}\label{lem:projection}
    Let $\mathcal{X} = \{\omega^{s(\mathbf{x})}\mathcal{W}_{\mathbf{x}}:\mathbf{x}\in\mathscr{X}\}$ be a partial stabiliser group and let $V_{\mathcal{X}} \triangleq \{|\Psi\rangle\in(\mathbb{C}^{d})^{\otimes n}: \omega^{s(\mathbf{x})}\mathcal{W}_{\mathbf{x}}|\Psi\rangle = |\Psi\rangle, \forall \mathbf{x}\in\mathscr{X}\}$. Then the projector onto $V_{\mathcal{X}}$ is
    \begin{align*}
        \Pi_{\mathcal{X}} = \frac{1}{|\mathscr{X}|}\sum_{\mathbf{x}\in\mathscr{X}}\omega^{s(\mathbf{x})}\mathcal{W}_{\mathbf{x}}.
    \end{align*}
    Moreover, $\operatorname{dim}(V_{\mathcal{X}}) = d^{n}/|\mathscr{X}|$. It then follows that any stabiliser group $\mathcal{S}$ has a unique stabiliser state, i.e., $\operatorname{dim}(V_{\mathcal{S}}) = 1$.
\end{lemma}
\begin{proof}
    See \Cref{app:stabiliser_groups}.
\end{proof}

Given a Lagrangian submodule $\mathscr{M}$, multiple phase functions lead to the same stabiliser state since replacing $s$ with $s'(\mathbf{x}) = s(\mathbf{x}) + [\mathbf{z},\mathbf{x}]$ for some $\mathbf{z}\in\mathbb{Z}_d^{2n}$ leads to $|\mathscr{M},s'\rangle = \mathcal{W}_{\mathbf{z}}|\mathscr{M},s\rangle$. More precisely, any two strings $\mathbf{z},\mathbf{z}'$ from a coset $C\in\mathbb{Z}_d^{2n}/\mathscr{M}$ yields the same stabiliser state $|\mathscr{M},s+[\mathbf{z},\cdot]\rangle$ since then $\mathbf{z} - \mathbf{z}' \in \mathscr{M}$. Given $\mathscr{M}$ and $s$, we can write the stabiliser state $|\mathscr{M},s+[\mathbf{z},\cdot]\rangle$ as $|\mathscr{M},s,C\rangle$ where $\mathbf{z}\in C$ and $C\in\mathbb{Z}_d^{2n}/\mathscr{M}$. Alternatively, given a Lagrangian submodule $\mathscr{M}$, any state that is a simultaneous eigenvector of $\{\mathcal{W}_{\mathbf{x}}\}_{\mathbf{x}\in\mathscr{M}}$ is a stabiliser state of some stabiliser group. The eigenvectors of $\{\mathcal{W}_{\mathbf{x}}\}_{\mathbf{x}\in\mathscr{M}}$ thus determine a stabiliser basis $\{|\mathscr{M},s\rangle\}_{s} = \{|\mathscr{M},s,C\rangle\}_{C\in\mathbb{Z}_d^{2n}/\mathscr{M}}$.

Consider a stabiliser group $\mathcal{S} = \{\omega^{s(\mathbf{x})}\mathcal{W}_{\mathbf{x}}:\mathbf{x}\in\mathscr{M}\}$. According to \cref{lem:projection}, the operator $d^{-n}\sum_{\mathbf{x}\in\mathscr{M}}\omega^{s(\mathbf{x})}\mathcal{W}_{\mathbf{x}}$ is the orthogonal projection onto the set $V_{\mathcal{S}}$ spanned by the stabiliser state $|\mathcal{S}\rangle$ of $\mathcal{S}$. Therefore, $|\mathcal{S}\rangle\langle\mathcal{S}| = \frac{1}{d^{n}}\sum_{\mathbf{x}\in\mathscr{M}}\omega^{s(\mathbf{x})}\mathcal{W}_{\mathbf{x}}$ and the characteristic function and probability of $|\mathcal{S}\rangle\langle\mathcal{S}|$ are
\begin{align}\label{eq:stabiliser_distribution}
    c_{\mathcal{S}}(\mathbf{x}) = \begin{cases}
        d^{-n/2}\omega^{s(\mathbf{x})} &\text{if}~\mathbf{x}\in\mathscr{M},\\
        0 &\text{otherwise},
    \end{cases} \qquad \qquad\text{and} \qquad \qquad 
    p_{\mathcal{S}}(\mathbf{x}) = \begin{cases}
        d^{-n} &\text{if}~\mathbf{x}\in\mathscr{M},\\
        0 &\text{otherwise}.
    \end{cases}
\end{align}
Thus $p_{\mathcal{S}}$ is the uniform distribution on the subspace $\mathscr{M}$, a fact which will be of paramount importance when studying Bell (difference) sampling in the next section. 

\subsection{Stabiliser size}

Sometimes we are interested in the Pauli operators that stabilise a given pure state $|\psi\rangle$ up to a phase. This is captured by the \emph{unsigned stabiliser group} of $|\psi\rangle$.
\begin{definition}[Unsigned stabiliser group and stabiliser size]
    Given $|\psi\rangle\in(\mathbb{C}^d)^{\otimes n}$, its \emph{unsigned stabiliser group} is ${\operatorname{Weyl}}(|\psi\rangle) \triangleq \{\mathbf{x}\in\mathbb{Z}_d^{2n}: |\langle\psi|\mathcal{W}_{\mathbf{x}}|\psi\rangle| = 1\}$ and its \emph{stabiliser size} is $|{\operatorname{Weyl}}(|\psi\rangle)|$.
\end{definition}
\begin{remark}
    Note that for all $|\psi\rangle\in(\mathbb{C}^d)^{\otimes n}$, ${\operatorname{Weyl}}(|\psi\rangle)$ is a partial stabiliser group. On the other hand, for any partial stabiliser group $\mathcal{X}$, there always is $|\psi\rangle\in(\mathbb{C}^d)^{\otimes n}$ such that ${\operatorname{Weyl}}(|\psi\rangle) = \mathcal{X}$.
\end{remark}

\begin{lemma}
    For any non-zero $|\psi\rangle\in(\mathbb{C}^d)^{\otimes n}$, ${\operatorname{Weyl}}(|\psi\rangle)$ is an isotropic submodule.
\end{lemma}
\begin{proof}
    See \Cref{app:stabiliser_groups}.
\end{proof}

The stabiliser size is invariant under Clifford transformations as a corollary of the next~fact.
\begin{definition}[Symplectic matrix]
    A matrix $\Gamma\in\mathbb{Z}_d^{2n\times 2n}$ is called \emph{symplectic} if it preserves the symplectic product, i.e., $[\Gamma\mathbf{x},\Gamma\mathbf{y}] = [\mathbf{x},\mathbf{y}]$ for all $\mathbf{x},\mathbf{y}\in\mathbb{Z}_d^{2n}$. Equivalently, $\Gamma$ satisfies $\Gamma^\top \Omega \Gamma = \Omega$, where $\Omega \triangleq \bigl(\begin{smallmatrix} \mathbf{0}_{n} & \mathbf{I}_{n}\\ -\mathbf{I}_{n} & \mathbf{0}_n \end{smallmatrix} \bigr)$. Let $\operatorname{Sp}(2n,d)$ be the set of symplectic matrices over $\mathbb{Z}_d^{2n}$.
\end{definition}

\begin{fact}[{\cite[Theorem~5]{beaudrap2013linearized}}]
    Any Clifford operator $\mathcal{C}\in\mathscr{C}_d^n$ is of the form $\mathcal{C} = \mathcal{W}_{\mathbf{a}}\mu(\Gamma)$ for some $\mathbf{a}\in\mathbb{Z}_d^{2n}$ and a unitary operator $\mu(\Gamma)\in (\mathbb{C}^d)^{\otimes n}$ such that $\mu(\Gamma)\mathcal{W}_{\mathbf{x}}\mu(\Gamma)^\dagger =  \mathcal{W}_{\Gamma\mathbf{x}}$ for all $\mathbf{x}\in\mathbb{Z}_d^{2n}$, where $\Gamma\in\operatorname{Sp}(2n,d)$.
\end{fact}

\begin{corollary}\label{cor:invariant}
    For all $\mathcal{C}\in\mathscr{C}_d^n$ and $|\psi\rangle\in(\mathbb{C}^d)^{\otimes n}$, $|{\operatorname{Weyl}}(\mathcal{C}|\psi\rangle)| = |{\operatorname{Weyl}}(|\psi\rangle)|$.
\end{corollary}
\begin{proof}
    See \Cref{app:stabiliser_groups}.
\end{proof}

The next lemma covers another important property of unsigned stabiliser groups: the stabiliser size is at most $d^n$ and is maximised for stabiliser states. 
\begin{lemma}\label{lem:stabiliser_dimension}
    For any $|\psi\rangle\in (\mathbb{C}^d)^{\otimes n}$, $|{\operatorname{Weyl}}(|\psi\rangle)| \leq d^n$, with equality if and only if $|\psi\rangle$ is a stabiliser state.
\end{lemma}
\begin{proof}
    A simple consequence of ${\operatorname{Weyl}}(|\psi\rangle)$ being isotropic and any isotropic set of elements having cardinality at most $d^n$, with equality if and only if the set is Lagrangian.
\end{proof}

\subsection{Stabiliser fidelity}

Alongside its stabiliser size, another important measure of the ``stabiliser complexity'' of a quantum state is its \emph{stabiliser fidelity}~\cite{bravyi2019simulation}. Here we propose a slight generalisation called ``$K$-sized stabiliser fidelity''. Recall the definitions of $\EuScript{X}_d^n(K)$ and $\EuScript{S}_d^n$ from \Cref{def:partial_stabiliser_state,def:stabiliser_state}.
\begin{definition}[($K$-sized) Stabiliser fidelity]
    Given an $n$-qudit quantum state $|\psi\rangle\in(\mathbb{C}^d)^{\otimes n}$, its \emph{$K$-sized stabiliser fidelity} is $F_{K}(|\psi\rangle) \triangleq \max_{|\Psi\rangle \in \EuScript{X}_d^n(K)} |\langle \Psi|\psi\rangle|^2$, while its \emph{stabiliser fidelity} is $F_{\mathcal{S}}(|\psi\rangle) \triangleq \max_{|\mathcal{S}\rangle \in \EuScript{S}_d^n} |\langle \mathcal{S}|\psi\rangle|^2$.
\end{definition}

It is possible to relate the $K$-sized stabiliser fidelity $F_{K}(|\psi\rangle)$ of a quantum state $|\psi\rangle$ to its characteristic distribution $p_{\psi}$ as shown next. 
\begin{lemma}\label{lem:fidelity_inequality}
    Given $|\psi\rangle\in(\mathbb{C}^d)^{\otimes n}$, let $\mathscr{X}\subset\mathbb{Z}_d^{2n}$ be the isotropic submodule characterising the partial stabiliser state that maximises $F_K(|\psi\rangle)$. Then
    \begin{align*}
        \sum_{\mathbf{x}\in \mathscr{X}^{\independent}} p_\psi(\mathbf{x}) \leq F_K(|\psi\rangle) \leq \sqrt{\sum_{\mathbf{x}\in \mathscr{X}^{\independent}} p_\psi(\mathbf{x})}.
    \end{align*}
\end{lemma}
\begin{proof}
    Let $|\Psi\rangle = \arg\max_{|\phi\rangle\in\EuScript{X}_d^n(K)} |\langle\phi|\psi\rangle|^2$ and $\mathcal{X} = \{\omega^{s(\mathbf{x})} \mathcal{W}_{\mathbf{x}}:\mathbf{x}\in\mathscr{X}\}$ be its partial stabiliser group. Let $\Pi_{\mathcal{X}} = |\mathscr{X}|^{-1}\sum_{\mathbf{x}\in\mathscr{X}} \omega^{s(\mathbf{x})} \mathcal{W}_{\mathbf{x}}$ be the projection onto the set of partial stabiliser states of $\mathcal{X}$, i.e., $V_{\mathcal{X}} = \{|\Psi\rangle\in(\mathbb{C}^d)^{\otimes n} : \omega^{s(\mathbf{x})} \mathcal{W}_{\mathbf{x}}|\Psi\rangle = |\Psi\rangle, \forall \mathbf{x}\in\mathscr{X}\}$. For the upper bound,
    \begin{align*}
        F_K(|\psi\rangle) &= \langle \psi|\Pi_{\mathcal{X}}|\psi\rangle \\
        &= \frac{1}{|\mathscr{X}|}\sum_{\mathbf{x}\in\mathscr{X}}\omega^{s(\mathbf{x})}\langle \psi|\mathcal{W}_{\mathbf{x}}|\psi\rangle  \tag{by \Cref{lem:projection}}\\
        &\leq \sqrt{\frac{1}{|\mathscr{X}|}\sum_{\mathbf{x}\in \mathscr{X}} |\langle \psi|\mathcal{W}_{\mathbf{x}}|\psi\rangle|^2} \tag{by Cauchy-Schwarz}\\
        &= \sqrt{\frac{d^n}{|\mathscr{X}|}\sum_{\mathbf{x}\in \mathscr{X}} p_\psi(\mathbf{x})} \tag{definition of $p_\psi(\mathbf{x})$}\\
        &= \sqrt{\sum_{\mathbf{x}\in \mathscr{X}^{\independent}} p_\psi(\mathbf{x})}. \tag{by \Cref{lem:properties_chateristic_Weyl_1}}
    \end{align*}
    For the lower bound, define the projections $\Pi_{C} = |\mathscr{X}|^{-1}\sum_{\mathbf{x}\in\mathscr{X}}\omega^{s(\mathbf{x}) + [\mathbf{b},\mathbf{x}]}\mathcal{W}_{\mathbf{x}}$ from the proof of \Cref{lem:projection}, where $C\in\mathbb{Z}_d^{2n}/\mathscr{X}^{\independent}$ is a coset and $\mathbf{b}\in\mathbb{Z}_d^{2n}$ is any element from $C$. Then 
    \begin{align*}
        F_K(|\psi\rangle) &= \max_{C\in\mathbb{Z}_d^{2n}/\mathscr{X}^{\independent}}\langle \psi|\Pi_{C}|\psi\rangle \\
        &\geq \sum_{C\in\mathbb{Z}_d^{2n}/\mathscr{X}^{\independent}} \langle \psi|\Pi_{C}|\psi\rangle^2 \tag{by $\sum_{C\in\mathbb{Z}_d^{2n}/\mathscr{X}^{\independent}} \Pi_C = \mathbf{I}$}\\
        &= \frac{1}{|\mathscr{X}|^{2}}\sum_{C\in\mathbb{Z}_d^{2n}/\mathscr{X}^{\independent}} \sum_{\mathbf{x},\mathbf{y}\in\mathscr{X}}\omega^{s(\mathbf{x}) + s(\mathbf{y}) +[\mathbf{b}_{C},\mathbf{x}+\mathbf{y}]}\langle \psi|\mathcal{W}_{\mathbf{x}}|\psi\rangle\langle\psi|\mathcal{W}_{\mathbf{y}}|\psi\rangle, \tag{by \Cref{lem:projection}}
    \end{align*}
    where $\mathbf{b}_{C}$ is any element of the coset $C$. By taking the average over all elements $\mathbf{b}\in C$, then
    \begin{align*}
        F_K(|\psi\rangle) &\geq \frac{1}{|\mathscr{X}|^{2}|\mathscr{X}^{\independent}|}\sum_{C\in\mathbb{Z}_d^{2n}/\mathscr{X}^{\independent}} \sum_{\mathbf{b}\in C}\sum_{\mathbf{x},\mathbf{y}\in\mathscr{X}}\omega^{s(\mathbf{x})+s(\mathbf{y}) + [\mathbf{b},\mathbf{x}+\mathbf{y}]}\langle \psi|\mathcal{W}_{\mathbf{x}}|\psi\rangle\langle\psi|\mathcal{W}_{\mathbf{y}}|\psi\rangle\\
        &= \frac{d^{-2n}}{|\mathscr{X}|} \sum_{\mathbf{b}\in\mathbb{Z}_d^{2n}}\sum_{\mathbf{x},\mathbf{y}\in\mathscr{X}}\omega^{s(\mathbf{x})+s(\mathbf{y})+[\mathbf{b},\mathbf{x}+\mathbf{y}]}\langle \psi|\mathcal{W}_{\mathbf{x}}|\psi\rangle\langle\psi|\mathcal{W}_{\mathbf{y}}|\psi\rangle \\
        &= \frac{1}{|\mathscr{X}|} \sum_{\mathbf{x}\in\mathscr{X}} \omega^{s(\mathbf{x}) + s(-\mathbf{x}\operatorname{mod}d)} \langle \psi|\mathcal{W}_{\mathbf{x}}|\psi\rangle\langle\psi|\mathcal{W}_{-\mathbf{x}\operatorname{mod}d}|\psi\rangle \tag{by \cref{lem:sum} over $\mathbf{b}$}\\
        &= \frac{1}{|\mathscr{X}|} \sum_{\mathbf{x}\in\mathscr{X}}  \langle \psi|\mathcal{W}_{\mathbf{x}}|\psi\rangle\langle\psi|\mathcal{W}_{-\mathbf{x}}|\psi\rangle \tag{by $\omega^{s(-\mathbf{x}\operatorname{mod}d)}\mathcal{W}_{-\mathbf{x}\operatorname{mod}d} = \omega^{s(-\mathbf{x})}\mathcal{W}_{-\mathbf{x}}$ and $s(-\mathbf{x}) = -s(\mathbf{x})$} \\
        &= \frac{d^n}{|\mathscr{X}|}\sum_{\mathbf{x}\in \mathscr{X}} p_\psi(\mathbf{x}) \tag{definition of $p_\psi(\mathbf{x})$}\\
        &= \sum_{\mathbf{x}\in\mathscr{X}^{\independent}} p_\psi(\mathbf{x}). \tag*{(by \Cref{lem:properties_chateristic_Weyl_1}) \qquad\qedhere}
    \end{align*}
\end{proof}

\begin{corollary}\label{lem:stabiliser_fidelity_inequality}
    Given a quantum state $|\psi\rangle\in(\mathbb{C}^d)^{\otimes n}$, let $|\mathscr{M},s\rangle = \argmax_{|\mathcal{S}\rangle \in \EuScript{S}_d^n}|\langle \mathcal{S}|\psi\rangle|^2$ be the stabiliser state that maximises the stabiliser fidelity $F_{\mathcal{S}}(|\psi\rangle)$. Then
    \begin{align*}
        \sum_{\mathbf{x}\in \mathscr{M}} p_\psi(\mathbf{x}) \leq F_{\mathcal{S}}(|\psi\rangle) \leq \sqrt{\sum_{\mathbf{x}\in \mathscr{M}} p_\psi(\mathbf{x})}.
    \end{align*}
\end{corollary}

\begin{remark}\label{remark:replacement}
    We note that the lower bounds in {\rm \Cref{lem:fidelity_inequality}} and {\rm \Cref{lem:stabiliser_fidelity_inequality}} are also valid if we replace $\mathscr{X}$ and $\mathscr{M}$ with an arbitrary isotropic and Lagrangian submodule, respectively.
\end{remark}

We also prove that given an isotropic submodule $\mathscr{X}$ with large enough $p_\psi$-mass, it is possible to find a state $|\phi\rangle$ with large fidelity with $|\psi\rangle$ such that $\mathscr{X}\subseteq{\operatorname{Weyl}}(|\phi\rangle)$.
\begin{lemma}\label{lem:existance_state_high_stabiliser_size}
    Let $\mathscr{X}\subset\mathbb{Z}_d^{2n}$ be an isotropic submodule with size $d^{n}/K$ and such that
    \begin{align*}
        \sum_{\mathbf{x}\in\mathscr{X}} p_\psi(\mathbf{x}) \geq \frac{1-\varepsilon}{K} \qquad\text{for}~\varepsilon\in[0,1].
    \end{align*}
    Then there is $|\phi\rangle\in(\mathbb{C}^d)^{\otimes n}$ with $\mathscr{X}\subseteq{\operatorname{Weyl}}(|\phi\rangle)$ such that the fidelity between $|\psi\rangle$ and $|\phi\rangle$ is at least $1-\varepsilon$, i.e., $|\langle \phi|\psi\rangle|^2 \geq 1-\varepsilon$.
\end{lemma}
\begin{proof}    
    For any suitable phase function $s:\mathbb{Z}_d^{2n}\to\mathbb{Z}_d$, consider the partial stabiliser group $\mathcal{X} = \{\omega^{s(\mathbf{x})}\mathcal{W}_{\mathbf{x}}:\mathbf{x}\in\mathscr{X}\}$. Define the projections $\Pi_{C} = |\mathscr{X}|^{-1}\sum_{\mathbf{x}\in\mathscr{X}}\omega^{s(\mathbf{x}) + [\mathbf{b},\mathbf{x}]}\mathcal{W}_{\mathbf{x}}$ from the proof of \Cref{lem:projection}, where $C\in\mathbb{Z}_d^{2n}/\mathscr{X}^{\independent}$ is a coset and $\mathbf{b}\in\mathbb{Z}_d^{2n}$ is any element from $C$. Let $V_{\mathcal{X}}(C) \triangleq \{|\Psi\rangle\in(\mathbb{C}^{d})^{\otimes n}: \omega^{s(\mathbf{x})+[\mathbf{b},\mathbf{x}]}\mathcal{W}_{\mathbf{x}}|\Psi\rangle = |\Psi\rangle, \forall \mathbf{x}\in\mathscr{X}\}$ for any $\mathbf{b}\in C$ and let $|\phi\rangle$ be the state that maximises $\max_{C\in\mathbb{Z}_d^{2n}/\mathscr{X}^{\independent}}\max_{|\Psi\rangle\in V_{\mathcal{X}}(C)} |\langle\Psi|\psi\rangle|^2$. Clearly $\mathscr{X}\subseteq \operatorname{Weyl}(|\phi\rangle)$. By following the exact same steps as the proof \Cref{lem:fidelity_inequality}, one can show that $\max_{C\in\mathbb{Z}_d^{2n}/\mathscr{X}^{\independent}}\max_{|\Psi\rangle\in V_{\mathcal{X}}(C)} |\langle\Psi|\psi\rangle|^2 = \max_{C\in\mathbb{Z}_d^{2n}/\mathscr{X}^{\independent}}\langle\psi|\Pi_C|\psi\rangle \geq \sum_{\mathbf{x}\in\mathscr{X}^{\independent}} p_\psi(\mathbf{x})$. Therefore, and also using \Cref{lem:properties_chateristic_Weyl_1}, 
    \begin{align*}
       |\langle\phi|\psi\rangle|^2 &\geq \sum_{\mathbf{x}\in\mathscr{X}^{\independent}} p_\psi(\mathbf{x}) = \frac{d^n}{|\mathscr{X}|}\sum_{\mathbf{x}\in\mathscr{X}} p_\psi(\mathbf{x}) = K \sum_{\mathbf{x}\in\mathscr{X}} p_\psi(\mathbf{x}) \geq 1-\varepsilon.\qedhere
    \end{align*}
\end{proof}

We now bound the value of $p_\psi(\mathbf{x})$ in terms of $F_{\mathcal{S}}(|\psi\rangle)$. 
\begin{lemma}\label{lem:lower_bound_p_psi}
    Let $|\psi\rangle\in(\mathbb{C}^d)^{\otimes n}$ and 
    $|\mathscr{M},s\rangle = \arg\max_{|\mathcal{S}'\rangle\in\EuScript{S}_d^n}|\langle \mathcal{S}'|\psi\rangle|^2$ such that $F_{\mathcal{S}}(|\psi\rangle) \geq \frac{1}{2}$. If $\mathbf{x}\in\mathscr{M}$, then $d^n p_\psi(\mathbf{x}) \geq (2F_{\mathcal{S}}(|\psi\rangle) - 1)^2$.
\end{lemma}
\begin{proof}
    Let $\theta \triangleq F_{\mathcal{S}}(|\psi\rangle)$ for simplicity and write $|\psi\rangle = \sqrt{\theta}|\mathcal{S}\rangle + \sqrt{1-\theta}|\mathcal{S}^\perp\rangle$, where $|\mathcal{S}\rangle$ is the stabiliser state that maximises $F_{\mathcal{S}}(|\psi\rangle)$ and $|\mathcal{S}^\perp\rangle$ is the part orthogonal to $|\mathcal{S}\rangle$. For $\mathbf{x}\in\mathscr{M}$,
    \begin{align*}
        \langle \psi|\mathcal{W}_{\mathbf{x}}|\psi\rangle &= \langle\psi|(\sqrt{\theta}\mathcal{W}_{\mathbf{x}}|\mathcal{S}\rangle + \sqrt{1-\theta}\mathcal{W}_{\mathbf{x}}|\mathcal{S}^\perp\rangle)\\
        &= \sqrt{\theta}\omega^{s(\mathbf{x})}\langle\psi|\mathcal{S}\rangle + \sqrt{1-\theta}\langle\psi|\mathcal{W}_{\mathbf{x}}|\mathcal{S}^\perp\rangle \tag{$\mathcal{W}_{\mathbf{x}}|\mathcal{S}\rangle = \omega^{s(\mathbf{x})}|\mathcal{S}\rangle$}\\
        &= \theta \omega^{s(\mathbf{x})} + \sqrt{1-\theta}\langle\psi|\mathcal{W}_{\mathbf{x}}|\mathcal{S}^\perp\rangle\\
        &= \theta \omega^{s(\mathbf{x})} + \sqrt{1-\theta}(\sqrt{\theta}\langle \mathcal{S}|\mathcal{W}_{\mathbf{x}}|\mathcal{S}^\perp\rangle + \sqrt{1-\theta}\langle\mathcal{S}^\perp|\mathcal{W}_{\mathbf{x}}|\mathcal{S}^\perp\rangle)\\
        &= \theta \omega^{s(\mathbf{x})} + (1-\theta)\langle\mathcal{S}^\perp|\mathcal{W}_{\mathbf{x}}|\mathcal{S}^\perp\rangle. \tag{$\mathcal{W}_{\mathbf{x}}^\dagger|\mathcal{S}\rangle = \omega^{s(-\mathbf{x})}|\mathcal{S}\rangle$ and $\langle\mathcal{S}|\mathcal{S}^\perp\rangle = 0$}
    \end{align*}
    Then
    \begin{align*}
        |\langle \psi|\mathcal{W}_{\mathbf{x}}|\psi\rangle|^2 &= \theta^2 + (1-\theta)^2|\langle \mathcal{S}^\perp|\omega^{-s(\mathbf{x})}\mathcal{W}_{\mathbf{x}}|\mathcal{S}^\perp\rangle|^2 + 2\theta(1-\theta)\operatorname{Re}[\langle\mathcal{S}|\omega^{-s(\mathbf{x})}\mathcal{W}_{\mathbf{x}}^\dagger|\mathcal{S}^\perp\rangle]\\
        &\geq \theta^2 + (1-\theta)^2 \operatorname{Re}[\langle\mathcal{S}|\omega^{-s(\mathbf{x})}\mathcal{W}_{\mathbf{x}}^\dagger|\mathcal{S}^\perp\rangle]^2 + 2\theta(1-\theta)\operatorname{Re}[\langle\mathcal{S}|\omega^{-s(\mathbf{x})}\mathcal{W}_{\mathbf{x}}^\dagger|\mathcal{S}^\perp\rangle] \\
        &\geq \theta^2 + (1-\theta)^2 - 2\theta(1-\theta) \tag{$\operatorname{Re}[\langle\mathcal{S}|\omega^{-s(\mathbf{x})}\mathcal{W}_{\mathbf{x}}^\dagger|\mathcal{S}^\perp\rangle]\in[-1,1]$}\\
        &= (2\theta - 1)^2. \qedhere
    \end{align*}
\end{proof}

\section{A New Bell Sampling Subroutine on Qudits}

Given a pure state of $2n$ qudits divided into systems $A_1,\dots,A_n$ and $B_1,\dots,B_n$, \emph{Bell sampling}~\cite{montanaro2017learning} is the operation of measuring each pair $A_i B_i$ of qudits in the generalised Bell basis, which returns a vector in $\mathbb{Z}_d^{2n}$. A similar operation, called \emph{Bell difference sampling}~\cite{gross2021schur}, performs Bell sampling twice and subtracts the results from each other. Performing Bell sampling on $|\psi_1\rangle|\psi_2\rangle$ returns $\mathbf{x}\in\mathbb{Z}_d^{2n}$ with probability $|\langle \mathcal{W}_{\mathbf{x}} |(|\psi_1\rangle|\psi_2\rangle)|^2 = d^{-n} |{\operatorname{Tr}}[\mathcal{W}_{\mathbf{x}}^\dagger |\psi_1\rangle\langle \psi_2^\ast| ]|^2 = d^{-n}|\langle \psi_1|\mathcal{W}_{\mathbf{x}}|\psi_2^\ast\rangle|^2$, since $|\psi_1\rangle|\psi_2\rangle = \operatorname{vec}(|\psi_1\rangle\langle\psi_2^\ast|)$. As an immediate corollary, Bell sampling on $|\psi\rangle|\psi^\ast\rangle$ is equivalent to sampling from $p_{\psi}(\mathbf{x})$. This means that Bell sampling on $|\mathcal{S}\rangle|\mathcal{S}^\ast\rangle$ returns an element of $\mathscr{M}$ according to \eqref{eq:stabiliser_distribution}.

However, without access to $|\psi^\ast\rangle$, the situation is more complex. It is well known that the transformation $|\psi\rangle \mapsto |\psi^\ast\rangle$ cannot be implemented by a physical process since it is not completely positive. There is thus little hope to sample from $p_{\psi}(\mathbf{x})$ using Bell sampling with only access to copies of $|\psi\rangle$ in general. The situation is more subtle if we restrict ourselves to stabiliser states $|\mathcal{S}\rangle$. First note that $|\mathcal{S}^\ast\rangle$ is itself a stabiliser state since $\mathcal{W}_{\mathbf{v},\mathbf{w}}^\ast = \mathcal{W}_{-\mathbf{v},\mathbf{w}} = (-1)^{(d+1)\langle (\mathbf{1}[v_i \neq 0])_{i=1}^n,\mathbf{w}\rangle} \mathcal{W}_{J(\mathbf{v},\mathbf{w})}$, where $(\mathbf{1}[v_i \neq 0])_{i=1}^n\in\{0,1\}^n$ is the vector whose $i$-th entry equals to $0$ if and only if $v_i = 0$.\footnote{Gross et al.~\cite{gross2021schur} claims that $\mathcal{W}_{\mathbf{v},\mathbf{w}}^\ast = (-1)^{(d+1)\langle \mathbf{v},\mathbf{w}\rangle}\mathcal{W}_{J(\mathbf{v},\mathbf{w})}$. We could not reproduce such equality.} This means that if $|\mathcal{S}\rangle = |\mathscr{M},s\rangle$, then $|\mathcal{S}^\ast\rangle = |J(\mathscr{M}),t\rangle$, where $\omega^{t(\mathbf{x})} = \omega^{-s(\mathbf{x})} (-1)^{(d+1)\langle (\mathbf{1}[v_i \neq 0])_{i=1}^n,\mathbf{w}\rangle}$.

For qubits ($d=2$), the involution operation $J$ is trivial, i.e., $J(\mathbf{x}) = \mathbf{x}$ for all $\mathbf{x}\in\mathbb{F}_2^{2n}$. As a consequence, the stabiliser state $|\mathcal{S}^\ast\rangle$ is characterised by the same Lagrangian subspace as $|\mathcal{S}\rangle$, possibly only with a different phase. Hence $|\mathcal{S}^\ast\rangle = \mathcal{W}_{\mathbf{z}}|\mathcal{S}\rangle$ for some $\mathbf{z}\in\mathbb{F}_2^{2n}$, so that $|\mathcal{S}^\ast\rangle$ is related to $|\mathcal{S}\rangle$ via a Weyl operator, which ultimately is the reason why a simple Bell sampling on $|\mathcal{S}\rangle^{\otimes 2}$ works for qubits. Allcock et al.~\cite{allcock2024beyond} explored Bell (difference) sampling on $|\mathcal{S}\rangle^{\otimes 4}$ for prime dimension $d$ and found that it returns a random string in $\mathscr{M} + J(\mathscr{M})$ which is the whole space $\mathbb{Z}_d^{2n}$ in the worst case. A new idea is thus needed to sample elements from $\mathscr{M}$.

Our main idea towards mitigating the action of the involution and thus proposing a new and useful Bell sampling operation is to ``mix'' several distinct and independent Bell sampling subroutines~\cite{montanaro2017learning,gross2021schur} using a global unitary. More precisely, such ``mixing'' is carried out by conjugating the underlying projectors of several independent Bell samplings with a single global unitary, as formalised below.

\begin{definition}[$U$-skewed Bell difference sampling]
    Let $U\in((\mathbb{C}^d)^{\otimes n})^{\otimes 4k}$ be a unitary. The $U$-skewed Bell difference sampling on $4k$ states in $(\mathbb{C}^d)^{\otimes n}$ is the projective measurement
    \begin{align*}
        U^\dagger\left(\bigotimes_{i\in[k]}\Pi_{\mathbf{x}_i}\right) U \quad\text{where}\quad
        \Pi_{\mathbf{x}_i} = \sum_{\mathbf{y} \in\mathbb{Z}_d^{2n}}\ |\mathcal{W}_{\mathbf{y}}\rangle\langle\mathcal{W}_{\mathbf{y}}| \otimes |\mathcal{W}_{\mathbf{x}_i+\mathbf{y}}\rangle\langle\mathcal{W}_{\mathbf{x}_i+\mathbf{y}}|, \quad\forall \mathbf{x}_1,\dots,\mathbf{x}_k\in\mathbb{Z}_d^{2n}.
    \end{align*}
    If $U$ is the identity, then the above recovers $k$ instances of the standard Bell difference sampling. 
\end{definition}

Our second main contribution is to actually prove the existence of a unitary $U\in((\mathbb{C}^d)^{\otimes n})^{\otimes 4k}$ that leads to a useful and interesting $U$-skewed Bell difference sampling. For such, we focus on unitaries that permute the computational basis of $((\mathbb{C}^d)^{\otimes n})^{\otimes k}$ using a linear map in $\mathbb{Z}_d^{n\times k}$.
\begin{definition}\label{def:unitary_B-d}
    For integers $d\geq 2$, $k\geq 1$, and matrix $\mathbf{R}\!\in\!\mathbb{Z}^{k\times k}$, let the unitary $\mathcal{B}_{\mathbf{R}}\in((\mathbb{C}^d)^{\otimes n})^{\otimes k}$
    \begin{align*}
         \mathcal{B}_{\mathbf{R}}:|\mathbf{Q}\rangle \mapsto |\mathbf{Q}\mathbf{R} ~(\operatorname{mod}d)\rangle, \quad\text{where} \quad \mathbf{Q} = [\mathbf{q}_1,\dots,\mathbf{q}_k]\in\mathbb{Z}_d^{n\times k}.
    \end{align*}
\end{definition}

The unitary $\mathcal{B}_{\mathbf{R}}$ maps the states $\bigotimes_{i\in[k]}|\mathbf{q}_i\rangle$ onto $\bigotimes_{i\in[k]}|\mathbf{q}'_i\rangle$ where $\mathbf{q}'_i$ is linear combination of $\{\mathbf{q}_i\}_{i\in[k]}$ described by the integer matrix $\mathbf{R}$. Note that $\mathcal{B}_{\mathbf{R}}$ depends only on $\mathbf{R}\operatorname{mod}d$, so that $\mathcal{B}_{\mathbf{R}} = \mathcal{B}_{\mathbf{R}\operatorname{mod}d}$. As will become clear below, the main property required from the matrix $\mathbf{R}$ is that $\mathbf{R}^\top\mathbf{R} = -\mathbf{I}\operatorname{mod}d$. Interestingly enough, such property can be enforced for any dimension $d$ using $k=4$ and Lagrange's four-square theorem, which states that every positive integer can be written as the sum of four integer squares.

%
\begin{fact}[Lagrange's four-square theorem]\label{fact:lagrange_four_square}
    For every $m\in\mathbb{N}$, there are $a_1,a_2,a_3,a_4\in\mathbb{N}\cup\{0\}$ such that $m = a_1^2 + a_2^2 + a_3^2 + a_4^2$. Moreover, there are randomised algorithms for computing a single representation $m = a_1^2 + a_2^2 + a_3^2 + a_4^2$ in expected time $O(\log^2{m}/\log\log{m})$~{\rm \cite{rabin1986randomized,pollack2018finding}}.
\end{fact}

\begin{lemma}\label{lem:matrix_R}
    For integer $d\geq 2$, let $a_1,a_2,a_3,a_4\in\mathbb{N}\cup\{0\}$ such that $a_1^2 + a_2^2 + a_3^2 + a_4^2 = d-1$. Then the matrix
    \begin{align*}
        \mathbf{R} = \begin{pmatrix}
            a_1 & a_2 & a_3 & a_4 \\
            a_2 & -a_1 & a_4 & -a_3 \\
            a_3 & -a_4 & -a_1 & a_2\\
            a_4 & a_3 & -a_2 & -a_1
        \end{pmatrix} (\operatorname{mod}d) \in\mathbb{Z}_d^{4\times 4}
    \end{align*}
    satisfies $\mathbf{R}^\top\mathbf{R} = \mathbf{R}\mathbf{R}^\top = -\mathbf{I} \operatorname{mod}d$.
\end{lemma}

In the following, we explore the unitary $\mathcal{B}_{\mathbf{R}}$ given a general matrix $\mathbf{R}\in\mathbb{Z}^{k\times k}$ such that $\mathbf{R}^\top\mathbf{R} = (D-1)\mathbf{I}$ or $\mathbf{R}^\top\mathbf{R} = (d-1)\mathbf{I}$. In practice, however, a smaller matrix $\mathbf{R}$ is always preferable, as it leads to smaller complexities. \Cref{lem:matrix_R} shows that one can always choose $k=4$ for any dimension $d$. We start by exploring the conjugation of Weyl operators by $\mathcal{B}_{\mathbf{R}}$. 
\begin{theorem}\label{thr:B_d_action_weyl_operators}
    Let $\mathbf{R}_D\in\mathbb{Z}_D^{k\times k}$ and $\mathbf{R}_d\in\mathbb{Z}_d^{k\times k}$ such that $\mathbf{R}_D^\top\mathbf{R}_D = -\mathbf{I} \operatorname{mod}D$ and $\mathbf{R}_d^\top\mathbf{R}_d = -\mathbf{I} \operatorname{mod}d$.
    Then for all $\mathbf{X}=[\mathbf{x}_1,\dots,\mathbf{x}_k]\in\mathbb{Z}_D^{2n\times k}$,
    \begin{align*}
        \mathcal{B}_{\mathbf{R}_D} \left(\bigotimes_{i=1}^k\mathcal{W}_{\mathbf{x}_i} \right) \mathcal{B}_{\mathbf{R}_D}^\dagger = \bigotimes_{i=1}^k \mathcal{W}_{(\mathbf{X}\mathbf{R}_D)_i}^\ast \qquad\text{and}\qquad \mathcal{B}_{\mathbf{R}_d} \left(\bigotimes_{i=1}^k\mathcal{W}_{\mathbf{x}_i} \right) \mathcal{B}_{\mathbf{R}_d}^\dagger = \pm\bigotimes_{i=1}^k \mathcal{W}_{(\mathbf{X}\mathbf{R}_d)_i}^\ast,
    \end{align*}
    where $(\mathbf{X}\mathbf{R})_i$ is the $i$-th column of $\mathbf{X}\mathbf{R}$.
\end{theorem}
\begin{proof}
    Write $\mathbf{x}_i = (\mathbf{v}_i,\mathbf{w}_i)$ for $i\in[k]$ and let $\mathbf{V} = [\mathbf{v}_1,\dots,\mathbf{v}_k],\mathbf{W} = [\mathbf{w}_1,\dots,\mathbf{w}_k]\in\mathbb{Z}_D^{n\times k}$. We start with matrix $\mathbf{R}_D$. For any $\mathbf{Q}\in\mathbb{Z}_d^{n\times k}$ (we omit $\operatorname{mod}d$ inside the kets for simplicity),
    \begin{align*}
        \mathcal{B}_{\mathbf{R}_D} \left(\bigotimes_{i=1}^k\mathcal{W}_{\mathbf{x}_i} \right) \mathcal{B}_{\mathbf{R}_D}^\dagger|\mathbf{Q}\rangle &= \mathcal{B}_{\mathbf{R}_D} \left(\bigotimes_{i=1}^k\mathcal{W}_{\mathbf{x}_i} \right) |-\mathbf{Q}\mathbf{R}_D^\top\rangle\\
        &=  \tau^{\sum_{i=1}^k \langle \mathbf{v}_i,\mathbf{w}_i\rangle} \omega^{-\sum_{i=1}^k \langle \mathbf{v}_i, (\mathbf{Q}\mathbf{R}_D^\top)_i\rangle} \mathcal{B}_{\mathbf{R}_D}|\mathbf{W} - \mathbf{Q}\mathbf{R}_D^\top\rangle\\
        &= \tau^{\sum_{i=1}^k \langle \mathbf{v}_i,\mathbf{w}_i\rangle} \omega^{-\sum_{i=1}^k \langle \mathbf{v}_i, (\mathbf{Q}\mathbf{R}_D^\top)_i\rangle}|\mathbf{W}\mathbf{R}_D + \mathbf{Q}\rangle \tag{$\mathbf{R}_D^\top\mathbf{R}_D = (D-1)\mathbf{I}$}\\
        &= \tau^{\operatorname{Tr}[\mathbf{W}\mathbf{V}^\top]} \omega^{-\operatorname{Tr}[\mathbf{Q}\mathbf{R}_D^\top\mathbf{V}^\top]}|\mathbf{W}\mathbf{R}_D + \mathbf{Q}\rangle\\
        &= \tau^{-\operatorname{Tr}[\mathbf{W}\mathbf{R}_D\mathbf{R}_D^\top\mathbf{V}^\top]} \omega^{-\operatorname{Tr}[\mathbf{Q}\mathbf{R}_D^\top\mathbf{V}^\top]}|\mathbf{W}\mathbf{R}_D + \mathbf{Q}\rangle \tag{$\mathbf{R}_D^\top\mathbf{R}_D = (D-1)\mathbf{I}$}\\
        &= \tau^{\operatorname{Tr}[(\mathbf{W}\mathbf{R}_D)(-\mathbf{V}\mathbf{R}_D)^\top]} \omega^{\operatorname{Tr}[\mathbf{Q}(-\mathbf{V}\mathbf{R}_D)^\top]}|\mathbf{W}\mathbf{R}_D + \mathbf{Q}\rangle \\
        &= \bigotimes_{i=1}^k \mathcal{W}_{(\mathbf{X}\mathbf{R}_D)_i}^\ast|\mathbf{Q}\rangle.
    \end{align*}
    The argument with the matrix $\mathbf{R}_d$ is very similar, the difference being that $\tau^{\operatorname{Tr}[\mathbf{W}\mathbf{V}^\top]}$ is equal to $\tau^{\operatorname{Tr}[(\mathbf{W}\mathbf{R}_d)(-\mathbf{V}\mathbf{R}_d)^\top]}$ up to a sign, i.e.,  $\tau^{\operatorname{Tr}[\mathbf{W}\mathbf{V}^\top]} = \pm\tau^{\operatorname{Tr}[(\mathbf{W}\mathbf{R}_d)(-\mathbf{V}\mathbf{R}_d)^\top]}$.
\end{proof}

Since $\mathcal{W}_{\mathbf{x}}^\ast = \pm \mathcal{W}_{J(\mathbf{x})}$, the usefulness of $\mathcal{B}_{\mathbf{R}}$ is already contained in 
\Cref{thr:B_d_action_weyl_operators}: it manages to map a submodule $\mathscr{X}$ into its involution $J(\mathscr{X})$, which is exactly the main obstacle behind the standard Bell (difference) sampling. As a consequence of the above result and further proof on the power of $\mathcal{B}_{\mathbf{R}}$, we now show that it can transform several copies of $|\mathcal{S}\rangle$ into copies of $|\mathcal{S}^\ast\rangle$ up to some Weyl operator and phase depending on $\mathcal{S}$. In the following, we denote the functional dependence of $\mathbf{x}\in\mathbb{Z}_d^{2n}$ on a stabiliser group $\mathcal{S}$ as $\mathbf{x}(\mathcal{S})$.

\begin{theorem}\label{thr:stabiliser_transformation}
    For an integer $d\geq 2$, consider a matrix $\mathbf{R}\in\mathbb{Z}_d^{k\times k}$ such that $\mathbf{R}^\top\mathbf{R} = -\mathbf{I} \operatorname{mod}d$. Then for any stabiliser group $\mathcal{S}$,
    \begin{align*}
        \mathcal{B}_{\mathbf{R}}|\mathcal{S}\rangle^{\otimes k} = \pm \left(\bigotimes_{i=1}^k \mathcal{W}_{\mathbf{x}_i(\mathcal{S})}\right)|\mathcal{S}^\ast\rangle^{\otimes k}
    \end{align*}
    for some  $\mathbf{x}_1(\mathcal{S}),\dots,\mathbf{x}_k(\mathcal{S})\in\mathbb{Z}_d^{2n}$ depending on $|\mathcal{S}\rangle$.
\end{theorem}
\begin{proof}
    Let $|\mathcal{S}\rangle = |\mathscr{M},s\rangle$ be the stabiliser state of $\mathcal{S} = \{\omega^{s(\mathbf{x})}\mathcal{W}_{\mathbf{x}}:\mathbf{x}\in\mathscr{M}\}$, where $\mathscr{M}\subset\mathbb{Z}_d^{2n}$ and $s:\mathbb{Z}_d^{2n}\to\mathbb{Z}_d$ are its underlying Lagrangian submodule and phase function, respectively. For any $\mathbf{X}=[\mathbf{x}_1,\dots,\mathbf{x}_k]\in \mathscr{M}^{k}$,
    \begin{align*}
        \mathcal{B}_{\mathbf{R}}|\mathcal{S}\rangle^{\otimes k} = \mathcal{B}_{\mathbf{R}} \left(\bigotimes_{i=1}^k \omega^{s(\mathbf{x}_i)} \mathcal{W}_{\mathbf{x}_i}\right) \mathcal{B}_{\mathbf{R}}^\dagger \mathcal{B}_{\mathbf{R}} |\mathcal{S}\rangle^{\otimes k} = \pm\left(\bigotimes_{i=1}^k \omega^{s(\mathbf{x}_i)} \mathcal{W}^\ast_{(\mathbf{X}\mathbf{R})_i} \right) \mathcal{B}_{\mathbf{R}} |\mathcal{S}\rangle^{\otimes k},
    \end{align*}
    where we used \Cref{thr:B_d_action_weyl_operators} in the last equality. Write $\mathbf{y}_i = (\mathbf{X}\mathbf{R})_i ~(\operatorname{mod}d)$ for $i\in[k]$. Since $\mathcal{W}^\ast_{(\mathbf{X}\mathbf{R})_i} = \pm \mathcal{W}_{J(\mathbf{y}_i)}$ and $\mathbf{y}_i \in\mathscr{M}$, the above expression means that $\mathcal{B}_{\mathbf{R}}|\mathcal{S}\rangle^{\otimes k}$ is stabilised (up to a phase) by $\bigotimes_{i\in[k]} \mathcal{W}_{\mathbf{x}_i}$ for all $\mathbf{x}_1,\dots,\mathbf{x}_k\in J(\mathscr{M})$. Therefore, $\mathcal{B}_{\mathbf{R}}|\mathcal{S}\rangle^{\otimes k}$ must be equal to the tensor product of stabiliser states with Lagrangian submodule $J(\mathscr{M})$. In other words, $\mathcal{B}_{\mathbf{R}}|\mathcal{S}\rangle^{\otimes k} = \bigotimes_{i\in[k]} |J(\mathscr{M}),s_i\rangle$ for phase functions $s_i:\mathbb{Z}_d^{2n} \to \mathbb{Z}_d$. The different phase functions $s_i$ can be transformed into the phase function $s_\ast$ of $|\mathcal{S}^\ast\rangle$ by means of some Weyl operator up to a sign, i.e., $|J(\mathscr{M}),s_i\rangle = \pm\mathcal{W}_{\mathbf{z}_i}|J(\mathscr{M}),s_\ast\rangle$ for some $\mathbf{z}_i\in\mathbb{Z}_d^{2n}$. This proves the result.
\end{proof}

\begin{remark}
    Since $\mathcal{B}_{\mathbf{R}}^\dagger = \mathcal{B}_{-\mathbf{R}^\top}$, then also $\mathcal{B}_{\mathbf{R}}^\dagger|\mathcal{S}\rangle^{\otimes k} = \mathcal{P}(\mathcal{S})|\mathcal{S}^\ast\rangle^{\otimes k}$ for some Pauli $\mathcal{P}(\mathcal{S})$ depending on $\mathcal{S}$.
\end{remark}

Given enough copies of a stabiliser state $|\mathcal{S}\rangle$, the unitary $\mathcal{B}_{\mathbf{R}}$ can successfully transform them into their complex conjugate $|\mathcal{S}^\ast\rangle$ up to some Pauli operator. By pairing these transformed states $|\mathcal{S}^\ast\rangle$ with the original states $|\mathcal{S}\rangle$, we can then sample from $p_{\mathcal{S}}(\mathbf{x})$, and thus $\mathscr{M}$, via the standard Bell difference sampling as mentioned above. The next result explores what happens when we consider general states $|\psi\rangle$ within skewed Bell difference sampling and not only stabiliser states.
\begin{theorem}\label{thr:skewed_bell_sampling}
    Let $|\psi\rangle\in(\mathbb{C}^d)^{\otimes n}$ be a pure state. Let $\mathcal{B}_{\mathbf{R}}\in((\mathbb{C}^d)^{\otimes n})^{\otimes k}$ be the unitary from {\rm \Cref{def:unitary_B-d}} with $\mathbf{R}\in\mathbb{Z}_d^{k\times k}$ such that $\mathbf{R}^\top\mathbf{R} = -\mathbf{I}\operatorname{mod}d$. Let $\mathcal{U}_{\mathbf{R}}\in((\mathbb{C}^d)^{\otimes n})^{\otimes 2k}$ be the unitary acting on $\bigotimes_{i\in[2k]}|\mathbf{q}_i\rangle$ that applies $\mathcal{B}_{\mathbf{R}}$ onto $\bigotimes_{i\in[k]}|\mathbf{q}_{2i}\rangle$ and the identity onto the remaining registers. Then $(\mathcal{U}_{\mathbf{R}}\otimes\mathcal{U}_{\mathbf{R}})$-skewed Bell difference sampling on $|\psi\rangle^{\otimes 4k}$ corresponds to sampling from the distribution
    \begin{align*}
        b_\psi(\mathbf{x}_1,\dots,\mathbf{x}_k) &\triangleq \operatorname{Tr}\!\left[\!(\mathcal{U}_{\mathbf{R}}\otimes\mathcal{U}_{\mathbf{R}})\!\left(\bigotimes_{i=1}^k \Pi_{\mathbf{x}_i}\right)\!(\mathcal{U}_{\mathbf{R}}^\dagger\otimes\mathcal{U}_{\mathbf{R}}^\dagger)\psi^{4k}\right] \\
        &= \sum_{\mathbf{Y}\in\mathbb{Z}_d^{2n\times k}} \prod_{i=1}^k \omega^{[\mathbf{x}_i,\mathbf{Y}_i]} p_\psi(\mathbf{Y}_i)p_\psi((\mathbf{Y}\mathbf{R})_i) = \sum_{\mathbf{Y}\in\mathbb{Z}_d^{2n\times k}} \prod_{i=1}^k p_\psi(\mathbf{x}_i + \mathbf{Y}_i) p_\psi((\mathbf{Y}\mathbf{R})_i).
    \end{align*}
    If $|\psi\rangle = |\mathcal{S}\rangle = |\mathscr{M},s\rangle$ is a stabiliser state, then
    \begin{align*}
        b_{\mathcal{S}}(\mathbf{x}_1,\dots,\mathbf{x}_k) = \begin{cases}
            d^{-kn} &\text{if}~\mathbf{x}_1,\dots,\mathbf{x}_k\in\mathscr{M},\\
            0 &\text{otherwise}.
        \end{cases}
    \end{align*}
\end{theorem}
We call the distribution $b_\psi(\mathbf{x})$ the $\mathbf{R}$-skewed (or simply skewed) Bell distribution of $|\psi\rangle$ and succinctly write $b_\psi(\mathbf{X})$ for $\mathbf{X} = [\mathbf{x}_1,\dots,\mathbf{x}_k]\in\mathbb{Z}_d^{2n\times k}$.
\begin{proof}
    First note that
    \begin{align*}
        &d^{4kn}(\mathcal{U}_{\mathbf{R}}\otimes\mathcal{U}_{\mathbf{R}})\left(\bigotimes_{i=1}^k \Pi_{\mathbf{x}_i}\right)(\mathcal{U}_{\mathbf{R}}^\dagger\otimes\mathcal{U}_{\mathbf{R}}^\dagger) \\
        &= \sum_{\mathbf{y}_1,\dots,\mathbf{y}_k\in\mathbb{Z}_d^{2n}} \!\!\mathcal{U}_{\mathbf{R}} \!\left(\bigotimes_{i=1}^k\sum_{\mathbf{z}'_i\in\mathbb{Z}_d^{2n}} \omega^{[\mathbf{y}_i,\mathbf{z}_i]}\mathcal{W}_{\mathbf{z}'_i}\otimes \mathcal{W}_{\mathbf{z}'_i}^\ast \!\right)\mathcal{U}_{\mathbf{R}}^\dagger\otimes \mathcal{U}_{\mathbf{R}} \!\left(\bigotimes_{i=1}^k\sum_{\mathbf{z}_i\in\mathbb{Z}_d^{2n}} \omega^{[\mathbf{x}_i+\mathbf{y}_i,\mathbf{z}_i]}\mathcal{W}_{\mathbf{z}'_i}\otimes \mathcal{W}_{\mathbf{z}_i}^\ast\!\right)\mathcal{U}_{\mathbf{R}}^\dagger \tag{by \Cref{lem:bell_basis}}\\
        &= \sum_{\mathbf{Y},\mathbf{Z},\mathbf{Z}'\in\mathbb{Z}_d^{2n\times k}} (-1)^{\phi(\mathbf{Z}) + \phi(\mathbf{Z}')} \omega^{\sum_{i=1}^k [\mathbf{Y}_i,\mathbf{Z}'_i + \mathbf{Z}_i]} \left(\bigotimes_{i=1}^k\mathcal{W}_{\mathbf{Z}'_i}\otimes \mathcal{W}_{(\mathbf{Z}'\mathbf{R})_i} \right) \!\otimes\! \left(\bigotimes_{i=1}^k \omega^{[\mathbf{x}_i,\mathbf{Z}_i]}\mathcal{W}_{\mathbf{Z}_i}\otimes \mathcal{W}_{(\mathbf{Z}\mathbf{R})_i} \right). \tag{by \Cref{thr:B_d_action_weyl_operators}}
    \end{align*}
    In the above, $\phi(\mathbf{Z}) = \operatorname{Tr}[\mathbf{W}\mathbf{F}\mathbf{V}^\top]$ is the phase in \Cref{thr:B_d_action_weyl_operators} for the matrix $\mathbf{F}\in\{0,d\}^{k\times k}$ given by $\mathbf{F} \equiv \mathbf{R}^\top \mathbf{R} ~(\operatorname{mod} D)$ where multiplication is over $\mathbb{Z}$, and with $\mathbf{V} = [\mathbf{v}_1,\dots,\mathbf{v}_k], \mathbf{W} = [\mathbf{w}_1,\dots,\mathbf{w}_k]\in\mathbb{Z}_d^{n\times k}$ by writing $\mathbf{Z} = [(\mathbf{v}_1,\mathbf{w}_1),\dots,(\mathbf{v}_k,\mathbf{w}_k)]$. Continuing with the calculation,
    \begin{align*}
        &d^{4kn}(\mathcal{U}_{\mathbf{R}}\otimes\mathcal{U}_{\mathbf{R}})\left(\bigotimes_{i=1}^k \Pi_{\mathbf{x}_i}\right)(\mathcal{U}_{\mathbf{R}}^\dagger\otimes\mathcal{U}_{\mathbf{R}}^\dagger) \\
        &= d^{2kn}\sum_{\mathbf{Z},\mathbf{Z}'\in\mathbb{Z}_d^{2n\times k}} (-1)^{\phi(\mathbf{Z}) + \phi(\mathbf{Z}')} \mathbf{1}[\mathbf{Z} = -\mathbf{Z}']  \left(\bigotimes_{i=1}^k\mathcal{W}_{\mathbf{Z}'_i}\otimes \mathcal{W}_{(\mathbf{Z}'\mathbf{R})_i} \right)\otimes \left(\bigotimes_{i=1}^k \omega^{[\mathbf{x}_i,\mathbf{Z}_i]}\mathcal{W}_{\mathbf{Z}_i}\otimes \mathcal{W}_{(\mathbf{Z}\mathbf{R})_i} \right) \tag{by \Cref{lem:sum} over $\mathbf{Y}_i$}\\
        &= d^{2kn}\sum_{\mathbf{Z}\in\mathbb{Z}_d^{2n\times k}}  \left(\bigotimes_{i=1}^k\mathcal{W}_{\mathbf{E}_i - \mathbf{Z}_i}\otimes \mathcal{W}_{(\mathbf{E}\mathbf{R})_i - (\mathbf{Z}\mathbf{R})_i} \right)\otimes \left(\bigotimes_{i=1}^k \omega^{[\mathbf{x}_i,\mathbf{Z}_i]}\mathcal{W}_{\mathbf{Z}_i}\otimes \mathcal{W}_{(\mathbf{Z}\mathbf{R})_i} \right) \tag{here $\mathbf{E}\in\{0,d\}^{2n\times k}$ is such that $(\mathbf{E}_{i})_j = \mathbf{1}[(\mathbf{Z}_{i})_j \neq 0]$ for $i\in[k],j\in[2n]$} \\
        &= d^{2kn}\! \sum_{\mathbf{Z}\in\mathbb{Z}_d^{2n\times k}} \!(-1)^{(d+1)\sum_{i=1}^k[\mathbf{E}_i,\mathbf{Z}_i]+[(\mathbf{E}\mathbf{R})_i,(\mathbf{Z}\mathbf{R})_i]} 
        \left(\bigotimes_{i=1}^k\mathcal{W}_{\mathbf{Z}_i}^\dagger\otimes \mathcal{W}_{(\mathbf{Z}\mathbf{R})_i}^\dagger \right) \!\otimes\! \left(\bigotimes_{i=1}^k\omega^{[\mathbf{x}_i,\mathbf{Z}_i]} \mathcal{W}_{\mathbf{Z}_i}\otimes \mathcal{W}_{(\mathbf{Z}\mathbf{R})_i} \right) \tag{by $\mathcal{W}_{-\mathbf{x}} = \mathcal{W}_{\mathbf{x}}^\dagger$ and $\mathcal{W}_{\mathbf{x} + d\mathbf{z}} = (-1)^{(d+1)[\mathbf{x},\mathbf{z}]}\mathcal{W}_{\mathbf{x}}$} \\
        &= d^{2kn}\sum_{\mathbf{Z}\in\mathbb{Z}_d^{2n\times k}} \left(\bigotimes_{i=1}^k\mathcal{W}_{\mathbf{Z}_i}^\dagger\otimes \mathcal{W}_{(\mathbf{Z}\mathbf{R})_i}^\dagger \right)\otimes \left(\bigotimes_{i=1}^k\omega^{[\mathbf{x}_i,\mathbf{Z}_i]} \mathcal{W}_{\mathbf{Z}_i}\otimes \mathcal{W}_{(\mathbf{Z}\mathbf{R})_i} \right) \tag{by $(-1)^{\sum_{i=1}^k[(\mathbf{E}\mathbf{R})_i, (\mathbf{Z}\mathbf{R})_i]} = (-1)^{\sum_{i=1}^k[\mathbf{E}_i, \mathbf{Z}_i]}$ since $\mathbf{R}^\top\mathbf{R} = -\mathbf{I}\operatorname{mod}d$} 
    \end{align*}
    Therefore,
    \begin{align*}
        b_\psi(\mathbf{x}_1,\dots,\mathbf{x}_k) &= \frac{1}{d^{2kn}} \sum_{\mathbf{Z}\in\mathbb{Z}_d^{2n\times k}} \!\operatorname{Tr}\!\left[ \left(\bigotimes_{i=1}^k\mathcal{W}_{\mathbf{Z}_i}^\dagger \psi\otimes \mathcal{W}_{(\mathbf{Z}\mathbf{R})_i}^\dagger \psi \right) \!\otimes \!\left(\bigotimes_{i=1}^k \omega^{ [\mathbf{x}_i,\mathbf{Z}_i] }\mathcal{W}_{\mathbf{Z}_i}\psi\otimes \mathcal{W}_{(\mathbf{Z}\mathbf{R})_i}\psi \right)\right] \\
        &= \sum_{\mathbf{Z}\in\mathbb{Z}_d^{2n\times k}} \prod_{i=1}^k \omega^{[\mathbf{x}_i,\mathbf{Z}_i]} p_\psi(\mathbf{Z}_i)p_\psi((\mathbf{Z}\mathbf{R})_i). 
    \end{align*}
    Note that $p_\psi$ is invariant modulo $d$, therefore we can interpret $\mathbf{Z}\mathbf{R}$ as being in $\mathbb{Z}_d^{2n\times k}$. The above expression is one part of the result and can be further simplified to
    \begin{align*}
        b_\psi(\mathbf{x}_1,\dots,\mathbf{x}_k) &= \sum_{\mathbf{Z},\mathbf{Y},\mathbf{Y}'\in\mathbb{Z}_d^{2n\times k}} \prod_{i=1}^k \omega^{[\mathbf{x}_i,\mathbf{Z}_i]}\omega^{[\mathbf{Z}_i,\mathbf{Y}_i] + [(\mathbf{Z}\mathbf{R})_i,\mathbf{Y}'_i]}  \widehat{p_\psi}(\mathbf{Y}_i) \widehat{p_\psi}(\mathbf{Y}'_i) \tag{Fourier expansion} \\
        &= \frac{1}{d^{2kn}}\sum_{\mathbf{Z},\mathbf{Y},\mathbf{Y}'\in\mathbb{Z}_d^{2n\times k}} \prod_{i=1}^k \omega^{[\mathbf{x}_i,\mathbf{Z}_i]}\omega^{[\mathbf{Z}_i,\mathbf{Y}_i] + [(\mathbf{Z}\mathbf{R})_i,(\mathbf{Y}'\mathbf{R})_i]}  p_\psi(\mathbf{Y}_i) p_\psi((\mathbf{Y}'\mathbf{R})_i) \tag{by \Cref{lem:invariant} and $\mathbf{Y}' \gets \mathbf{Y}'\mathbf{R}$} \\
        &= \frac{1}{d^{2kn}}\sum_{\mathbf{Z},\mathbf{Y},\mathbf{Y}'\in\mathbb{Z}_d^{2n\times k}} \prod_{i=1}^k \omega^{[\mathbf{x}_i,\mathbf{Z}_i]}\omega^{[\mathbf{Z}_i,\mathbf{Y}_i] - [\mathbf{Z}_i,\mathbf{Y}'_i]}  p_\psi(\mathbf{Y}_i) p_\psi((\mathbf{Y}'\mathbf{R})_i) \tag{due to $\mathbf{R}^\top\mathbf{R} = -\mathbf{I}\operatorname{mod}d$} \\
        &= \sum_{\mathbf{Y}\in\mathbb{Z}_d^{2n\times k}} \prod_{i=1}^k p_\psi(\mathbf{x}_i + \mathbf{Y}_i) p_\psi((\mathbf{Y}\mathbf{R})_i). \tag{by \Cref{lem:sum} over $\mathbf{Z}_i$}
    \end{align*}
    If $|\psi\rangle = |\mathcal{S}\rangle = |\mathscr{M},s\rangle$ is a stabiliser state, then we know that $p_{\mathcal{S}}(\mathbf{y}) = d^{-n}$ if $\mathbf{y}\in\mathscr{M}$ and $p_{\mathcal{S}}(\mathbf{y}) = 0$ if $\mathbf{y}\notin\mathscr{M}$. This means that $\prod_{i=1}^k p_{\mathcal{S}}(\mathbf{x}_i + \mathbf{Y}_i) p_{\mathcal{S}}((\mathbf{Y}\mathbf{R})_i) \neq 0$ if and only if $\mathbf{x}_i + \mathbf{Y}_i \in \mathscr{M}$ and $(\mathbf{Y}\mathbf{R})_i \in \mathscr{M}$ for $i\in[k]$. Since $\mathscr{M}$ is a submodule, this implies that $\mathbf{x}_i,\mathbf{Y}_i\in\mathscr{M}$ for $i\in[k]$ if and only if $\prod_{i=1}^k p_{\mathcal{S}}(\mathbf{x}_i + \mathbf{Y}_i) p_{\mathcal{S}}((\mathbf{Y}\mathbf{R})_i) \neq 0$. Therefore,
    \[
        b_{\mathcal{S}}(\mathbf{x}_1,\dots,\mathbf{x}_k) = \begin{cases}
            d^{-kn} &\text{if}~\mathbf{x}_1,\dots,\mathbf{x}_k\in\mathscr{M},\\
            0 &\text{otherwise}.
        \end{cases}\qedhere
    \]
\end{proof}

\Cref{thr:skewed_bell_sampling} is the much sought-after generalisation of Bell sampling: a properly skewed Bell difference sampling can successfully sample generators from a stabiliser group if sufficient copies of its stabiliser state are given. We compare the above result with the qubit case ($d=2$) $q_\psi(\mathbf{x}) = \sum_{\mathbf{y}\in\mathbb{F}_2^{2n}}p_\psi(\mathbf{x} + \mathbf{y})p_\psi(\mathbf{y})$ from Gross, Nezami, and Walter~\cite{gross2021schur}. Due to the action of the matrix $\mathbf{R}$ in order to map $|\mathcal{S}\rangle^{\otimes k} \mapsto |\mathcal{S}^\ast\rangle^{\otimes k}$, all characteristic distributions inside $b_\psi$ become non-trivially correlated.

From the above results, it should be clear that the number of copies of $|\psi\rangle$ required by the unitary $\mathcal{B}_{\mathbf{R}}$ is related to the size of the integer matrix $\mathbf{R}$. As shown in \Cref{lem:matrix_R}, a size $k=4$ is sufficient for all dimensions $d$, meaning that our skewed Bell difference sampling requires $4k = 16$ copies of $|\psi\rangle$ in order to return $k=4$ samples. Nonetheless, specific dimensions $d$ might require less copies of $|\psi\rangle$. As examples, the sum-of-two-squares theorem states that $m\in\mathbb{N}$ can be written as the sum of two squares of integers, $m = a_1^2 + a_2^2$, if and only if its prime decomposition contains no factor $p^\ell$ where prime $p\equiv 3~(\operatorname{mod}4)$ and $\ell$ is odd. More specifically for odd prime dimensions, it is well known that if $d\equiv 3~(\operatorname{mod}4)$, then there are $a_1,a_2\in\mathbb{F}_d$ such that $a_1^2 + a_2^2 \equiv -1~(\operatorname{mod}d)$. Also, $d\equiv 1~(\operatorname{mod}4)$ if and only if $-1$ is a quadratic residue modulo $d$, i.e., there is $a_1\in\mathbb{F}_d$ such that $a_1^2 \equiv -1~(\operatorname{mod}d)$. In this case, the unitary $\mathcal{B}_{\mathbf{R}}$ can be replaced with the identity. We summarise these remarks in the next result.
\begin{corollary}\label{cor:dimension_langrange}
    Let $|\mathcal{S}\rangle \in (\mathbb{C}^d)^{\otimes n}$ be any stabiliser state. Then:
    \begin{enumerate}
        \item If $p^\ell \nmid (d-1)$ for prime $p\equiv 3~(\operatorname{mod}4)$ and $\ell$ odd, then there is a unitary that maps $|\mathcal{S}\rangle^{\otimes 2} \mapsto \mathcal{P}_1\otimes\mathcal{P}_2|\mathcal{S}^\ast\rangle^{\otimes 2}$ for all $|\mathcal{S}\rangle\in(\mathbb{C}^d)^{\otimes n}$, where $\mathcal{P}_1,\mathcal{P}_2$ depend on $\mathcal{S}$; 
        \item If $d$ is prime and $d\equiv 3~(\operatorname{mod}4)$, there is a unitary that maps $|\mathcal{S}\rangle^{\otimes 2} \mapsto \mathcal{P}_1\otimes\mathcal{P}_2|\mathcal{S}^\ast\rangle^{\otimes 2}$ for all $|\mathcal{S}\rangle\in(\mathbb{C}^d)^{\otimes n}$, where $\mathcal{P}_1,\mathcal{P}_2$ depend on $\mathcal{S}$;
        \item If $d$ is prime and $d\equiv 1~(\operatorname{mod}4)$, then $|\mathcal{S}\rangle = \mathcal{P} |\mathcal{S}^\ast\rangle$ where $\mathcal{P}$ depends on $\mathcal{S}$.
    \end{enumerate}
\end{corollary}


\subsection{Further properties of distributions $p_\psi$ and $b_\psi$}

The distributions $p_\psi$ and $b_\psi$ are quite rich and we now explore them in more details. From \Cref{thr:skewed_bell_sampling,lem:sum}, we obtain the following simple corollaries. The first regards the mass on a submodule $\mathscr{X}\subseteq\mathbb{Z}_d^{2n}$ under $b_\psi$.
The distributions $p_\psi$ and $b_\psi$ are quite rich and we now derive some extra properties. From \Cref{thr:skewed_bell_sampling,lem:sum}, we obtain the following simple corollaries. The first regards the mass on a submodule $\mathscr{X}\subseteq\mathbb{Z}_d^{2n}$ under $b_\psi$.
\begin{corollary}\label{lem:properties_chateristic_Weyl}
    Let $\mathscr{X}_1,\dots,\mathscr{X}_k\subseteq\mathbb{Z}_d^{2n}$ be submodules and $\mathbf{X} = [\mathbf{x}_1,\dots,\mathbf{x}_k]\in \bigotimes_{i=1}^k\mathscr{X}_i$ be the matrix with columns $\mathbf{x}_i\in\mathscr{X}_i$, $i\in[k]$. Then
    \begin{align*}
        \frac{1}{\prod_{i=1}^k|\mathscr{X}_i|}\sum_{\mathbf{X}\in\bigotimes_{i=1}^k\mathscr{X}_i} b_\psi(\mathbf{X}) &= \sum_{\mathbf{X}\in\bigotimes_{i=1}^k\mathscr{X}_i^{\independent}} \prod_{i=1}^k p_\psi(\mathbf{X}_i) p_\psi((\mathbf{X}\mathbf{R})_i).
    \end{align*}
\end{corollary}
The second gives an expression for the marginal distribution of $b_\psi$.
\begin{corollary}\label{cor:marginal_distributions}
    For all $|\psi\rangle\in(\mathbb{C}^d)^{\otimes n}$,
    \begin{align*}
        \sum_{\mathbf{x}_2,\dots,\mathbf{x}_k\in\mathbb{Z}_d^{2n}} b_\psi(\mathbf{x}_1,\dots,\mathbf{x}_k) = d^{(k-1)n} \sum_{\mathbf{y}\in\mathbb{Z}_d^{2n}} \omega^{[\mathbf{x}_1,\mathbf{y}]} p_\psi(\mathbf{y})\prod_{i=1}^k p_\psi(a_i\mathbf{y}).
    \end{align*}
\end{corollary}

\Cref{lem:properties_chateristic_Weyl} has several important applications. The first one tells us that the mass of $b_\psi$ on a submodule can be upper bounded by the $p_\psi$-mass.
\begin{lemma}\label{lem:b_psi_mass}
    For any submodule $\mathscr{X}\subseteq\mathbb{Z}_d^{2n}$, $\sum_{\mathbf{X}\in\mathscr{X}^{\otimes k}} b_\psi(\mathbf{X}) \leq \big(\sum_{\mathbf{x}\in\mathscr{X}} p_\psi(\mathbf{x})\big)^k$.
\end{lemma}
\begin{proof}
    \begin{align*}
        \sum_{\mathbf{X}\in\mathscr{X}^{\otimes k}} b_\psi(\mathbf{X}) &= |\mathscr{X}|^k \sum_{\mathbf{X}\in(\mathscr{X}^{\independent})^{\otimes k}} \prod_{i=1}^k p_\psi(\mathbf{X}_i) p_\psi((\mathbf{X}\mathbf{R})_i) \tag{by \Cref{lem:properties_chateristic_Weyl}}\\
        &\leq \frac{|\mathscr{X}|^k}{d^{kn}} \sum_{\mathbf{X}\in(\mathscr{X}^{\independent})^{\otimes k}} \prod_{i=1}^k p_\psi(\mathbf{X}_i) \tag{$p_\psi((\mathbf{X}\mathbf{R})_i) \leq d^{-n}$}\\
        &= \left(\sum_{\mathbf{x}\in\mathscr{X}} p_\psi(\mathbf{x})\right)^k. \tag*{(by \Cref{lem:properties_chateristic_Weyl_1})\qquad\qedhere}
    \end{align*}
\end{proof}

We further obtain that $p_\psi$ and $b_\psi$ are supported on ${\operatorname{Weyl}}(|\psi\rangle)^{\independent}$, which is a generalisation of~\cite[Lemma~4.3 \& Corollary~4.4]{grewal2023improved}.
\begin{lemma}\label{lem:support}
    Let $|\psi\rangle\in(\mathbb{C}^d)^{\otimes n}$ and its characteristic and skewed Bell distributions $p_\psi$ and $b_\psi$. The support of $p_\psi$ and $b_\psi$ are contained in $\operatorname{Weyl}(|\psi\rangle)^{\independent}$ and $(\operatorname{Weyl}(|\psi\rangle)^{\independent})^{\otimes k}$, respectively.
\end{lemma}
\begin{proof}
    We show that the mass of $p_\psi$ on $\operatorname{Weyl}(|\psi\rangle)^{\independent}$ equals $1$.
    \begin{align*}
        \sum_{\mathbf{x}\in\operatorname{Weyl}(|\psi\rangle)^{\independent}} p_\psi(\mathbf{x}) &= \frac{|\operatorname{Weyl}(|\psi\rangle)^{\independent}|}{d^n}\sum_{\mathbf{y}\in \operatorname{Weyl}(|\psi\rangle)} p_\psi(\mathbf{y}) \tag{by \cref{lem:properties_chateristic_Weyl_1}}\\
        &= \frac{|\operatorname{Weyl}(|\psi\rangle)^{\independent}|}{d^n}\frac{|\operatorname{Weyl}(|\psi\rangle)|}{d^n} = 1. \tag{$p_\psi(\mathbf{x}) = d^{-n}$ iff $\mathbf{x}\in\operatorname{Weyl}(|\psi\rangle)$}
    \end{align*}
    Now, since $b_\psi(\mathbf{X}) = \sum_{\mathbf{Y}\in\mathbb{Z}_d^{2n\times k}} \prod_{i=1}^k p_\psi(\mathbf{X}_i + \mathbf{Y}_i) p_\psi((\mathbf{Y}\mathbf{R})_i)$ and $p_\psi$ is supported on $\operatorname{Weyl}(|\psi\rangle)^{\independent}$, which is a submodule, we must have that $b_\psi(\mathbf{X}) \neq 0 \iff \mathbf{X}_i + \mathbf{Y}_i, (\mathbf{Y}\mathbf{R})_i\in \operatorname{Weyl}(|\psi\rangle)^{\independent}$, and thus $\mathbf{X}\in(\operatorname{Weyl}(|\psi\rangle)^{\independent})^{\otimes k}$.
\end{proof}

We have previously seen that the characteristic function of a stabiliser state $|\mathcal{S}\rangle$ is uniformly distributed on its associated Lagrangian submodule $\mathscr{M}$, i.e., $p_{\mathcal{S}}(\mathbf{x}) = d^{-n}$ if $\mathbf{x}\in\mathscr{M}$ and $p_{\mathcal{S}}(\mathbf{x}) = 0$ otherwise (see \eqref{eq:stabiliser_distribution}). Such fact can be recovered from the above \Cref{lem:support}, since ${\operatorname{Weyl}}(|\mathcal{S}\rangle)^{\independent} = \mathscr{M}^{\independent} = \mathscr{M}$ and $p_{\mathcal{S}}(\mathbf{x}) = d^{-n}$ if and only if $\mathbf{x}\in{\operatorname{Weyl}}(|\mathcal{S}\rangle) = \mathscr{M}$.

We now prove that a submodule whose symplectic complement has high $b_\psi$-mass must be isotropic. For such, we shall need the following fact.
\begin{fact}[{\cite[Lemma~3.10]{gross2021schur}}]\label{fact:commutativity}
    Let $|\psi\rangle\in(\mathbb{C}^d)^{\otimes n}$ and $\mathbf{x},\mathbf{y}\in\mathbb{Z}_d^{2n}$ such that $p_{\psi}(\mathbf{x}),p_{\psi}(\mathbf{y}) \geq d^{-n}\big(1 - \frac{1}{4d^2}\big)$. Then $[\mathbf{x},\mathbf{y}]_d = 0$.
\end{fact}
\begin{lemma}\label{lem:isotropic}
    Let $\mathscr{X}\subseteq\mathbb{Z}_d^{2n}$ be a submodule such that
    \begin{align*}
        \sum_{\mathbf{X}\in(\mathscr{X}^{\independent})^{\otimes k}} b_\psi(\mathbf{X}) > \left(1 - \frac{1}{2d^2}\left(1-\frac{1}{p_1}\right)\left(1 - \frac{1}{8d^2}\right) \right)^k,
    \end{align*}
    where $p_1$ is the smallest prime that divides $d$. Then $\mathscr{X}$ is isotropic.
\end{lemma}
\begin{proof}
    Consider the set $\mathscr{A}\triangleq \{\mathbf{x}\in \mathscr{X}:d^n p_\psi(\mathbf{x}) \geq 1-\frac{1}{4d^2}\}$. According to \Cref{fact:commutativity}, $[\mathbf{x},\mathbf{y}]_d = 0$ for all $\mathbf{x},\mathbf{y}\in\mathscr{A}$. Therefore, the submodule $\langle\mathscr{A}\rangle$ generated by the span of all elements in $\mathscr{A}$ is isotropic. We now show that actually $\langle \mathscr{A}\rangle = \mathscr{X}$. For such, we assume by contradiction that $\langle \mathscr{A}\rangle \subset \mathscr{X}$ is a proper submodule of $\mathscr{X}$. We then have that $|\mathscr{A}| \leq \frac{|\mathscr{X}|}{p_1}$. Hence
    \begin{align*}
        \sum_{\mathbf{X}\in(\mathscr{X}^{\independent})^{\otimes k}} b_\psi(\mathbf{X}) &= |\mathscr{X}^{\independent}|^k \sum_{\mathbf{X}\in\mathscr{X}^{\otimes k}} \prod_{i=1}^k p_\psi(\mathbf{X}_i) p_\psi((\mathbf{X}\mathbf{R})_i) \tag{by \Cref{lem:properties_chateristic_Weyl}}\\
        &\leq  |\mathscr{X}^{\independent}|^k \sum_{\mathbf{X}\in\mathscr{X}^{\otimes k}} \prod_{i=1}^k p_\psi(\mathbf{X}_i)^2 \tag{by Cauchy-Schwarz}\\
        &= |\mathscr{X}^{\independent}|^k\left(\sum_{\mathbf{x}\in\mathscr{A}} p_\psi(\mathbf{x})^2 + \sum_{\mathbf{x}\in\mathscr{X}\setminus\mathscr{A}} p_\psi(\mathbf{x})^2 \right)^k \\
        &\leq |\mathscr{X}^{\independent}|^k \left(\frac{|\mathscr{A}|}{d^{2n}} + \frac{|\mathscr{X}|-|\mathscr{A}|}{d^{2n}}\left(1-\frac{1}{4d^2}\right)^2 \right)^k \\
        &\leq \frac{|\mathscr{X}|^k|\mathscr{X}^{\independent}|^k}{d^{2kn}} \left(\frac{1}{p_1} + \left(1-\frac{1}{p_1}\right)\left(1-\frac{1}{4d^2}\right)^2 \right)^k \tag{$|\mathscr{A}| \leq \frac{|\mathscr{X}|}{p_1}$}\\
        &= \left(\frac{1}{p_1} + \left(1-\frac{1}{p_1}\right)\left(1-\frac{1}{4d^2}\right)^2 \right)^k, \tag{$|\mathscr{X}||\mathscr{X}^{\independent}| = d^{2n}$}
    \end{align*}
    which contradicts the assumption of the lemma. Hence $\langle \mathscr{A}\rangle = \mathscr{X}$ and so $\mathscr{X}$ is isotropic.

    From now on, we shall consider the integer matrix $\mathbf{R}\in\mathbb{Z}_d^{4\times 4}$ from \Cref{lem:matrix_R} with $k=4$.
\end{proof}

\section{Hidden Stabiliser Group Problem / Stabiliser State Learning}
\label{sec:hidden_stabiliser_group}

The fundamental problem of identifying an unknown stabiliser state $|\mathcal{S}\rangle\in(\mathbb{C}^d)^{\otimes n}$ can be framed as an instance of the so-called \emph{Hidden Stabiliser Group Problem} defined by Hinsche, Eisert, and Carrasco~\cite{hinsche2025abelianstatehiddensubgroup}, who narrowed down the \emph{State Hidden Subgroup Problem} (StateHSP) introduced by Bouland, Giurgică-Tiron, and Wright~\cite{bouland2025state} to the case when the hidden group is a partial stabiliser group.\footnote{\cite{hinsche2025abelianstatehiddensubgroup} defines the Hidden Stabiliser Group Problem by supposing that $|\psi\rangle\in(\mathbb{C}^d)^{\otimes n}$ has a non-trivial partial stabiliser group $\mathcal{X}\subseteq\mathscr{P}_d^n$ such that $\mathcal{P}|\psi\rangle = |\psi\rangle$ for all $\mathcal{P}\in \mathcal{X}$ and $|\langle\psi|\mathcal{P}|\psi\rangle| \leq 1-\varepsilon$ for all $\mathcal{P}\in\mathscr{P}_d^n\setminus \mathcal{X}$. We believe that this is not an appropriate definition since for any $\mathcal{P}\in\mathcal{X}$ and $s\in\mathbb{Z}_d\setminus\{0\}$, $\omega^s \mathcal{P}\notin \mathcal{X}$ and still $|\langle\psi|\omega^s \mathcal{P}|\psi\rangle| = 1$.}
\begin{definition}[Hidden Stabiliser Group Problem]\label{def:hidden_stabiliser_group_problem}
    Given $\varepsilon>0$, let $|\psi\rangle\in(\mathbb{C}^d)^{\otimes n}$ such that $|\langle\psi|\mathcal{W}_{\mathbf{x}}|\psi\rangle| < 1 - \varepsilon$ for all $\mathbf{x}\notin {\operatorname{Weyl}}(|\psi\rangle)$. The \emph{hidden stabiliser group problem} asks to identify ${\operatorname{Weyl}}(|\psi\rangle)$ given access to copies of $|\psi\rangle$.
\end{definition}

Hinsche, Eisert, and Carrasco~\cite{hinsche2025abelianstatehiddensubgroup} proposed a quantum algorithm to solve the Hidden Stabiliser Group Problem that uses $O(n\log{d}\max\{d,\varepsilon^{-1}\})$ copies of $|\psi\rangle$ and runs in polynomial time.
Their algorithm uses the POVM given by $\Pi_{\mathbf{x}}^{\rm HEC} = d^{-2n}\sum_{\mathbf{y}\in\mathbb{Z}_d^{2n}} \omega^{[\mathbf{y},\mathbf{x}]} \mathcal{W}_{\mathbf{y}}^{\otimes D}$ to sample elements from, and thus learn, ${\operatorname{Weyl}}(|\psi\rangle)^{\independent}$. A similar idea can be employed for all dimensions $d\geq 2$ but now using our generalised Bell difference sampling subroutine: by sampling enough elements from the distribution $b_\psi$, their span will generate ${\operatorname{Weyl}}(|\psi\rangle)^{\independent}$ with high probability, from which ${\operatorname{Weyl}}(|\psi\rangle)$ can be obtained. In order to analyse the probability of successfully generating the whole submodule ${\operatorname{Weyl}}(|\psi\rangle)^{\independent}$, we shall need the following lemma, which generalises~\cite[Lemma~2.3]{grewal2023efficient} and~\cite[Lemma~4.21]{chen2025stabilizer}.

\begin{lemma}\label{lem:sampling_distribution_fact}
    Let $\delta\in(0,1)$, $\varepsilon>0$, and $\mathcal{D}$ be a distribution over $\mathbb{Z}_d^{n}$. Let $d = \prod_{i=1}^\ell p_i^{k_i}$ be the prime decomposition of $d$. Let $\mathscr{X}\subseteq\mathbb{Z}_d^{n}$ be the submodule spanned by $m$ i.i.d.\ samples from $\mathcal{D}$. If $m \geq \frac{2}{\varepsilon}\big(\!\ln\frac{1}{\delta} + n\sum_{i=1}^\ell k_i\big)$, then $\mathcal{D}(\mathscr{X}) \triangleq \sum_{\mathbf{x}\in\mathscr{X}} \mathcal{D}(\mathbf{x}) \geq 1 - \varepsilon$ with probability at least $1-\delta$.
\end{lemma}
\begin{proof}
    For $0\leq i \leq m$, let $\mathscr{X}_i \triangleq \langle \mathbf{x}_1,\dots,\mathbf{x}_i\rangle$ with $\mathscr{X}_0 \triangleq \{0^{n}\}$. Let the indicator random variable
    \begin{align*}
        Z_i = \begin{cases}
            1 &\text{if}~\mathbf{x}_i\in\mathbb{Z}_d^{n}\setminus \mathscr{X}_{i-1} ~\text{or}~ \mathcal{D}(\mathscr{X}_{i-1}) \geq 1- \varepsilon,\\
            0 &\text{otherwise}.
        \end{cases}
    \end{align*}
    The value $Z_i=1$ indicates that either the new sample $\mathbf{x}_i$ has increased the span of $\mathscr{X}_i$ or that $\mathscr{X}_{i-1}$ already accounts for a $(1-\varepsilon)$-fraction of the mass of $\mathcal{D}$. The following facts are true:
    \begin{enumerate}
        \item For any $\mathbf{x}_1,\dots,\mathbf{x}_{i-1}$, $\operatorname{Pr}[Z_i=1|\mathbf{x}_1,\dots,\mathbf{x}_{i-1}] \geq \varepsilon$. Indeed, if $\mathcal{D}(\mathscr{X}_{i-1}) > 1-\varepsilon$, then already $Z_i = 1$, while if $\mathcal{D}(\mathscr{X}_{i-1}) \leq 1-\varepsilon$, then the probability that $Z_i = 1$ is $1-\mathcal{D}(\mathscr{X}_{i-1}) \geq \varepsilon$.
        \item As a consequence of the above fact, $\mathbb{E}[Z_i] \geq \varepsilon$.
        \item Whenever $\sum_{i=1}^m Z_i \geq n\sum_{i=1}^\ell k_i$, we have $\mathcal{D}(\mathscr{X}_m) \geq 1-\varepsilon$. To see that, assume by contradiction that $\mathcal{D}(\mathscr{X}_m) < 1-\varepsilon$, in which case $\mathcal{D}(\mathscr{X}_i) < 1-\varepsilon$ for all $i\leq m$, and so $Z_i=1$ implies that $\mathbf{x}_i \in \mathbb{Z}_d^{n}\setminus \mathscr{X}_{i-1}$. There are at most $n\sum_{i=1}^\ell k_i$ such $i$, which is the length of the module $\mathbb{Z}_d^{n}$, therefore $\mathbf{x}_1,\dots,\mathbf{x}_m$ must span the whole space $\mathbb{Z}_d^{n}$, which contradicts the assumption $\mathcal{D}(\mathscr{X}_m) < 1-\varepsilon$.
    \end{enumerate}
    Building on the above facts, we now prove the statement via a Chernoff bound, the only caveat being the fact that the $Z_i$'s not independent. In order to deal with that, consider the new random variables $Z_i'$ (for $i\in[m]$) as $m$ i.i.d.\ samples from the Bernoulli distribution $\operatorname{Pr}[Z_i'=1] = \varepsilon$. Then $\operatorname{Pr}[Z_i' = 1|Z_1',\dots,Z_{i-1}'] = \varepsilon \leq \operatorname{Pr}[Z_i=1|Z_1,\dots,Z_{i-1}]$ from the first fact, and thus $\operatorname{Pr}[\sum_{i=1}^m < n\sum_{i=1}^\ell k_i] \leq \operatorname{Pr}[\sum_{i=1}^m Z_i' < n\sum_{i=1}^\ell k_i]$. Let $\theta \triangleq 1 - \frac{n}{m\varepsilon}\sum_{i=1}^\ell k_i$. Then
    \begin{align*}
        \operatorname{Pr}[\mathcal{D}(\mathscr{X}_m) \geq 1-\varepsilon] &\geq \operatorname{Pr}\left[\sum_{i=1}^m Z_i \geq n\sum_{i=1}^\ell k_i\right] \tag{by third fact}\\
        &\geq 1 - \operatorname{Pr}\left[\sum_{i=1}^m Z_i' < n\sum_{i=1}^\ell k_i\right] \tag{by first fact}\\
        &= 1 - \operatorname{Pr}\left[\sum_{i=1}^m Z_i' < (1-\theta)m\varepsilon\right]\\
        &\geq 1 - \exp\left(-\frac{\theta^2m\varepsilon}{2}\right) \tag{Chernoff bound}\\
        &\geq 1 - \exp\left(-\frac{m\varepsilon}{2} + n\sum_{i=1}^\ell k_i \right) \geq 1-\delta. \qedhere
    \end{align*}
\end{proof}

\begin{theorem}\label{thr:hidden_stabiliser_group}
    Let $\varepsilon>0$ and $\delta\in(0,1)$. Let $|\psi\rangle\in(\mathbb{C}^d)^{\otimes n}$ be an unknown state such that $|\langle\psi|\mathcal{W}_{\mathbf{x}}|\psi\rangle| < 1-\varepsilon$ for all $\mathbf{x}\notin {\operatorname{Weyl}}(|\psi\rangle)$. Let $d = \prod_{i=1}^\ell p_i^{k_i}$ be the prime factorisation of $d$ with largest prime factor $p_\ell$. {\rm \cref{algo:hidden_stabiliser_group}} identifies ${\operatorname{Weyl}}(|\psi\rangle)$ with probability $1-\delta$ using
    \begin{align*}
        &O\left(\frac{1}{\varepsilon}{\min}\!\left\{n\sum_{i=1}^\ell k_i + \log\frac{1}{\delta}, n\log{p_\ell} + \log\frac{\ell}{\delta}\right\}\right) \quad\text{copies of}~ |\psi\rangle,\\
        &O\left(\frac{n^2}{\varepsilon}{\min}\!\left\{n\sum_{i=1}^\ell k_i + \log\frac{1}{\delta}, n\log{p_\ell} + \log\frac{\ell}{\delta}\right\}\right) \quad\text{time}, 
    \end{align*}
    and acting on at most $8$ copies of $|\psi\rangle$ at a time.
\end{theorem}

\begin{algorithm}[t]
\caption{Hidden Stabiliser Group Problem}
\DontPrintSemicolon
\label{algo:hidden_stabiliser_group}
\SetKwInput{KwPromise}{Promise}

    \KwIn{$8m+8$ copies of $|\psi\rangle\in(\mathbb{C}^d)^{\otimes n}$,  $m\triangleq \min\big\{\big\lceil\frac{4\ln\frac{1}{\delta} + 32n\sum_{i=1}^\ell k_i}{\varepsilon}, \frac{\ln\frac{\ell}{\delta} + 2n\ln{p_\ell}}{2\varepsilon} \big\rceil$ and $d = \prod_{i=1}^\ell p_i^{k_i}$ is the prime decomposition of $d$ and $p_\ell$ its largest prime factor.}
    \KwPromise{$|\langle\psi|\mathcal{W}_{\mathbf{x}}|\psi\rangle| < 1 - \varepsilon$ for all $\mathbf{x}\notin{\operatorname{Weyl}}(|\psi\rangle)$.}
    \KwOut{a succinct description of ${\operatorname{Weyl}}(|\psi\rangle)$.}

   \For{$i\gets -1$ \KwTo $m-1$}
   {    
        Perform Bell sampling on $(\mathbf{I}\otimes\mathcal{B}_{\mathbf{R}}^\dagger)|\mathcal{S}\rangle^{\otimes 8}$ to obtain $\mathbf{x}_{4i+1},\mathbf{x}_{4i+2},\mathbf{x}_{4i+3},\mathbf{x}_{4i+4}\in\mathbb{Z}_d^{2n}$
        
        \If{$i\neq -1$}
        {
            $\mathbf{x}_{4i+j} \gets \mathbf{x}_{4i+j}-\mathbf{x}_{-4+j}$ for $j=1,2,3,4$
        }
    }

    Determine $\mathscr{W} = \langle \{\mathbf{x}_i\}_{i\in[4m]}\rangle^{\independent}$ by putting the matrix $[\mathbf{x}_1,\dots,\mathbf{x}_{4m}]\in\mathbb{Z}_d^{2n\times 4m}$ into its Smith normal form

    \Return $\mathscr{W}$
\end{algorithm}

\begin{proof}
    By performing skewed Bell difference sampling $m$ times (which requires $8m+8$ copies of $|\psi\rangle$ since one can subtract the first $4$-tuple sample from all the subsequent ones), we obtain $\mathbf{x}_1,\dots,\mathbf{x}_{4m}\in {\operatorname{Weyl}}(|\psi\rangle)^{\independent}$. The time complexity is simply $O(mn^2)$. \Cref{algo:hidden_stabiliser_group} fails if all $4m$ samples are contained in a maximal proper submodule of ${\operatorname{Weyl}}(|\psi\rangle)^{\independent}$. 
    
    We provide two error analyses. For the first one, consider a submodule $\mathscr{X}^{\independent}\subset{\operatorname{Weyl}}(|\psi\rangle)^{\independent}$ (${\operatorname{Weyl}}(|\psi\rangle) \subset \mathscr{X}$) such that ${\operatorname{Weyl}}(|\psi\rangle)^{\independent}/\mathscr{X}^{\independent} \cong \mathbb{Z}/p_i\mathbb{Z}$. The probability that all samples $\mathbf{x}_1,\dots,\mathbf{x}_{4m}$ are in $\mathscr{X}^{\independent}$ is $\big(\sum_{\mathbf{X}\in(\mathscr{X}^{\independent})^{\otimes 4}} b_\psi(\mathbf{X})\big)^m$. We can bound
    \begin{align}
        \sum_{\mathbf{X}\in (\mathscr{X}^{\independent})^{\otimes 4}} b_\psi(\mathbf{X}) &= |\mathscr{X}^{\independent}|^4 \sum_{\mathbf{X}\in \mathscr{X}^{\otimes 4}} \prod_{i=1}^4 p_\psi(\mathbf{X}_i) p_\psi((\mathbf{X}\mathbf{R})_i) \tag{by \Cref{lem:properties_chateristic_Weyl}} \nonumber\\
        &\leq |\mathscr{X}^{\independent}|^4\sqrt{\sum_{\mathbf{X}\in\mathscr{X}^{\otimes 4}} \prod_{i=1}^4 p_\psi(\mathbf{X}_i)^2 } \sqrt{\sum_{\mathbf{X}\in\mathscr{X}^{\otimes 4}} \prod_{i=1}^4 p_\psi((\mathbf{XR})_i)^2} \tag{Cauchy-Schwarz} \nonumber\\
        &= |\mathscr{X}^{\independent}|^4\left(\sum_{\mathbf{x}\in\mathscr{X}} p_\psi(\mathbf{x})^2 \right)^4 \tag{$\mathbf{X} \gets \mathbf{X}\mathbf{R}^\top$} \nonumber\\
        &= |\mathscr{X}^{\independent}|^4\left(\sum_{\mathbf{x}\in{\operatorname{Weyl}}(|\psi\rangle)} p_\psi(\mathbf{x})^2 + \sum_{\mathbf{x}\in\mathscr{X}\setminus{\operatorname{Weyl}}(|\psi\rangle)} p_\psi(\mathbf{x})^2\right)^4 \nonumber\\
        &< |\mathscr{X}^{\independent}|^4\left(d^{-2n}|{\operatorname{Weyl}}(|\psi\rangle)| + d^{-2n}(1-\varepsilon)^4 (|\mathscr{X}| - |{\operatorname{Weyl}}(|\psi\rangle)|) \right)^4 \nonumber\\
        &\leq \left(\frac{1}{p_i} + (1-\varepsilon)^4\left(1 - \frac{1}{p_i}\right)\right)^4 \nonumber \tag{$|{\operatorname{Weyl}}(|\psi\rangle)| \leq \frac{|\mathscr{X}|}{p_i}$} \\
        &\leq (1-\varepsilon/2)^4. \label{eq:b_psi_mass}
    \end{align}
    Hence, by a union bound over all $\leq \frac{p_i^{2n}-1}{p_i-1}$ submodules $\mathscr{X}^{\independent}$ such that ${\operatorname{Weyl}}(|\psi\rangle)^{\independent}/\mathscr{X}^{\independent} \cong \mathbb{Z}/p_i\mathbb{Z}$ (\Cref{lem:number_submodules}) and all $i\in[\ell]$, the failure probability is, already taking $m\geq \frac{2n\ln{p_\ell} + \ln(\ell/\delta)}{2\varepsilon}$,  at most
    \begin{align*}
        \sum_{i=1}^\ell \frac{p_i^{2n} - 1}{p_i - 1} (1 - \varepsilon/2)^{4m} \leq \sum_{i=1}^\ell \e^{2n\ln{p_i} - 2\varepsilon m} \leq \sum_{i=1}^\ell \e^{-\ln(\ell/\delta)} = \delta.
    \end{align*}

    For our second error analysis, we consider a slightly weaker algorithm wherein one samples $\mathbf{X}^{(1)},\dots,\mathbf{X}^{(m)}\sim b_\psi$, computes the four separate spans $\mathscr{W}_j \triangleq \langle \{\mathbf{X}_j^{(i)}\}_{i\in[m]}\rangle$ for $j\in[4]$, and considers the largest one. In practice there is no need to separate the four samples coming from one Bell difference sampling, but since \Cref{lem:sampling_distribution_fact} requires i.i.d.\ samples and we cannot guarantee their independence, we opted to separate the samples. We know that all $\mathbf{X}^{(1)},\dots,\mathbf{X}^{(m)}\sim b_\psi$ belong to $({\operatorname{Weyl}}(|\psi\rangle)^{\independent})^{\otimes 4}$. By collecting $m \geq \frac{2}{\varepsilon}(2\ln\frac{1}{\delta} + 16n\sum_{i=1}^\ell k_i)$ samples $\mathbf{X}^{(1)},\dots,\mathbf{X}^{(m)}$ ($b_\psi$ is over $(\mathbb{Z}_d^{2n})^{\otimes 4}$), then according to \Cref{lem:sampling_distribution_fact} the $b_\psi$-mass of $\langle\mathbf{X}^{(1)},\dots,\mathbf{X}^{(m)}\rangle$ is at least $1-\varepsilon/2$. On the other hand, by a very similar calculation to \eqref{eq:b_psi_mass}, the $b_\psi$-mass of $\bigotimes_{i=1}^4\mathscr{X}_i^{\independent}$ for any proper submodules $\mathscr{X}_1^{\independent},\dots,\mathscr{X}_4^{\independent}\subset{\operatorname{Weyl}}(|\psi\rangle)^{\independent}$ is at most $(1-\varepsilon/2)^4$. This means that the largest span out of $\mathscr{W}_1,\mathscr{W}_2,\mathscr{W}_3,\mathscr{W}_4$ must equal ${\operatorname{Weyl}}(|\psi\rangle)^{\independent}$ (notice that we do not need to consider proper submodules of $({\operatorname{Weyl}}(|\psi\rangle)^{\independent})^{\otimes 4}$ but only of ${\operatorname{Weyl}}(|\psi\rangle)^{\independent}$ since we pick the largest span out of $\mathscr{W}_1,\mathscr{W}_2,\mathscr{W}_3,\mathscr{W}_4$).
\end{proof}

The quantum algorithm of~\cite{hinsche2025abelianstatehiddensubgroup} has sample complexity $O(n\log{d}\max\{d,\varepsilon^{-1}\})$. On the other hand, \Cref{algo:hidden_stabiliser_group} has sample complexity $O\big(\frac{n}{\varepsilon}\min\{\log{p_\ell},\sum_{i=1}^\ell k_i\}\big)$, or simply $O\big(\frac{n}{\varepsilon}\big)$ for prime dimensions, a substantial improvement. Moreover, our quantum algorithm acts on at most $8$ copies of $|\psi\rangle$ at a time ($4$ copies for prime dimensions and $2$ if further $d\equiv 1~(\operatorname{mod}4)$), while $\{\Pi_{\mathbf{x}}^{\rm HEC}\}_{\mathbf{x}}$ requires $D$ copies of $|\psi\rangle$ at a time, which can be impractical for large $d$.

As previously mentioned, the problem of learning an unknown stabiliser state $|\mathcal{S}\rangle = |\mathscr{M},s\rangle$ can be framed as an instance of Hidden Stabiliser Group Problem where one is interested in learning ${\operatorname{Weyl}}(|\psi\rangle) = \mathscr{M}$. Under the promise that $|\psi\rangle$ is a stabiliser state, $\varepsilon = 1$ in \Cref{def:hidden_stabiliser_group_problem}. The next result is thus a straight corollary of \Cref{thr:hidden_stabiliser_group}.
\begin{corollary}\label{thr:algorithm_bell_sampling}
    Let $\delta\in(0,1)$ and $|\mathcal{S}\rangle\in(\mathbb{C}^{d})^{\otimes n}$ an unknown stabiliser state. Let $d = \prod_{i=1}^\ell p_i^{k_i}$ be the prime factorisation of $d$ with smallest prime factor $p_1$. {\rm \cref{algo:hidden_stabiliser_group}} identifies $|\mathcal{S}\rangle$ with probability $1-\delta$ using $O(n+\log_{p_1}\frac{\ell}{\delta})$ copies of $|\mathcal{S}\rangle$ in $O(n^3 + n^2\log_{p_1}\frac{\ell}{\delta})$ time.
\end{corollary}
\begin{proof}
    Note that \eqref{eq:b_psi_mass} reduces to $\sum_{\mathbf{X}\in (\mathscr{X}^{\independent})^{\otimes 4}} b_\psi(\mathbf{X}) \leq 1/p_i^4$ if $\varepsilon = 1$. In this case, the failure probability becomes, already taking $m = \lceil\frac{n}{2} + \frac{1}{4}\log_{p_1}(\ell/\delta) \rceil$,
    \[
        \sum_{i=1}^\ell \frac{p_i^{2n}-1}{p_i-1} p_i^{-4m} \leq \sum_{i=1}^\ell p_i^{2n-4m} \leq \sum_{i=1}^\ell p_i^{-\log_{p_1}(\ell/\delta)} \leq \ell\cdot p_1^{-\log_{p_1}(\ell/\delta)} = \delta. \qedhere
    \]
\end{proof}

\section{Property Testing Stabiliser Size}

In this section, we focus on testing whether a given state $|\psi\rangle$ has stabiliser size at least $d^t$ or is $\varepsilon$-far from any state with stabiliser size at least $d^t$, promised that one of these cases holds. As we shall see, this section serves as a warm-up to \Cref{sec:pseudorandomness}. The results presented here generalise~\cite{grewal2023efficient} to qudits, who proposed a property testing algorithm for stabiliser dimension. Since the concept of dimension is not well defined for general $d$, the right concept to focus on is that of stabiliser size.

The idea of \Cref{alg:testing_stabiliser_size} is quite simple: sample several values from the skewed Bell distribution $b_\psi$ and compute the size of their span. If it is at most $d^{2n-t}$, then we output the case where $|\psi\rangle$ has stabiliser size at least $d^t$. If, on the other hand, the span has size greater than $d^{2n-t}$, then we output the case where $|\psi\rangle$ is $\varepsilon$-far from any state with stabiliser size at least $d^t$. Implicit in the algorithm is learning the submodule ${\operatorname{Weyl}}(|\psi\rangle)$, which was the task behind the Hidden Stabiliser Group Problem from \Cref{sec:hidden_stabiliser_group}. As a consequence, \Cref{alg:testing_stabiliser_size,algo:hidden_stabiliser_group} are quite similar. The correctness of \Cref{alg:testing_stabiliser_size} is proven in the next theorem. 

\begin{algorithm}[h]
\caption{Property testing stabiliser size}
\DontPrintSemicolon
\label{alg:testing_stabiliser_size}
\SetKwInput{KwPromise}{Promise}

    \KwIn{$8m+8$ copies of $|\psi\rangle\in(\mathbb{C}^d)^{\otimes n}$, where $m \triangleq \big\lceil\frac{2}{3\varepsilon}\big({\ln}\frac{1}{\delta} + 8n\sum_{i=1}^\ell k_i\big)\big\rceil$.}
    \KwPromise{$|\psi\rangle$ has stabiliser size at least $d^t$ or is $\varepsilon$-far in fidelity from all such states.}
    \KwOut{$1$ if $|\psi\rangle$ has stabiliser size at least $d^t$ and $0$ otherwise, with probability $\geq 1-\delta$.}
    
   \For{$i\gets -1$ \KwTo $m-1$}
   {    
        Perform Bell sampling on $(\mathbf{I}\otimes\mathcal{B}_{\mathbf{R}}^\dagger)|\psi\rangle^{\otimes 8}$ to obtain $\mathbf{x}_{4i+1},\mathbf{x}_{4i+2},\mathbf{x}_{4i+3},\mathbf{x}_{4i+4}\in\mathbb{Z}_d^{2n}$
        
        \If{$i\neq -1$}
        {
            $\mathbf{x}_{4i+j} \gets \mathbf{x}_{4i+j}-\mathbf{x}_{-4+j}$ for $j=1,2,3,4$
        }
    }

    Compute the size $K$ of $\langle\{\mathbf{x}_i\}_{i\in[4m]} \rangle$ by putting the matrix $[\mathbf{x}_1,\dots,\mathbf{x}_{4m}]\in\mathbb{Z}_d^{2n\times 4m}$ into its Smith normal form
    
    \Return $1$ if $K \leq d^{2n-t}$ and $0$ otherwise
\end{algorithm}

\begin{theorem}\label{thr:property_testing_stabiliser_size}
    Let $d = \prod_{i=1}^\ell p_i^{k_i}$ be the prime decomposition of $d$ and $p_1$ the smallest prime to divide $d$. Let $\delta\in(0,1)$ and $0<\varepsilon\leq\frac{1}{6d^2}\big(1-\frac{1}{p_1}\big)\big(1 - \frac{1}{8d^2}\big)$. Let $|\psi\rangle\in(\mathbb{C}^d)^{\otimes n}$ be a state promised to either have stabiliser size at least $d^t$ or be $\varepsilon$-far in fidelity from all such states. {\rm \cref{alg:testing_stabiliser_size}} distinguishes the two cases with probability at least $1-\delta$, uses $8\big\lceil\frac{2}{3\varepsilon}\big({\ln}\frac{1}{\delta} + 8n\sum_{i=1}^\ell k_i\big)\big\rceil + 8$ copies of $|\psi\rangle$, and has runtime $O\big(\frac{n^2}{\varepsilon}\big(\!\log\frac{1}{\delta} + n\sum_{i=1}^\ell k_i\big)\big)$.
\end{theorem}
\begin{proof}
    Assume first that $|\psi\rangle$ has stabiliser size at least $d^t$. According to \Cref{lem:support}, the distribution $b_\psi$ is supported on $({\operatorname{Weyl}}(|\psi\rangle)^{\independent})^{\otimes 4}$. Therefore, $\mathscr{X}^{\independent} \triangleq \langle \{\mathbf{x}_i\}_{i\in[4m]}\rangle$ is supported on ${\operatorname{Weyl}}(|\psi\rangle)^{\independent}$, which has size at most $d^{2n-t}$, and so \Cref{alg:testing_stabiliser_size} accepts with probability $1$.

    Now assume that $|\psi\rangle$ is $\varepsilon$-far in fidelity from all states with stabiliser size at least $d^t$. Since $m \geq \frac{1}{3\varepsilon}\big(2\ln\frac{1}{\delta} + 16n\sum_{i=1}^\ell k_i\big)$ (recall that $b_\psi$ is over $(\mathbb{Z}_d^{2n})^{\otimes 4}$), \Cref{lem:sampling_distribution_fact,lem:b_psi_mass} imply that, with probability at least $1-\delta$,
    \begin{align*}
        \sum_{\mathbf{x}\in\mathscr{X}^{\independent}} p_\psi(\mathbf{x}) \geq \left(\sum_{\mathbf{X}\in(\mathscr{X}^{\independent})^{\otimes 4}} b_\psi(\mathbf{X}) \right)^{\frac{1}{4}} \geq (1-3\varepsilon)^{\frac{1}{4}} \geq 1-\varepsilon,
    \end{align*}
    where we used that $\varepsilon<\frac{1}{2}$ in the last step. Under this event, \Cref{lem:properties_chateristic_Weyl_1} leads to
    \begin{align}\label{eq:mass_p_psi}
        \sum_{\mathbf{x}\in\mathscr{X}} p_\psi(\mathbf{x}) = \frac{|\mathscr{X}|}{d^n}\sum_{\mathbf{x}\in\mathscr{X}^{\independent}} p_\psi(\mathbf{x}) \geq \frac{1-\varepsilon}{K}.
    \end{align}
    On the other hand, since
    \begin{align*}
        \sum_{\mathbf{X}\in(\mathscr{X}^{\independent})^{\otimes 4}} b_\psi(\mathbf{X}) \geq 1-3\varepsilon \geq 1 - \frac{1}{2d^2}\left(1-\frac{1}{p_1}\right)\left(1 - \frac{1}{8d^2}\right),
    \end{align*}
    \Cref{lem:isotropic} implies that $\mathscr{X}$ is isotropic, which together with \eqref{eq:mass_p_psi} allows us to use \Cref{lem:existance_state_high_stabiliser_size} to argue that there is a state with stabiliser size at least $d^{2n}/K$ and fidelity at least $1-\varepsilon$ with $|\psi\rangle$. By assumption, $|\psi\rangle$ is $\varepsilon$-far from states with stabiliser size at least $d^t$, therefore $d^{2n}/K < d^t \implies K>d^{2n-t}$ and thus \Cref{alg:testing_stabiliser_size} rejects with probability at least $1-\delta$.

    Regarding the complexities, \Cref{alg:testing_stabiliser_size} uses $8m + 8$ copies of $|\psi\rangle$ and requires $O(mn^2)$ time to compute the Smith normal form of $\mathscr{X}^{\independent}$.
\end{proof}

\section{Testing Doped Clifford Circuits and Pseudorandomness Bounds}
\label{sec:pseudorandomness}

In this section, we prove that the output of any Clifford circuit augmented with a few non-Clifford single-qudit gates can be efficiently distinguished from a Haar-random quantum state, and therefore Clifford circuits require several non-Clifford single-qudit gates in order to generate pseudorandom quantum states (\cref{def:pseudorandomness}). These results generalise~\cite{grewal2022low,grewal2023improved} from qubits to qudits. Moreover, as we shall see, our final lower bound is exponentially better compared to the prior work of~\cite{allcock2024beyond}, who also focused on qudits. The ideas behind our proof are similar to the ones from~\cite{grewal2023improved}: a Haar-random state has vanishing stabiliser size with very high probability, while the output of a Clifford circuit with just a few non-Clifford single-qudit gates has a high stabiliser size. Before presenting our algorithm, though, we prove a few auxiliary results regarding Haar-random states.

\subsection{Anti-concentration of Haar-random states}

In this section, we show that the stabiliser size of a Haar random state $|\psi\rangle$ is $1$ with high probability, or, equivalently, $b_\psi$ is supported on the entire space $\mathbb{Z}_d^{2n\times 4}$ with high probability. To do so, we will make use of L\'evy's lemma~\cite{milman1986asymptotic,ledoux2001concentration}.
\begin{fact}[L\'evy's lemma]
    Let $f: \mathbb{S}^d\to \mathbb{R}$ be a function defined on the $d$-dimensional hypersphere $\mathbb{S}^d$. Assume $f$ is $K$-Lipschitz, meaning that $|f(\psi) - f(\phi)| \leq K\|\psi - \phi\|$. Then, for every $\epsilon > 0$,
    \begin{align*}
        \operatorname*{\mathbb{P}}_{\psi\sim\mu_{\rm Haar}}[|f(\psi) - \mathbb{E}[f]| \geq \epsilon] \leq 2\exp\left(-\frac{(d+1)\epsilon^2}{9\pi^3 K^2}\right).
    \end{align*}
\end{fact}

We will apply L\'evy's lemma to functions of the form $\langle\psi|\mathcal{W}_{\mathbf{x}}|\psi\rangle$, which are $2$-Lipschitz as shown next. The next result is basically~\cite[Lemma~20]{grewal2022low}.
\begin{lemma}\label{lem:Lipschitz}
    For any Weyl operator $\mathcal{W}_{\mathbf{x}}\in\mathscr{P}^n_d$, the function $f_{\mathbf{x}}:\mathbb{S}^{d^n}\to\mathbb{R}$ defined as $f_{\mathbf{x}}(|\psi\rangle) = \langle \psi|\mathcal{W}_{\mathbf{x}}|\psi\rangle$ is $2$-Lipschitz.
\end{lemma}
\begin{proof}
    $\begin{aligned}[t]
        |\langle\psi|\mathcal{W}_{\mathbf{x}}|\psi\rangle - \langle\phi|\mathcal{W}_{\mathbf{x}}|\phi\rangle| &= |\langle\psi|\mathcal{W}_{\mathbf{x}}|\psi\rangle - \langle\phi|\mathcal{W}_{\mathbf{x}}|\psi\rangle + \langle\phi|\mathcal{W}_{\mathbf{x}}|\psi\rangle -\langle\phi|\mathcal{W}_{\mathbf{x}}|\phi\rangle| \\ 
        &\leq |(\langle\psi| - \langle\phi|)\mathcal{W}_{\mathbf{x}}|\psi\rangle| + |\langle\phi|\mathcal{W}_{\mathbf{x}}(|\psi\rangle -|\phi\rangle)| \\
        &\leq \|\mathcal{W}_{\mathbf{x}} |\psi\rangle\|\||\psi\rangle - |\phi\rangle\| + \|\mathcal{W}_{\mathbf{x}} |\phi\rangle\|\||\psi\rangle - |\phi\rangle\| \\
        &= 2\||\psi\rangle - |\phi\rangle\|. \hspace{7.6cm}\qedhere
    \end{aligned}$
\end{proof}
Putting the above two results together, we show that all Weyl operators $\mathcal{W}_{\mathbf{x}}$ for $\mathbf{x}\neq \mathbf{0}$ have very low expectation under Haar random states.
\begin{lemma}\label{lem:haar_random_low_expectation}
    For any $\epsilon > 0$,
    \begin{align*}
        \operatorname*{\mathbb{P}}_{|\psi\rangle\sim\mu_{\rm Haar}}\left[\exists \mathbf{x}\in\mathbb{Z}_d^{2n}\setminus\{\mathbf{0}\}:|\langle \psi|\mathcal{W}_{\mathbf{x}}|\psi\rangle| \geq \epsilon \right] \leq 2d^{2n}\exp\left(-\frac{d^n\epsilon^2}{36\pi^3}\right).
    \end{align*}
\end{lemma}
\begin{proof}
    Consider the function $f_{\mathbf{x}}(|\psi\rangle) = \langle \psi|\mathcal{W}_{\mathbf{x}}|\psi\rangle$ for $\mathbf{x}\in\mathbb{Z}_d^{2n}\setminus\{\mathbf{0}\}$, which is $2$-Lipschitz according to \Cref{lem:Lipschitz}. Moreover, $\mathbb{E}[f_{\mathbf{x}}] = 0$ over the Haar measure because $d^{-1}$-fraction of the eigenvalues of $\mathcal{W}_{\mathbf{x}}$ are $\omega^s$, $s\in\mathbb{Z}_d$. The result thus follows from L\'evy's lemma and a union bound over all $d^{2n}$ possible Weyl operators.
\end{proof}

As a consequence of the above result, $b_\psi$ is supported on the entire space $\mathbb{Z}_d^{2n\times 4}$ with high probability if $|\psi\rangle$ is Haar random, or more precisely, that $b_\psi$ has a bounded fraction of mass on maximal proper submodules of $\mathbb{Z}_d^{2n\times 4}$, which generalises~\cite[Lemma~4.7]{grewal2023improved}.
\begin{lemma}\label{lem:upper_bound_bpsi_haar}
    Let $\epsilon>0$ and let $|\psi\rangle\in(\mathbb{C}^d)^{\otimes n}$ be a Haar random $n$-qudit state. Let $d = \prod_{i=1}^\ell p_i^{k_i}$ be the prime decomposition of $d$. Then, for all maximal proper submodules $\mathscr{X}\subset\mathbb{Z}_d^{2n}$, $\sum_{\mathbf{X}\in \mathscr{X}^4} b_\psi(\mathbf{X}) \leq \big(\frac{1}{p_j} + \epsilon^4 \big)^4$ with probability at least $1 - 2d^{2n}{\exp}\big({-}\frac{d^n\epsilon^2}{36\pi^3}\big)$, where the prime $p_j|d$ is such that $\mathbb{Z}_d^{2n}/\mathscr{X} \cong \mathbb{Z}/p_j\mathbb{Z}$.
\end{lemma}
\begin{proof}
    Consider any maximal proper submodule $\mathscr{X}\subset \mathbb{Z}_d^{2n}$. Then $\mathscr{X}^{\independent} \cong \mathbb{Z}/p_j\mathbb{Z}$ must be simple for some $p_j|d$. By \Cref{lem:properties_chateristic_Weyl},
    \begin{align*}
        \sum_{\mathbf{X}\in \mathscr{X}^4} b_\psi(\mathbf{X}) &= \frac{d^{8n}}{p_j^4} \sum_{\mathbf{X}\in (\mathscr{X}^{\independent})^4} \prod_{i=1}^4 p_\psi(\mathbf{X}_i) p_\psi((\mathbf{X}\mathbf{R})_i) = \frac{1}{p_j^4} \sum_{\mathbf{X}\in (\mathscr{X}^{\independent})^4} \prod_{i=1}^4 |\langle\psi|\mathcal{W}_{\mathbf{X}_i}|\psi\rangle\langle\psi|\mathcal{W}_{(\mathbf{X}\mathbf{R})_i}|\psi\rangle|^2.
    \end{align*}
    Note that, in the above expression, there are $(p_j-1)^4\binom{4}{0}$ terms of the form $\prod_{i=1}^8 |\langle\psi|\mathcal{W}_{\mathbf{x}_i}|\psi\rangle|^2$, $(p_j-1)^3\binom{4}{1}$ terms of the form $\prod_{i=1}^6 |\langle\psi|\mathcal{W}_{\mathbf{x}_i}|\psi\rangle|^2$, $(p_j-1)^2\binom{4}{2} $ terms of the form $\prod_{i=1}^4 |\langle\psi|\mathcal{W}_{\mathbf{x}_i}|\psi\rangle|^2$, and $(p_j-1)\binom{4}{3}$ terms of the form $\prod_{i=1}^2 |\langle\psi|\mathcal{W}_{\mathbf{x}_i}|\psi\rangle|^2$, all with $\mathbf{x}_1,\dots,\mathbf{x}_8\neq \mathbf{0}$. Finally, there is $1$ term $\langle\psi|\mathcal{W}_{\mathbf{0}}|\psi\rangle^8 = 1$. Conditioning on the event that $|\langle \psi|\mathcal{W}_{\mathbf{x}}|\psi\rangle| \leq \epsilon$ for all $\mathbf{x}\in\mathbb{Z}_d^{2n}\setminus\{\mathbf{0}\}$, which happens with probability at least $1 - 2d^{2n}{\exp}\big({-}\frac{d^n\epsilon^2}{36\pi^3}\big)$ (\Cref{lem:haar_random_low_expectation}), the mass of $b_\psi$ over $\mathscr{X}^4$ is
    \begin{align*}
        \sum_{\mathbf{X}\in \mathscr{X}^4} b_\psi(\mathbf{X}) \leq \frac{1}{p_j^4}\sum_{i=0}^4 \binom{4}{i} (p_j-1)^i \epsilon^{16i} = \frac{(1 + \epsilon^4(p_j-1))^4}{p_j^4} \leq \left(\frac{1}{p_j} + \epsilon^4 \right)^4.
    \end{align*}
    The above is valid for any maximal proper submodule $\mathscr{X}\subset \mathbb{Z}_d^{2n}$, which concludes the proof.
\end{proof}

\subsection{Stabiliser size of $t$-doped Clifford circuits}

Following~\cite{leone2023learning,salvatore2024unscrambling,leone2024learning}, we first define $t$-doped Clifford circuits over qudits. Recall that a Clifford gate is any element belonging to the Clifford group $\mathscr{C}^n_d$.
\begin{definition}[$t$-doped Clifford circuits]
    A $t$-doped Clifford circuit is a quantum circuit composed of Clifford gates and at most $t$ non-Clifford single-qudit gates that starts in the state $|0\rangle^{\otimes n}$.
\end{definition}
The main technical lemma from this section is the fact that every non-Clifford gate in a Clifford circuit decreases the stabiliser length by at most $1$.
\begin{lemma}\label{lem:t-doped}
    Let $|\psi\rangle\in(\mathbb{C}^d)^{\otimes n}$ be the output state of a $t$-doped Clifford circuit. Then the stabiliser size of $|\psi\rangle$ is at least $d^{n-2t}$, i.e., $|{\operatorname{Weyl}}(|\psi\rangle)| \geq d^{n-2t}$.
\end{lemma}
\begin{proof}
    The proof is by induction on $t$. For $t=0$, the output state is a stabiliser state, which has stabiliser size $d^n$. Assume then the induction hypothesis for $t-1$. Write $|\psi\rangle = \mathcal{C}\mathcal{U}|\phi\rangle$, where $|\phi\rangle$ is the output of a $(t-1)$-doped Clifford circuit, $\mathcal{U}$ is a single-qudit gate, and $\mathcal{C}$ is a Clifford circuit. Since the stabiliser size is unchanged by Clifford circuits (\Cref{cor:invariant}), we just need to show that the stabiliser size of $\mathcal{U}|\phi\rangle$ is at least $d^{n-2t}$. For such, consider ${\operatorname{Weyl}}(|\phi\rangle)$ with $|{\operatorname{Weyl}}(|\phi\rangle)| \geq d^{n- 2(t-1)}$ by the induction assumption. Note that, for any $\mathbf{x}\in{\operatorname{Weyl}}(|\phi\rangle)$, if $\mathcal{W}_{\mathbf{x}}$ commutes with $\mathcal{U}$, then
    \begin{align*}
        \langle \phi|\mathcal{U}^\dagger \mathcal{W}_{\mathbf{x}}\mathcal{U}|\phi\rangle = \langle \phi| \mathcal{W}_{\mathbf{x}}|\phi\rangle \in \{1,\omega,\dots,\omega^{d-1}\}.
    \end{align*}
    Consider the set of commuting elements with $\mathcal{U}$, $\mathscr{E} \triangleq \{\mathbf{x}\in{\operatorname{Weyl}}(|\phi\rangle) : \mathcal{U}^\dagger \mathcal{W}_{\mathbf{x}}\mathcal{U} = \mathcal{W}_{\mathbf{x}}\}$. Then the stabiliser size of $\mathcal{U}|\phi\rangle$ is at least the size of $\mathscr{E}$, but $|\mathscr{E}| \geq |{\operatorname{Weyl}}(|\phi\rangle)|/d^2$, because $\mathscr{E}$ contains all elements $\mathbf{x}\in{\operatorname{Weyl}}(|\phi\rangle)$ for which $\mathcal{W}_{\mathbf{x}}$ restricts to the identity on the qudit to which $\mathcal{U}$ applies, i.e., $x_i = (v_i,w_i) = (0,0) \in \mathbb{Z}_d^{2}$ if $\mathcal{U}$ acts on the $i$-th qudit. Therefore, the stabiliser size of $\mathcal{U}|\phi\rangle$ is at least $d^{n-2t}$.
\end{proof}

In \Cref{lem:support} we showed that the skewed Bell distribution $b_\psi$ is supported on $({\operatorname{Weyl}}(|\psi\rangle)^{\independent})^4$. Together with the above lemma, we conclude that if $|\psi\rangle$ is the output of a $t$-doped Clifford circuit, then $b_\psi$ is supported on a submodule of size at most $d^{4n + 8t}$.
\begin{corollary}
    If $|\psi\rangle\in(\mathbb{C}^d)^{\otimes n}$ is the output of a $t$-doped Clifford circuit, then the support of the skewed Bell distribution $b_\psi$ is a cartesian product of four submodules, each of size at most~$d^{n+2t}$.
\end{corollary}
\begin{proof}
    According to \Cref{lem:support}, the support of $b_\psi$ is contained in $({\operatorname{Weyl}}(|\psi\rangle)^{\independent})^4$, each of which with size $d^{2n}/|{\operatorname{Weyl}}(|\psi\rangle)| \leq d^{n+2t}$, where the last inequality follows from \Cref{lem:t-doped}.
\end{proof}

\subsection{Distinguishing $t$-doped Clifford circuits from Haar randomness}

In hold of all the results from the previous sections, we now prove that, given enough copies of $|\psi\rangle\in(\mathbb{C}^d)^{\otimes n}$, it is possible to distinguish whether (i) $|\psi\rangle$ is a Haar random state (ii) or $|\psi\rangle$ is the output of a $(n/2-1)$-doped Clifford circuit, under the promise that one case is the correct one. The idea of the algorithm is very simple: after performing skewed Bell difference sampling a number of $O(n)$ times, we check whether their span generates the whole space $\mathbb{Z}_d^{2n}$ or not. If $|\psi\rangle$ is Haar-random, then the sampled elements generate $\mathbb{Z}_d^{2n}$ with high probability, which can never happen if $|\psi\rangle$ is the output of a $t$-doped Clifford circuit for $t < n/2$, since $b_{\psi}$ is supported on a submodule of size at most $d^{4(n+2t)}$.

\begin{algorithm}[h]
\caption{Distinguishing $(n/2-1)$-doped Clifford circuits from Haar-random}
\DontPrintSemicolon
\label{alg:clifford}
\SetKwInput{KwPromise}{Promise}

    \KwIn{$8m+8$ copies of $|\psi\rangle\in(\mathbb{C}^d)^{\otimes n}$, where $m \triangleq \lceil \frac{n}{2}\frac{\ln{p_1}}{\ln(2p_1/3)} + \frac{\ln(\ell/\delta)}{4\ln(2p_1/3)} \rceil$ and $d=\prod_{i=1}^\ell p_i^{k_i}$ is the prime decomposition of $d$ with smallest prime factor.}
    \KwPromise{$|\psi\rangle$ is Haar-random or the output of a $(n/2-1)$-doped Clifford circuit}
    \KwOut{$0$ if $|\psi\rangle$ is Haar-random and $1$ otherwise, with probability at least $1-\delta$.}
    
   \For{$i\gets -1$ \KwTo $m-1$}
   {    
        Perform Bell sampling on $(\mathbf{I}\otimes\mathcal{B}_{\mathbf{R}}^\dagger)|\psi\rangle^{\otimes 8}$ to obtain $\mathbf{x}_{4i+1},\mathbf{x}_{4i+2},\mathbf{x}_{4i+3},\mathbf{x}_{4i+4}\in\mathbb{Z}_d^{2n}$
        
        \If{$i\neq -1$}
        {
            $\mathbf{x}_{4i+j} \gets \mathbf{x}_{4i+j}-\mathbf{x}_{-4+j}$ for $j=1,2,3,4$
        }
    }

    Compute the size $K$ of $\langle\{\mathbf{x}_i\}_{i\in[4m]} \rangle$ by putting the matrix $[\mathbf{x}_1,\dots,\mathbf{x}_{4m}]\in\mathbb{Z}_d^{2n\times 4m}$ into its Smith normal form
    
    \Return $0$ if $K = d^{2n}$ and $1$ otherwise
\end{algorithm}

\begin{theorem}\label{thr:algorithm_pseudorandomness}
    Let $\delta\in(0,1)$ and $|\psi\rangle\in(\mathbb{C}^d)^{\otimes n}$ a state promised to be either Haar-random or the output of a $t$-doped Clifford circuit for $t< \frac{n}{2}$. Let $d=\prod_{i=1}^\ell p_i^{k_i}$ be the prime decomposition of $d$ with smallest and largest prime factors $p_1$ and $p_\ell$, respectively. {\rm \cref{alg:clifford}} distinguishes the two cases with probability at least $1-\delta-2d^{2n}\exp\big({-}\frac{d^n}{36\pi^3\sqrt{2p_\ell}}\big)$ using $O(n + \log_{p_1}\frac{\ell}{\delta})$ copies of $|\psi\rangle$ and in $O(n^3 + n^2\log_{p_1}\frac{\ell}{\delta})$ time.
\end{theorem}
\begin{proof}
    If $|\psi\rangle$ is the output of a $t$-doped Clifford circuit with $t< n/2$, then $\langle\{\mathbf{x}_i\}_{i\in[4m]}\rangle$ always has size less than $d^{2n}$ according to \Cref{cor:dimension_langrange}, where $\mathbf{x}_1,\dots,\mathbf{x}_{4m}\in {\operatorname{Weyl}}(|\psi\rangle)^{\independent}$ are skewed Bell difference samples. This means that \Cref{alg:clifford} can only fail when $|\psi\rangle$ is Haar random, which happens if $\mathbf{x}_1,\dots,\mathbf{x}_{4m}\in {\operatorname{Weyl}}(|\psi\rangle)^{\independent}$ all fall in the same maximal proper submodule $\mathscr{X}\subset \mathbb{Z}_d^{2n}$. Assume that $\mathbb{Z}_d^{2n}/\mathscr{X} \cong \mathbb{Z}/p_i\mathbb{Z}$ for some prime $p_i|d$. The probability that all samples $\mathbf{x}_1,\dots,\mathbf{x}_{4m}$ are in $\mathscr{X}$ is $\big(\sum_{\mathbf{X}\in\mathscr{X}^4}b_\psi(\mathbf{X})\big)^m$. According to \Cref{lem:upper_bound_bpsi_haar}, with probability at least $1 - 2d^{2n}{\exp}\big({-}\frac{d^n\epsilon^2}{36\pi^3}\big)$, we have that $\big(\sum_{\mathbf{X}\in\mathscr{X}^4}b_\psi(\mathbf{X})\big)^m \leq \big(\frac{1}{p_i} + \epsilon^4\big)^{4m}$. Take $\epsilon^4 = \frac{1}{2p_\ell}$ so that $\frac{1}{p_i} + \epsilon^4 \leq \frac{3}{2p_i}$. There are at most $\frac{p_i^{2n} - 1}{p_i-1}$ submodules such that $\mathbb{Z}_d^{2n}/\mathscr{X} \cong \mathbb{Z}/p_i\mathbb{Z}$ according to \Cref{lem:number_submodules}. A union bound over all such submodules and $i\in[\ell]$ leads to
    \begin{align*}
        \sum_{i=1}^\ell \frac{p_i^{2n} - 1}{p_i-1}\left(\frac{1}{p_i} + \epsilon^4\right)^{4m} \leq \sum_{i=1}^\ell p_i^{2n}\left(\frac{3}{2p_i}\right)^{4m} = \sum_{i=1}^\ell \e^{2n\ln{p_i} - 4m\ln(2p_i/3)} \leq \sum_{i=1}^\ell \e^{-\ln(\ell/\delta)} = \delta
    \end{align*}
    taking $m \geq \frac{n}{2}\frac{\ln{p_1}}{\ln(2p_1/3)} + \frac{\ln(\ell/\delta)}{4\ln(2p_1/3)}$. The total failure probability also needs to take into account the Haar measure failure probability $2d^{2n}{\exp}\big({-}\frac{d^n}{36\pi^3\sqrt{2p_\ell}}\big)$.    
\end{proof}

The above results generalises~\cite[Lemma~4.8 \& Theorem~4.9]{grewal2023improved} to qudits and also exponentially improves the range of $t$ compared to~\cite{allcock2024beyond}, who provided a $\operatorname{poly}(n)$-sample-and-time-complexity algorithm for distinguishing Haar random states in $(\mathbb{C}^d)^{\otimes n}$ from $t$-doped Clifford circuits with $t=O(\log{n}/\log{d})$.

As an immediate corollary of the above result, no $t$-doped Clifford circuit with $t<n/2$ can produce pseudorandom quantum states, otherwise they would be distinguishable from Haar random states in $\operatorname{poly}(n)$ time.

\begin{definition}[{\cite[Definition~2]{ji2018pseudorandom}}]\label{def:pseudorandomness}
    Given a security parameter $\kappa$, a keyed family of $n$-qudit quantum states $\{|\phi_k\rangle\in(\mathbb{C}^d)^{\otimes n}\}_{k\in\{0,1\}^\kappa}$ is pseudorandom if (i) there is a polynomial-time quantum algorithm that generates $|\phi_k\rangle$ on input $k$, and (ii) for any $\operatorname{poly}(\kappa)$-time quantum adversary $\mathcal{A}$,
    \begin{align*}
        \left|\operatorname*{\mathbb{P}}_{k\sim\{0,1\}^\kappa}[\mathcal{A}(|\phi_k\rangle^{\otimes \operatorname{poly}(\kappa)}) = 1] - \operatorname*{\mathbb{P}}_{|\psi\rangle\sim\mu_{\rm Haar}}[\mathcal{A}(|\psi\rangle^{\otimes \operatorname{poly}(\kappa)}) = 1]\right| = \operatorname{negl}(\kappa),
    \end{align*}
    where $\mu_{\rm Haar}$ is the $n$-qudit Haar measure and $\operatorname{negl}(\kappa)$ is an arbitrary negligible function of $\kappa$, i.e., $\operatorname{negl}(\kappa) = o(\kappa^{-c})$ $\forall c>0$.
\end{definition}

\begin{corollary}\label{cor:pseudorandomness_lower bound}
    Any doped Clifford circuit that uses at most $n/2-1$ non-Clifford single-qudit gates cannot produce an ensemble of pseudorandom quantum states in $(\mathbb{C}^d)^{\otimes n}$.
\end{corollary}

\section{Tolerant Stabiliser State Testing}
\label{sec:stabiliser_testing}

Gross, Nezami, and Walter~\cite{gross2021schur} developed a stabiliser testing algorithm for qudits (for all $d\geq 2$) which distinguishes between a stabiliser state and a state with stabiliser fidelity at most $1-\varepsilon$ using $O(1/\varepsilon)$ copies. Here we extend their results to the tolerant version where one wants to test whether $F_{\mathcal{S}}(|\psi\rangle) \geq 1 - \varepsilon_1$ or $F_{\mathcal{S}}(|\psi\rangle) \leq 1 - \varepsilon_2$ for $\varepsilon_1 < \varepsilon_2$. For such, we employ two different strategies: one based on directly extending the non-tolerant algorithm of~\cite{gross2021schur}, and another based on our skewed Bell difference sampling procedure.


\subsection{Approach 1: employing the POVM of Gross, Nezami, and Walter}

For qubits, the stabiliser-testing algorithm of~\cite{gross2021schur} is quite simple: perform Bell difference sampling to obtain $\mathbf{x}\in\mathbb{F}_2^{2n}$, measure the Weyl operator $\mathcal{W}_{\mathbf{x}}$ twice, and check whether the outcomes are equal or not. For large enough $\varepsilon$, their algorithm thus employs only $6$ copies of $|\psi\rangle$, which is known to be almost optimal~\cite{damanik2018optimality}. Their algorithm for higher dimensions, however, does not have a simple interpretation. The authors introduced the operator
\begin{align*}
    \mathcal{V}_r \triangleq \frac{1}{d^{n}} \sum_{\mathbf{x}\in\mathbb{Z}_d^{2n}}(\mathcal{W}_{\mathbf{x}}\otimes \mathcal{W}_{\mathbf{x}}^\dagger)^{\otimes r} \qquad\text{for}~r\in\mathbb{N}
\end{align*}
and showed that it possesses all the required properties for stabiliser testing. We are interested in exploring such operator in more details, so we review some of its properties.
\begin{lemma}[\cite{gross2021schur}]
    Let $\mathcal{V}_r \triangleq d^{-n}\sum_{\mathbf{x}\in\mathbb{Z}_d^{2n}}(\mathcal{W}_{\mathbf{x}}\otimes \mathcal{W}_{\mathbf{x}}^\dagger)^{\otimes r}$. For $r\geq 2$ such that $\operatorname{gcd}(d,r) = 1$, $\mathcal{V}_r$ is a Hermitian unitary. As a consequence, the operator $\Pi_{r}^+ = \frac{1}{2}(\mathbf{I} + \mathcal{V}_r)$ is a projector and the binary POVM $\{\Pi_{r}^+, \mathbf{I} - \Pi_{r}^+\}$ can be implemented in $O(nr)$ time.
\end{lemma}
\begin{proof}
    It is easy to see that the operator $\mathcal{V}_r$ is Hermitian by a change of index. Moreover, $\mathcal{V}_r$ is unitary, since
    \begin{align*}
        \mathcal{V}_r^2 &= \frac{1}{d^{2n}}\sum_{\mathbf{x},\mathbf{y}\in\mathbb{Z}_d^{2n}}(\mathcal{W}_{\mathbf{x}}\mathcal{W}_{\mathbf{y}}\otimes \mathcal{W}_{\mathbf{x}}^\dagger \mathcal{W}_{\mathbf{y}}^\dagger)^{\otimes r} = \frac{1}{d^{2n}}\sum_{\mathbf{x},\mathbf{y}\in\mathbb{Z}_d^{2n}}\omega^{r[\mathbf{x},\mathbf{y}]}(\mathcal{W}_{\mathbf{x}+\mathbf{y}}\otimes \mathcal{W}_{\mathbf{x}+\mathbf{y}}^\dagger)^{\otimes r}\\
        &= \frac{1}{d^{2n}}\sum_{\mathbf{x},\mathbf{z}\in\mathbb{Z}_d^{2n}}\omega^{r[\mathbf{x},\mathbf{z}]}(\mathcal{W}_{\mathbf{z}}\otimes \mathcal{W}_{\mathbf{z}}^\dagger)^{\otimes r} = \frac{1}{d^{2n}}\sum_{\mathbf{x},\mathbf{z}\in\mathbb{Z}_d^{2n}}\omega^{[\mathbf{x},\mathbf{z}]}(\mathcal{W}_{\mathbf{z}}\otimes \mathcal{W}_{\mathbf{z}}^\dagger)^{\otimes r} = \mathbf{I},
    \end{align*}
    where we used that $\operatorname{gcd}(d,r) = 1$ to remove $r$ from $\omega^{r[\mathbf{x},\mathbf{z}]}$.
    Therefore, $\Pi_{r}^+ = \frac{1}{2}(\mathbf{I} + \mathcal{V}_r)$ is a projector, since $(\Pi_{r}^+)^2 = \frac{1}{4}(\mathbf{I} + 2\mathcal{V}_r + \mathcal{V}_r^2) = \Pi_{r}^+$.

    It is possible to implement the binary POVM $\{\Pi_{r}^+, \mathbf{I} - \Pi_{r}^+\}$ by employing quantum phase estimation with $O(1)$ ancillae, $O(1)$ controlled versions of $\mathcal{V}_r$, and $O(1)$ auxiliary operations (we do not require a high accuracy in determining the eigenvalues of $\mathcal{V}_r$ since they are simply $\pm 1$). And since we can write $\mathcal{V}_r = \big(d^{-1}\sum_{\mathbf{x}\in\mathbb{Z}_d^{2}}(\mathcal{W}_{\mathbf{x}}\otimes \mathcal{W}_{\mathbf{x}}^\dagger)^{\otimes r}\big)^{\otimes n}$, then a controlled version of $\mathcal{V}_r$ is equal to a composition of $n$ controlled versions of $d^{-1}\sum_{\mathbf{x}\in\mathbb{Z}_d^{2}}(\mathcal{W}_{\mathbf{x}}\otimes \mathcal{W}_{\mathbf{x}}^\dagger)^{\otimes r}$, each of which acts on $2r$ qudits. Therefore, the POVM $\{\Pi_{r}^+, \mathbf{I} - \Pi_{r}^+\}$ can be implemented in time $O(nr)$.
\end{proof}

Define the binary POVM $\{\Pi_{r}^+, \mathbf{I} - \Pi_{r}^+\}$. The accept probability $P_r(|\psi\rangle)$ of measuring $\Pi_{r}^+$ is
\begin{align}\label{eq:accepting_probability_gross}
    P_r(|\psi\rangle) \triangleq \operatorname{Tr}[\psi^{\otimes 2r}\Pi_{r}^+] = \frac{1}{2} + \frac{1}{2d^n}\sum_{\mathbf{x}\in\mathbb{Z}_d^{2n}}|{\operatorname{Tr}}[\psi\mathcal{W}_{\mathbf{x}}]|^{2r} = \frac{1}{2} +  \frac{1}{2}d^{(r-1)n}\sum_{\mathbf{x}\in\mathbb{Z}_d^{2n}}p_\psi(\mathbf{x})^{r}.
\end{align}
It is possible to relate $P_r(|\psi\rangle)$ with the stabiliser fidelity $F_{\mathcal{S}}(|\psi\rangle)$ using \Cref{fact:commutativity}.
\begin{lemma}\label{lem:expected_variable_stabiliser_fidelity}
    Let $r\in\mathbb{N}$, $r\geq 2$, such that $\operatorname{gcd}(d,r) = 1$ and $C_{d,r} \triangleq \frac{1}{2}\big(1-(1-\frac{1}{4d^2})^{r-1}\big)$. Then
    \begin{align*}
         \frac{1}{2} + \frac{1}{2} F_{\mathcal{S}}(|\psi\rangle)^{2r} \leq P_r(|\psi\rangle) \leq 1 - C_{d,r}(1-F_{\mathcal{S}}(|\psi\rangle)) \qquad\forall |\psi\rangle\in(\mathbb{C}^d)^{\otimes n}.
    \end{align*}
\end{lemma}
\begin{proof}
    For the upper bound, let the set $\mathscr{X}_0 = \{\mathbf{x}\in\mathbb{Z}_d^{2n}:d^n p_\psi(\mathbf{x}) \geq 1 - \frac{1}{4d^2}\}$, which according to \Cref{fact:commutativity} is isotropic. It can thus be extended to a Lagrangian (maximal) submodule. Then
    \begin{align*}
        F_{\mathcal{S}}(|\psi\rangle) &\geq \sum_{\mathbf{x}\in\mathscr{X}_0} p_\psi(\mathbf{x}) \tag{by \Cref{lem:stabiliser_fidelity_inequality,remark:replacement}} \\
        &= 1 - \sum_{\mathbf{x}\in\mathbb{Z}_d^{2n} } p_\psi(\mathbf{x}) \mathbf{1}\!\left[d^n p_\psi(\mathbf{x}) < 1 - \frac{1}{4d^2} \right] \\
        &\geq 1 - \frac{\sum_{\mathbf{x}\in\mathbb{Z}_d^{2n}} p_\psi(\mathbf{x})(1- d^{(r-1)n}p_\psi(\mathbf{x})^{r-1})}{1 - (1-\frac{1}{4d^2})^{r-1}} \tag{$(r-1)$-th moment Markov's inequality}\\
        &= 1 - \frac{1 - P_r(|\psi\rangle)}{C_{d,r}}. \tag{by \eqref{eq:accepting_probability_gross}}
    \end{align*}
    Regarding the lower bound,
    \begin{align*}
        P_r(|\psi\rangle) &= \frac{1}{2} + \frac{1}{2} d^{(r-1)n}\sum_{\mathbf{x}\in\mathbb{Z}_d^{2n}} p_\psi(\mathbf{x})^r \tag{by \eqref{eq:accepting_probability_gross}}\\
        &\geq\frac{1}{2} + \frac{1}{2} d^{(r-1)n}\sum_{\mathbf{x}\in\mathscr{M}} p_\psi(\mathbf{x})^r\\
        &\geq \frac{1}{2} + \frac{1}{2} \left(\sum_{\mathbf{x}\in\mathscr{M}} p_\psi(\mathbf{x}) \right)^{r} \tag{H\"older's inequality}\\
        &\geq \frac{1}{2} + \frac{1}{2} F_{\mathcal{S}}(|\psi\rangle)^{2r}. \tag*{(by \Cref{lem:stabiliser_fidelity_inequality})\qquad\qedhere}
    \end{align*}
\end{proof}

Using the above theorem, we can extend the stabiliser testing algorithm from~\cite{gross2021schur} to the tolerant setting: measure the POVM $\{\Pi_{r}^+, \mathbf{I} - \Pi_{r}^+\}$ several times and estimate the quantity $P_r(|\psi\rangle)$. If the estimate of $P_r(|\psi\rangle)$ is large enough, then $F_{\mathcal{S}}(|\psi\rangle) \geq 1 - \varepsilon_1$ with high probability, otherwise $F_{\mathcal{S}}(|\psi\rangle) \leq 1 - \varepsilon_2$. Due to \Cref{lem:expected_variable_stabiliser_fidelity},
%
%
the cases $F_{\mathcal{S}}(|\psi\rangle) \geq 1 - \varepsilon_1$ and $F_{\mathcal{S}}(|\psi\rangle) \leq 1 - \varepsilon_2$ can be distinguished as long as
\begin{align*}
    \frac{1}{2}(1-\varepsilon_1)^{2r} - \frac{1}{2} + C_{d,r}\varepsilon_2 \geq \frac{1}{\operatorname{poly}(n)}.
\end{align*}

\begin{algorithm}[h]
\caption{Tolerant stabiliser testing algorithm based on \cite{gross2021schur}}
\DontPrintSemicolon
\label{algo:robust_testing_stabiliser_states}
\SetKwInput{KwPromise}{Promise}

    \KwIn{$2rm$ copies of  $|\psi\rangle\in(\mathbb{C}^d)^{\otimes n}$ where $m\triangleq \lceil\frac{8(1-\gamma_2)}{(\gamma_1-\gamma_2)^2}\ln\frac{1}{\delta}\rceil$, $r\geq 2$ is an integer such that $\operatorname{gcd}(d,r) = 1$, $\gamma_1 \triangleq \frac{1}{2} + \frac{(1-\varepsilon_1)^{2r}}{2}$, and $\gamma_2 \triangleq 1 -\frac{1}{2}\big(1-(1-\frac{1}{4d^2})^{r-1}\big)\varepsilon_2$.}
    \KwPromise{either $F_{\mathcal{S}}(|\psi\rangle) \geq 1 - \varepsilon_1$ or $F_{\mathcal{S}}(|\psi\rangle) \leq 1 - \varepsilon_2$.}
    \KwOut{$1$ if $F_{\mathcal{S}}(|\psi\rangle) \geq 1 - \varepsilon_1$ and $0$ if $F_{\mathcal{S}}(|\psi\rangle) \leq 1 - \varepsilon_2$.}

   \For{$i=1$ \KwTo $m$}
   {    
        Measure the POVM $\{\Pi_{r}^+, \mathbf{I} - \Pi_{r}^+\}$ on $|\psi\rangle^{\otimes 2r}$ and obtain the outcome $Z_i\in\{0,1\}$, where $\Pi_r^+ = \frac{\mathbf{I}}{2} + \frac{1}{2d^n}\sum_{\mathbf{x}\in\mathbb{Z}_d^{2n}}(\mathcal{W}_{\mathbf{x}}\otimes\mathcal{W}_{\mathbf{x}}^\dagger)^{\otimes r}$

    }

    \Return $1$ if $\frac{1}{m}\sum_{i=1}^m Z_i > \frac{\gamma_1 + \gamma_2}{2}$ and $0$ otherwise
\end{algorithm}

\begin{theorem}\label{thr:tolerant_stabiliser_testing_1}
    Let $|\psi\rangle \in (\mathbb{C}^d)^{\otimes n}$, $\delta\in(0,1)$, $r\geq 2$ an integer such that $\operatorname{gcd}(d,r) = 1$, and $C_{d,r} \triangleq \frac{1}{2}\big(1-(1-\frac{1}{4d^2})^{r-1}\big)$. Let $\varepsilon_2 > \varepsilon_1 \geq 0$ and $\gamma_1 \triangleq \frac{1}{2} + \frac{(1-\varepsilon_1)^{2r}}{2}$ and $\gamma_2 \triangleq 1 - C_{d,r}\varepsilon_2$ such that $\gamma_1 > \gamma_2$. {\rm \Cref{algo:robust_testing_stabiliser_states}} accepts if $F_{\mathcal{S}}(|\psi\rangle) \geq 1 - \varepsilon_1$ and rejects if $F_{\mathcal{S}}(|\psi\rangle) \leq 1 - \varepsilon_2$ with probability at least $1-\delta$. It uses 
    \begin{align*}
        2r\left\lceil\frac{8(1-\gamma_2)}{(\gamma_1-\gamma_2)^2}\ln\frac{1}{\delta}\right\rceil = 2r\left\lceil \frac{32C_{d,r}\varepsilon_2}{((1-\varepsilon_1)^{2r} - 1 + 2C_{d,r}\varepsilon_2)^2}\ln\frac{1}{\delta} \right\rceil \quad\text{copies of}~|\psi\rangle.
    \end{align*}
\end{theorem}
\begin{proof}
    The measurement outcome of $\{\Pi_{r}^+, \mathbf{I} - \Pi_{r}^+\}$ on $|\psi\rangle^{\otimes 2r}$ is a Bernoulli random variable $Z\in\{0,1\}$ such that $\operatorname{Pr}[Z=1] = P_r(|\psi\rangle)$ given by \eqref{eq:accepting_probability_gross}. Consider $m\triangleq \lceil\frac{8(1-\gamma_2)}{(\gamma_1-\gamma_2)^2}\ln\frac{1}{\delta}\rceil$ i.i.d.\ copies $Z_1,\dots,Z_m\in\{0,1\}$ of the random variable $Z$ and the quantity $\widetilde{P}_r \triangleq \frac{1}{m}\sum_{i=1}^m Z_i$. Note that $\mathbb{E}[\widetilde{P}_r] = P_r(|\psi\rangle)$.  If $F_{\mathcal{S}}(|\psi\rangle) \geq 1 - \varepsilon_1$, then $P_r(|\psi\rangle) \geq \gamma_1$, while $P_r(|\psi\rangle) \leq \gamma_2$ if $F_{\mathcal{S}}(|\psi\rangle) \leq 1 - \varepsilon_2$. Therefore, \Cref{algo:robust_testing_stabiliser_states} fails (i) if $\widetilde{P}_r \leq \frac{\gamma_1 + \gamma_2}{2}$ when $F_{\mathcal{S}}(|\psi\rangle) \geq 1 - \varepsilon_1$ and fails (ii) if $\widetilde{P}_r \geq \frac{\gamma_1 + \gamma_2}{2}$ when $F_{\mathcal{S}}(|\psi\rangle) \leq 1 - \varepsilon_2$. Consider auxiliary Bernoulli random variables $X,Y\in\{0,1\}$ such that $\operatorname{Pr}[X = 1] = \gamma_1$ and $\operatorname{Pr}[Y=1] = \gamma_2$. According to Chernoff’s bound,
    \begin{align*}
        F_{\mathcal{S}}(|\psi\rangle) \geq 1 - \varepsilon_1 &: \quad \operatorname{Pr}\!\left[\widetilde{P}_r \leq \frac{\gamma_1 + \gamma_2}{2}\right] \leq \operatorname{Pr}\!\left[\frac{1}{m}\sum_{i=1}^m X_i \leq \frac{\gamma_1 + \gamma_2}{2}\right] \leq \e^{-\frac{m(\gamma_1-\gamma_2)^2}{8\gamma_1(1-\gamma_1)}} \leq \delta,\\
        F_{\mathcal{S}}(|\psi\rangle) \leq 1 - \varepsilon_2 &: \quad \operatorname{Pr}\!\left[\widetilde{P}_r \geq \frac{\gamma_1 + \gamma_2}{2}\right] \leq \operatorname{Pr}\!\left[\frac{1}{m}\sum_{i=1}^m Y_i \geq \frac{\gamma_1 + \gamma_2}{2}\right] \leq \e^{-\frac{m(\gamma_1-\gamma_2)^2}{8\gamma_2(1-\gamma_2)}} \leq \delta.
    \end{align*}
    The number of employed copies of $|\psi\rangle$ is $2r\lceil\frac{8(1-\gamma_2)}{(\gamma_1-\gamma_2)^2}\ln\frac{1}{\delta}\rceil$.
\end{proof}

The complexity of \Cref{algo:robust_testing_stabiliser_states} is $O\big(\frac{rC_{d,r}\varepsilon_2}{(C_{d,r}\varepsilon_2 - r\varepsilon_1)^2}\log\frac{1}{\delta}\big)$, which reduces to $O\big(\frac{r}{C_{d,r}\varepsilon_2}\log\frac{1}{\delta}\big)$ for $\varepsilon_1=0$, thus recovering the result of~\cite{gross2021schur}. If further $r=O(d^2)$, so that $C_{d,r} = \Theta(\frac{r}{d^2})$, then it becomes $O\big(\frac{d^2}{\varepsilon_2}\log\frac{1}{\delta}\big)$. We further note that for $d=2$, the complexity of \Cref{algo:robust_testing_stabiliser_states} is $O\big(\frac{\varepsilon_2}{(\varepsilon_2 - \varepsilon_1)^2}\log\frac{1}{\delta}\big)$, an improvement over the complexity of~\cite{grewal2023improved} by a factor of $O(\varepsilon_2)$.

\subsection{Approach 2: employing skewed Bell difference sampling}

We now develop an alternative testing procedure: get $\mathbf{X}\in\mathbb{Z}_d^{2n\times 4}$ through skewed Bell difference sampling, measure\footnote{We abuse the language and say that an observable is measured on a quantum state to refer to measuring this quantum state using the eigenstates of the observable as projective measurements.} $|\psi\rangle^{\otimes 2}$ with (the eigenstates of) $\mathcal{W}_{\mathbf{X}_i}\otimes\mathcal{W}_{\mathbf{X}_i}^\dagger + \mathcal{W}_{\mathbf{X}_i}^\dagger\otimes\mathcal{W}_{\mathbf{X}_i}$ for each $i\in[4]$, and check if all outcomes are $1$. The next result provides a close formula for such probability.

\begin{definition}\label{def:testing_procedure}
    Given $|\psi\rangle\in(\mathbb{C}^d)^{\otimes n}$, define the Bernoulli distribution $(T(|\psi\rangle), 1 - T(|\psi\rangle))$ where 
    \begin{align*}
        T(|\psi\rangle) &\triangleq \operatorname*{Pr}_{\mathbf{X}\sim b_\psi}\left[\text{measurement of}~\mathcal{W}_{\mathbf{X}_i}\otimes\mathcal{W}_{\mathbf{X}_i}^\dagger + \mathcal{W}_{\mathbf{X}_i}^\dagger\otimes\mathcal{W}_{\mathbf{X}_i}~\text{on}~|\psi\rangle^{\otimes 2}~\text{equals}~1~\forall i\in[4] \right].
    \end{align*}
\end{definition}

\begin{lemma}\label{lem:expression_T-psi}
    Let $|\psi\rangle\in(\mathbb{C}^d)^{\otimes n}$ and let the Bernoulli distribution $(T(|\psi\rangle), 1 - T(|\psi\rangle))$ from {\rm \Cref{def:testing_procedure}}. Let $\mathbf{R}\in\mathbb{Z}_d^{4\times 4}$ be such that $\mathbf{R}^\top\mathbf{R} = -\mathbf{I}\operatorname{mod}d$. Then
    \begin{align*}
        T(|\psi\rangle) &= d^{8n-4} \sum_{\mathbf{X}\in\mathbb{Z}_d^{2n\times 4}} \sum_{\mathbf{t}\in\mathbb{Z}_d^4} \prod_{i=1}^4  p_\psi(\mathbf{X}_i) p_\psi(t_i\mathbf{X}_i) p_\psi((\mathbf{X}\operatorname{diag}(\mathbf{t})\mathbf{R})_i).
    \end{align*}
\end{lemma}
\begin{proof}
    To analyse the success probability $T(|\psi\rangle)$, first note that, given $\mathcal{W}_{\mathbf{x}}$ for some $\mathbf{x}\in\mathbb{Z}_d^{2n}$, the projection onto its $\omega^k$-eingespace for $k\in\mathbb{Z}_d$ is
    \begin{align*}
        \Pi_k^{\mathbf{x}} = \frac{1}{d}\sum_{t\in\mathbb{Z}_d} \omega^{tk}(\mathcal{W}_{\mathbf{x}}^\dagger)^t.
    \end{align*}
    Hence, the projection onto the $\omega^k$-eingespace of $\mathcal{W}_{\mathbf{x}}\otimes\mathcal{W}_{\mathbf{x}}^\dagger$ is
    \begin{align*}
        \Xi_k^{\mathbf{x}} = \sum_{t\in\mathbb{Z}_d} \Pi_t^{\mathbf{x}} \otimes \Pi_{k-t}^{-\mathbf{x}} &= \frac{1}{d^2}\sum_{t,t_1,t_2\in\mathbb{Z}_d} \omega^{tt_1 + (k-t)t_2} (\mathcal{W}_{\mathbf{x}}^\dagger)^{t_1}\otimes (\mathcal{W}_{-\mathbf{x}}^\dagger)^{t_2} = \frac{1}{d} \sum_{t\in\mathbb{Z}_d} \omega^{k t} \mathcal{W}_{-t\mathbf{x}}\otimes \mathcal{W}_{t\mathbf{x}}
    \end{align*}
    Since the projector onto the $+1$-eigenspace of $\mathcal{W}_{\mathbf{x}}\otimes\mathcal{W}_{\mathbf{x}}^\dagger + \mathcal{W}_{\mathbf{x}}^\dagger\otimes\mathcal{W}_{\mathbf{x}}$ is also $\Xi_0^{\mathbf{x}}$, the probability of measuring $\mathcal{W}_{\mathbf{x}}\otimes\mathcal{W}_{\mathbf{x}}^\dagger + \mathcal{W}_{\mathbf{x}}^\dagger\otimes\mathcal{W}_{\mathbf{x}}$ and obtaining $1$ is thus
    \begin{align*}
        \operatorname{Tr}[ \Xi_0^{\mathbf{x}}\psi^{\otimes 2} ] &= \frac{1}{d}\sum_{t\in\mathbb{Z}_d} \operatorname{Tr}[\mathcal{W}_{-t\mathbf{x}} \psi]\operatorname{Tr}[\mathcal{W}_{t\mathbf{x}} \psi] = d^{n-1}\sum_{t\in\mathbb{Z}_d} p_\psi(t\mathbf{x}). \tag{by $p_\psi(t\mathbf{x}) = d^{-n}|\langle \psi|\mathcal{W}_{t\mathbf{x}}|\psi\rangle|^2$}
    \end{align*}
    If $\mathbf{X}\in\mathbb{Z}_d^{2n\times 4}$ is thus sampled from $b_\psi$, the probability of measuring $\mathcal{W}_{\mathbf{X}_i}\otimes\mathcal{W}_{\mathbf{X}_i}^\dagger + \mathcal{W}_{\mathbf{X}_i}^\dagger\otimes\mathcal{W}_{\mathbf{X}_i}$ on $|\psi\rangle^{\otimes 2}$ and obtaining $+1$ for each $i\in[4]$ is
    \begin{align*}
        T(|\psi\rangle) &= \sum_{\mathbf{X}\in\mathbb{Z}_d^{2n\times 4}} b_\psi(\mathbf{X}) \prod_{i=1}^4 \left(\sum_{t\in\mathbb{Z}_d} d^{n-1}p_\psi(t\mathbf{X}_i) \right) \\
        &= d^{4n-4} \sum_{\mathbf{X},\mathbf{Z}\in\mathbb{Z}_d^{2n\times 4}} \sum_{\mathbf{t}\in\mathbb{Z}_d^4} \prod_{i=1}^4 \omega^{[\mathbf{X}_i,\mathbf{Z}_i]} p_\psi(\mathbf{Z}_i) p_\psi((\mathbf{Z}\mathbf{R})_i) p_\psi(t_i\mathbf{X}_i) \tag{by \Cref{thr:skewed_bell_sampling}} \\
        &= \frac{1}{d^{4}} \sum_{\mathbf{X},\mathbf{Z},\mathbf{Z}'\in\mathbb{Z}_d^{2n\times 4}} \sum_{\mathbf{t}\in\mathbb{Z}_d^4} \prod_{i=1}^4 \omega^{[\mathbf{X}_i, \mathbf{Z}_i]} p_\psi(\mathbf{Z}_i) p_\psi((\mathbf{Z}\mathbf{R})_i) \omega^{[t_i\mathbf{X}_i,\mathbf{Z}_i']} p_\psi(\mathbf{Z}'_i) \tag{Fourier expansion and \Cref{lem:invariant}}\\
        &= d^{8n-4} \sum_{\mathbf{Z}\in\mathbb{Z}_d^{2n\times 4}} \sum_{\mathbf{t}\in\mathbb{Z}_d^4} \prod_{i=1}^4  p_\psi(\mathbf{Z}_i) p_\psi(t_i\mathbf{Z}_i) p_\psi((\mathbf{Z}\operatorname{diag}(\mathbf{t})\mathbf{R})_i) .  \tag*{(by \Cref{lem:sum} and $p_\psi(\mathbf{Z}'_i) = p_\psi(-\mathbf{Z}'_i)$)\qquad\qedhere}
    \end{align*}
\end{proof}

In the same spirit of \Cref{lem:expected_variable_stabiliser_fidelity}, we now relate $T(|\psi\rangle)$ with the stabiliser fidelity $F_{\mathcal{S}}(|\psi\rangle)$ by giving upper and lower bounds.
\begin{lemma}\label{lem:bounds_Tpsi}
    Let $\varphi(d)$ be Euler's totient function. For all $|\psi\rangle\in(\mathbb{C}^d)^{\otimes n}$ such that $F_{\mathcal{S}}(|\psi\rangle) \geq \frac{1}{2}$,
    \begin{align*}
        T(|\psi\rangle) &\leq \left(1 - \frac{\varphi(d)}{4d^3}(1-F_{\mathcal{S}}(|\psi\rangle))\right)^4,\\
        T(|\psi\rangle) &\geq (2F_{\mathcal{S}}(|\psi\rangle) - 1)^8 F_{\mathcal{S}}(|\psi\rangle)^8 \left(\frac{1+F_{\mathcal{S}}(|\psi\rangle)^2}{d} + \frac{d-2}{d}(2F_{\mathcal{S}}(|\psi\rangle) - 1)^2 \right)^4.
    \end{align*}
\end{lemma}
\begin{proof}
    For the upper bound, let $A_d\subseteq \mathbb{Z}_d\setminus\{0\}$ be the set of coprime numbers to $d$. Then
    \begin{align*}
        T(|\psi\rangle) &= d^{8n-4} \sum_{\mathbf{X}\in\mathbb{Z}_d^{2n\times 4}} \sum_{\mathbf{t}\in\mathbb{Z}_d^4} \prod_{i=1}^4  p_\psi(\mathbf{X}_i) p_\psi(t_i\mathbf{X}_i) p_\psi((\mathbf{X}\operatorname{diag}(\mathbf{t})\mathbf{R})_i) \tag{by \Cref{lem:expression_T-psi}}\\
        &\leq d^{4n-4} \sum_{\mathbf{X}\in\mathbb{Z}_d^{2n\times 4}} \sum_{\mathbf{t}\in\mathbb{Z}_d^4} \prod_{i=1}^4  p_\psi(\mathbf{X}_i) p_\psi(t_i\mathbf{X}_i) \tag{$p_\psi((\mathbf{X}\operatorname{diag}(\mathbf{t})\mathbf{R})_i) \leq d^{-n}$}\\
        &= \left(\frac{d - \varphi(d)}{d} + d^{n-1}\sum_{\mathbf{x}\in\mathbb{Z}_d^{2n}}\sum_{t\in A_d} p_\psi(\mathbf{x}) p_\psi(t\mathbf{x}) \right)^4 \tag{$p_\psi(t\mathbf{x}) \leq d^{-n}$ for $t\in\mathbb{Z}_d\setminus A_d$}\\
        &\leq \left(\frac{d - \varphi(d)}{d} + d^{n-1}\sum_{\mathbf{x}\in\mathbb{Z}_d^{2n}}\sum_{t\in A_d} \frac{p_\psi(\mathbf{x})^2 +  p_\psi(t\mathbf{x})^2}{2} \right)^4 \\
        &= \left(\frac{d - \varphi(d)}{d} + d^{n-1}\sum_{\mathbf{x}\in\mathbb{Z}_d^{2n}}\sum_{t\in A_d} p_\psi(\mathbf{x})^2 \right)^4 \tag{$\mathbf{x}\gets t^{-1}\mathbf{x}$}\\
        &\leq \left(1 - \frac{\varphi(d)}{4d^3}(1-F_{\mathcal{S}}(|\psi\rangle))\right)^4. \tag{by \Cref{lem:expected_variable_stabiliser_fidelity} with $r=2$}
    \end{align*}
    %
    
    For the lower bound, let $|\mathscr{M},s\rangle$ be the stabiliser state that maximises $F_{\mathcal{S}}(|\psi\rangle)$. Then
    \begin{align*}
        T(|\psi\rangle) &\geq d^{8n-4} \sum_{\mathbf{X}\in\mathscr{M}^{4}} \sum_{\mathbf{t}\in\mathbb{Z}_d^4} \prod_{i=1}^4  p_\psi(\mathbf{X}_i) p_\psi(t_i\mathbf{X}_i) p_\psi((\mathbf{X}\operatorname{diag}(\mathbf{t})\mathbf{R})_i) \tag{by \Cref{lem:expression_T-psi} and $\mathscr{M}\subset\mathbb{Z}_d^{2n}$} \\
        &\geq (2F_{\mathcal{S}}(|\psi\rangle) - 1)^8 \left(d^{n-1} \sum_{t\in\mathbb{Z}_d}\sum_{\mathbf{x}\in\mathscr{M}} p_\psi(\mathbf{x})p_\psi(t\mathbf{x}) \right)^4 \tag{$d^n p_\psi((\mathbf{X}\operatorname{diag}(\mathbf{t})\mathbf{R})_i) \geq (2F_{\mathcal{S}}(|\psi\rangle) - 1)^2$ by \Cref{lem:lower_bound_p_psi}} \\
        &= (2F_{\mathcal{S}}(|\psi\rangle) - 1)^8 \left(\frac{1}{d} \sum_{\mathbf{x}\in\mathscr{M}} p_\psi(\mathbf{x}) + d^{n-1} \sum_{\mathbf{x}\in\mathscr{M}} p_\psi(\mathbf{x})^2 + d^{n-1}\sum_{t\in\mathbb{Z}_d\setminus\{0,1\}} \sum_{\mathbf{x}\in\mathscr{M}} p_\psi(\mathbf{x})p_\psi(t\mathbf{x}) \right)^4 \\
        &\geq (2F_{\mathcal{S}}(|\psi\rangle) - 1)^8 \left(\left(\frac{1}{d} + \frac{d-2}{d}(2F_{\mathcal{S}}(|\psi\rangle) - 1)^2\right) \sum_{\mathbf{x}\in\mathscr{M}} p_\psi(\mathbf{x}) + d^{n-1} \sum_{\mathbf{x}\in\mathscr{M}} p_\psi(\mathbf{x})^2 \right)^4 \tag{$d^n p_\psi(t\mathbf{x}) \geq (2F_{\mathcal{S}}(|\psi\rangle) - 1)^2$ by \Cref{lem:lower_bound_p_psi}} \\
        &\geq (2F_{\mathcal{S}}(|\psi\rangle) - 1)^8 \left(\left(\frac{1}{d} + \frac{d-2}{d}(2F_{\mathcal{S}}(|\psi\rangle) - 1)^2\right) \sum_{\mathbf{x}\in\mathscr{M}} p_\psi(\mathbf{x}) + \frac{1}{d}\left(\sum_{\mathbf{x}\in\mathscr{M}} p_\psi(\mathbf{x})\right)^2 \right)^4 \tag{by H\"older inequality} \\
        &\geq (2F_{\mathcal{S}}(|\psi\rangle) - 1)^8 \left(\left(\frac{1}{d} + \frac{d-2}{d}(2F_{\mathcal{S}}(|\psi\rangle) - 1)^2\right) F_{\mathcal{S}}(|\psi\rangle)^2 + \frac{1}{d}F_{\mathcal{S}}(|\psi\rangle)^4 \right)^4. \tag*{(by \Cref{lem:stabiliser_fidelity_inequality})\quad\qedhere}
    \end{align*}
\end{proof}

Our second tolerant stabiliser testing algorithm (\Cref{algo:robust_testing_stabiliser_states2}) employs skewed Bell difference sampling to estimate the expected value $T(|\psi\rangle)$. Once again, if the estimate of $T(|\psi\rangle)$ is large enough, then $F_{\mathcal{S}}(|\psi\rangle) \geq 1 - \varepsilon_1$ with high probability, otherwise $F_{\mathcal{S}}(|\psi\rangle) \leq 1 - \varepsilon_2$. According to \Cref{lem:bounds_Tpsi}, the cases $F_{\mathcal{S}}(|\psi\rangle) \geq 1 - \varepsilon_1$ and $F_{\mathcal{S}}(|\psi\rangle) \leq 1 - \varepsilon_2$ can be distinguished as long as
\begin{align*}
    (1-2\varepsilon_1)^8 (1-\varepsilon_1)^8 \left(\frac{1+(1-\varepsilon_1)^2}{d} + \frac{d-2}{d}(1-2\varepsilon_1)^2 \right)^4 - \left(1 - \frac{\varphi(d)}{4d^3}\varepsilon_2\right)^4 \geq \frac{1}{\operatorname{poly}(n)}.
\end{align*}

\begin{algorithm}[h]
\caption{Alternative tolerant stabiliser testing algorithm}
\DontPrintSemicolon
\label{algo:robust_testing_stabiliser_states2}
\SetKwInput{KwPromise}{Promise}

    \KwIn{$16m + 8$ copies of $|\psi\rangle\in(\mathbb{C}^d)^{\otimes n}$ where $m \triangleq \lceil \frac{8(1-\alpha_2)}{(\alpha_1 - \alpha_2)^2}\ln\frac{1}{\delta} \rceil$, $\alpha_1 \triangleq (1-2\varepsilon_1)^8(1-\varepsilon_1)^8 \big(\frac{1+(1-\varepsilon_1)^2}{d} + \frac{d-2}{d}(1-2\varepsilon_1)^2 \big)^4$, and $\alpha_2 \triangleq \big(1- \frac{\varphi(d)}{4d^3}\varepsilon_2\big)^4$.}
    \KwPromise{either $F_{\mathcal{S}}(|\psi\rangle) \geq 1 - \varepsilon_1$ or $F_{\mathcal{S}}(|\psi\rangle) \leq 1 - \varepsilon_2$.}
    \KwOut{$1$ if $F_{\mathcal{S}}(|\psi\rangle) \geq 1-\varepsilon_1$ and $0$ if $F_{\mathcal{S}}(|\psi\rangle) \leq 1 - \varepsilon_2$.}

    \For{$i\gets 0$ \KwTo $m$}
    {    
        Perform Bell sampling on $(\mathbf{I}\otimes\mathcal{B}_{\mathbf{R}}^\dagger)|\psi\rangle^{\otimes 8}$ to obtain $\mathbf{X}^{(i)}\in\mathbb{Z}_d^{2n\times 4}$
        
        \If{$i\neq 0$}
        {
            $\mathbf{X}^{(i)} \gets \mathbf{X}^{(i)}-\mathbf{X}^{(0)}$

            Measure $\mathcal{W}_{\mathbf{X}^{(i)}_1}\otimes\mathcal{W}_{\mathbf{X}^{(i)}_1}^\dagger + \mathcal{W}_{\mathbf{X}^{(i)}_1}^\dagger\otimes\mathcal{W}_{\mathbf{X}^{(i)}_1}, \dots, \mathcal{W}_{\mathbf{X}^{(i)}_4}\otimes\mathcal{W}_{\mathbf{X}^{(i)}_4}^\dagger + \mathcal{W}_{\mathbf{X}^{(i)}_4}^\dagger\otimes\mathcal{W}_{\mathbf{X}^{(i)}_4}$. Let $Z_i = 1$ if all four measurements return $1$ and $Z_i = 0$ otherwise
        }
    }

    \Return $1$ if $\frac{1}{m}\sum_{i=1}^m Z_i > \frac{\alpha_1+\alpha_2}{2}$ and $0$ otherwise
\end{algorithm}

\begin{theorem}\label{thr:tolerant_stabiliser_testing_2}
    Let $|\psi\rangle \in (\mathbb{C}^d)^{\otimes n}$, $\delta\in(0,1)$, and $\varphi(d)$ be Euler's totient function. Let $\varepsilon_2 > \varepsilon_1 \geq 0$ and $\alpha_1 \triangleq (1-2\varepsilon_1)^8(1-\varepsilon_1)^8 \big(\frac{1+(1-\varepsilon_1)^2}{d} + \frac{d-2}{d}(1-2\varepsilon_1)^2 \big)^4$ and $\alpha_2 \triangleq \big(1- \frac{\varphi(d)}{4d^3}\varepsilon_2\big)^4$ such that $\alpha_1 > \alpha_2$. {\rm \Cref{algo:robust_testing_stabiliser_states2}} accepts if $F_{\mathcal{S}}(|\psi\rangle) \geq 1 - \varepsilon_1$ and rejects if $F_{\mathcal{S}}(|\psi\rangle) \leq 1 - \varepsilon_2$ with probability at least $1-\delta$. It uses
    \begin{align*}
        16\left\lceil \frac{8(1-\alpha_2)}{(\alpha_1 - \alpha_2)^2}\ln\frac{1}{\delta} \right\rceil + 8 \quad\text{copies of}~|\psi\rangle.
    \end{align*}
\end{theorem}
\begin{proof}
    The proof is the same as in \Cref{thr:tolerant_stabiliser_testing_1}, just replace $\gamma_1$ and $\gamma_2$ with $\alpha_1$ and $\alpha_2$, respectively. Instead of using $16$ copies of $|\psi\rangle$ per skewed Bell difference sampling, the outcome of one skewed Bell sampling can be subtracted from all the others, reducing the number of copies to $8$ per sample. Another $8$ copies are used to measure $\mathcal{W}_{\mathbf{X}^{(i)}_1}\otimes\mathcal{W}_{\mathbf{X}^{(i)}_1}^\dagger + \mathcal{W}_{\mathbf{X}^{(i)}_1}^\dagger\otimes\mathcal{W}_{\mathbf{X}^{(i)}_1}$, $\dots, \mathcal{W}_{\mathbf{X}^{(i)}_4}\otimes\mathcal{W}_{\mathbf{X}^{(i)}_4}^\dagger + \mathcal{W}_{\mathbf{X}^{(i)}_4}^\dagger\otimes\mathcal{W}_{\mathbf{X}^{(i)}_4}$. The number of employed copies of $|\psi\rangle$ is thus $16m + 8$ with $m \triangleq \lceil \frac{8(1-\alpha_2)}{(\alpha_1 - \alpha_2)^2}\ln\frac{1}{\delta} \rceil$.
\end{proof}

The complexity of \Cref{algo:robust_testing_stabiliser_states2} is $O\big(\frac{\varepsilon_2\varphi(d)/d^3}{(\varepsilon_2\varphi(d)/d^3 - \varepsilon_1)^2}\log\frac{1}{\delta}\big)$, which reduces to $O\big(\frac{d^3}{\varphi(d)\varepsilon_2}\log\frac{1}{\delta}\big)$ for $\varepsilon_1 = 0$. It is known that $\varphi(d) = \Omega(d/\log\log{d})$, so in the worst case the final complexity is $O\big(\frac{d^2\log\log{d}}{\varepsilon_2}\log\frac{1}{\delta}\big)$, a factor $O(\log\log{d})$ worse than \Cref{algo:robust_testing_stabiliser_states}. For prime dimensions $\varphi(d) = d-1$, so both complexities coincide.

\begin{figure}
    \centering
    \begin{subfigure}[b]{0.48\textwidth}
         \centering
        \includegraphics[width=0.7\textwidth]{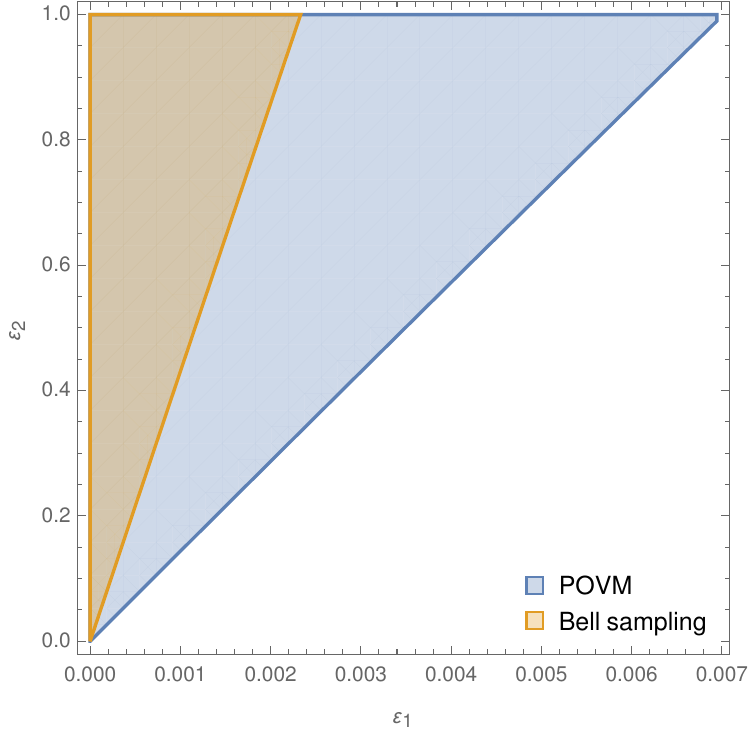}
        \caption{Range comparison for dimension $d=3$. The POVM-based algorithm uses $r=2$.}
    \end{subfigure}
    ~
    \begin{subfigure}[b]{0.48\textwidth}
        \centering
        \vstretch{1.2}{\includegraphics[width=\textwidth]{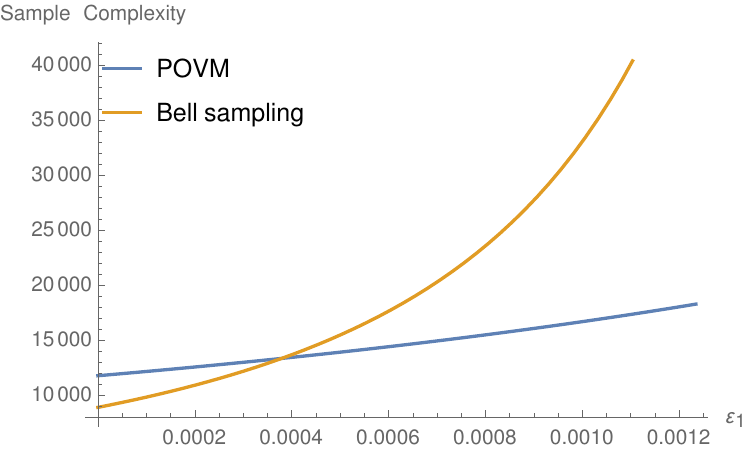}}
        \caption{Sample complexity comparison for $d=3$. The POVM-based algorithm uses $r=2$.}
    \end{subfigure}
    \begin{subfigure}[b]{0.48\textwidth}
        \centering
        \includegraphics[width=0.7\textwidth]{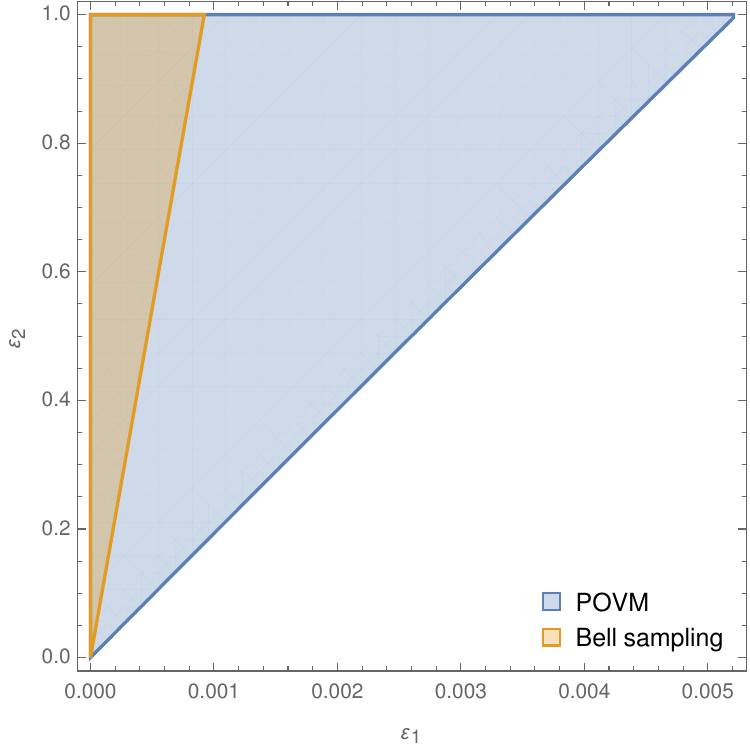}
        \caption{Range comparison for dimension $d=4$. The POVM-based algorithm uses $r=3$.}
    \end{subfigure}
    ~
    \begin{subfigure}[b]{0.48\textwidth}
        \centering
        \vstretch{1.2}{\includegraphics[width=\textwidth]{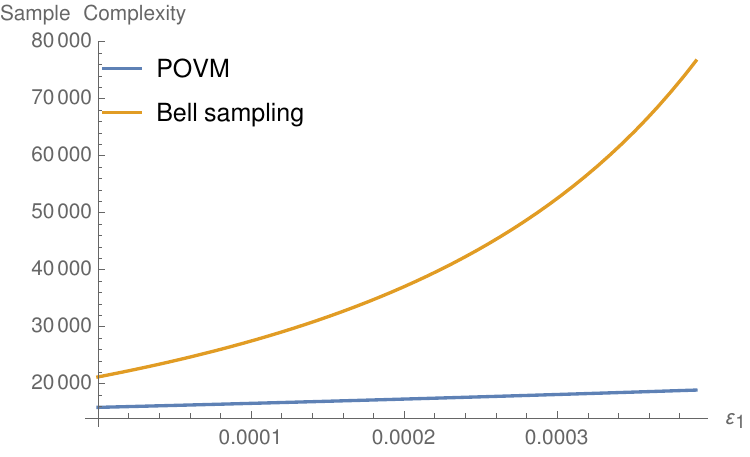}}
        \caption{Sample complexity comparison for $d=4$. The POVM-based algorithm uses $r=3$.}
    \end{subfigure}
    \begin{subfigure}[b]{0.48\textwidth}
        \centering
        \includegraphics[width=0.7\textwidth]{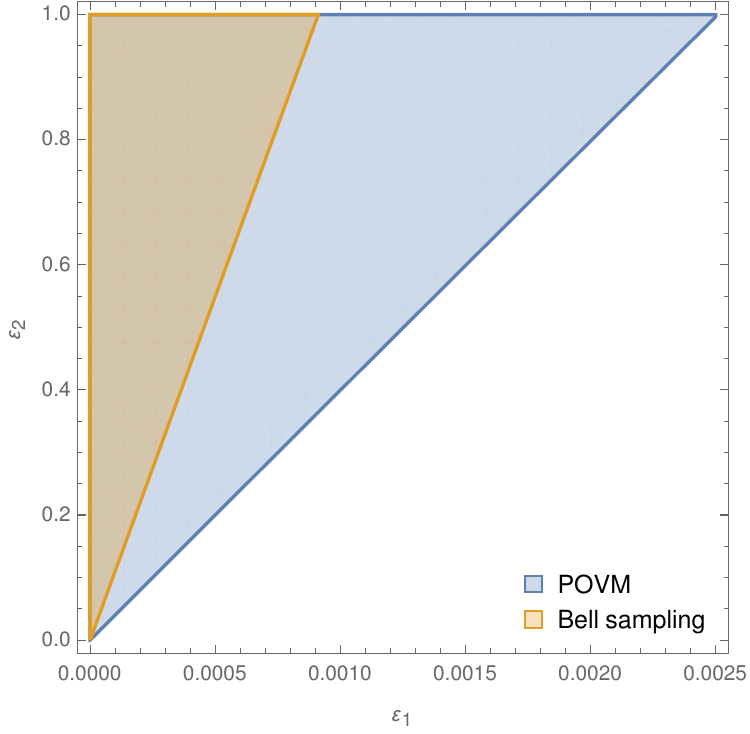}
        \caption{Range comparison for dimension $d=5$. The POVM-based algorithm uses $r=2$.}
    \end{subfigure}
    ~
    \begin{subfigure}[b]{0.48\textwidth}
        \centering
        \vstretch{1.2}{\includegraphics[width=\textwidth]{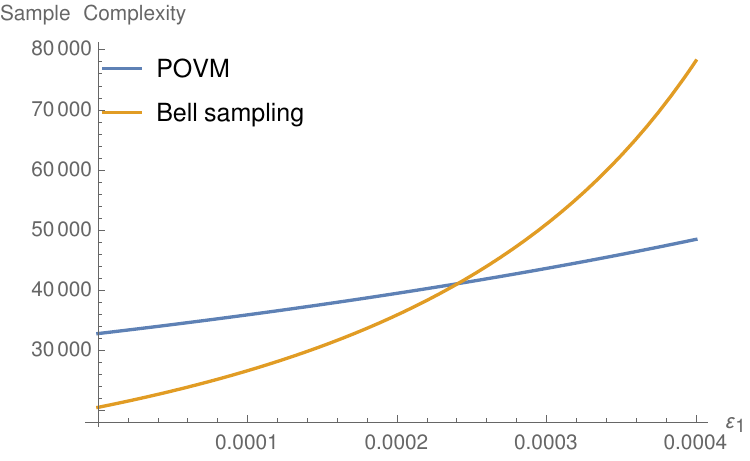}}
        \caption{Sample complexity comparison for $d=5$. The POVM-based algorithm uses $r=2$.}
    \end{subfigure}
    \caption{Comparison of sample complexities and range of parameters $\varepsilon_1,\varepsilon_2$ between \Cref{algo:robust_testing_stabiliser_states} (POVM-based) and \Cref{algo:robust_testing_stabiliser_states2} (Bell-sampling-based). For the sample complexity, $\varepsilon_2 = 0.9$ and $\delta=0.01$.}\label{fig:robust_testing_range_parameters}
\end{figure}

We compare the range of parameters $\varepsilon_2 > \varepsilon_1 \geq 0$ for which \Cref{algo:robust_testing_stabiliser_states,algo:robust_testing_stabiliser_states2} work in \Cref{fig:robust_testing_range_parameters}, i.e., the values of $\varepsilon_1,\varepsilon_2$ for which $\gamma_1 > \gamma_2$ and $\alpha_1>\alpha_2$. We also compare their sample complexities. The POVM-based algorithm \Cref{algo:robust_testing_stabiliser_states} can reach a larger range of parameters, specially for high values of $r$ at the expense of performing joint measurements across $2r$ qudits. \Cref{algo:robust_testing_stabiliser_states2}, in comparison, only requires measurements across at most $8$ qudits. Moreover, the Bell-sampling-based \Cref{algo:robust_testing_stabiliser_states2} can have a better sample complexity for smaller values of $\varepsilon_1$ in some dimensions. The choice of one algorithm over the other is highly dependent on the specific dimension.

\paragraph*{Acknowledgments.} We thank David Gross for very useful discussions and ideas that led to the development of the unitary $\mathcal{B}_{\mathbf{R}}$. This research is funded by ERC grant No.\ 810115-DYNASNET. This work was done in part while JFD and MS were visiting the Simons Institute for the Theory of Computing.

\bibliographystyle{alpha}
\bibliography{Pauli.bib}

\appendix
\section{Mathematical Background on Stabiliser Groups}

In this appendix, we provide a more extensive and self-contained introduction to stabiliser groups, starting with a quick review on modules.

\subsection{Background on modules}
\label{app:modules}


Given a commutative ring $R$, an $R$-module $M$ is an abelian group $(M,+)$ together with an operation $\cdot : R\times M\mapsto M$ (whose images we denote by $rx$) which satisfies the conditions $r(x+y) = rx + ry$, $(r+s)x = rx + sx$, $(rs)x = r(sx)$, and $1x = x$ for all $r,s\in R$ and $x,y\in M$. 
We call the cardinality of a module $M$ its \emph{size}. A set $\mathscr{B} \subseteq M$ is called a generating set of $M$ if every element of $M$ can be expressed as a linear combination of elements in $\mathscr{B}$. If $M$ has a finite generating set, it is said to be finitely generated. A finite subset $\{x_{i_1},\dots,x_{i_k}\}$ is \emph{linearly independent} over $R$ if $\sum_{j=1}^k r_j x_{i_j} = 0 \implies r_1=\cdots= r_k = 0$ for all $r_1,\dots,r_k\in R$. A generating set $\mathscr{B}$ that is linearly independent over $R$ is called a basis of $M$. An $R$-module with a basis is said to be free.  The rank of a free module is the cardinality of any of its bases. The $\mathbb{Z}_d$-module $\mathbb{Z}_d^n$ is free and has rank $n$. However, submodules of $\mathbb{Z}_d^n$ are not necessarily free.

Given an $R$-module $M$, a non-empty set $N\subseteq M$ is called a \emph{submodule} if $N$ is closed under addition and multiplication. A submodule $N\subseteq M$ is called \emph{maximal} if there are no other submodules in between $N$ and $M$.  A chain of submodules of $M$ of the form $M_1\subset M_2 \subset \cdots \subset M_{\ell}$ is said to have \emph{length} $\ell$, and the \emph{length} of $M$ is largest length of any of its chains. If an $R$-module $M$ has finite length, then it is finitely generated. 

Given submodules $N_1,N_2\subseteq M$, let 
\begin{align*}
    N_1 + N_2 \triangleq \{x_1 + x_2 : x_1\in N_1, x_2\in N_2 \} \qquad\text{and}\qquad N_1\times N_2 \triangleq \{(x_1,x_2): x_1\in N_1, x_2\in N_2\}.
\end{align*}
If $M = \sum_{i=1}^k N_i$ and $N_j\cap \sum_{i\in[k]:i\neq j} N_i = 0$ for all $j\in[k]$, then $M = \bigoplus_{i=1}^k N_i$ is said to be the direct sum of $N_1,\dots, N_k$.
\begin{fact}[Fundamental theorem of finitely generated abelian groups]\label{fact:fundamental_theorem}
    Given a module $M$ with size $|M| = d$ where $d = \prod_{i=1}^\ell p_i^{k_i}$ is the prime decomposition of $d$, then $M \cong \bigoplus_{i=1}^\ell M_i$ where $|M_i| = p_i^{k_i}$ and $M_i$ decomposes as $M_i \cong \bigoplus_{j=1}^{r_i} \mathbb{Z}/p_i^{\lambda_{ij}}\mathbb{Z}$ such that $1\leq \lambda_{ij} \leq k_i$ and $\sum_{j=1}^{r_i} \lambda_{ij} = k_i$. If $M$ is a submodule of a free module with rank $n$, then $r_i \leq n$ in the decomposition of $M$.
\end{fact}

\begin{lemma}\label{lem:number_submodules}
    Let $\mathscr{M}\subseteq\mathbb{Z}_d^{2n}$ be a module of size $d^n$. There are at most $\frac{p^{2n}-1}{p-1}$ maximal proper submodules $\mathscr{X}\subset\mathscr{M}$ such that $\mathscr{M}/\mathscr{X} \cong \mathbb{Z}/p\mathbb{Z}$ for any prime $p|d$.
\end{lemma}
\begin{proof}
    Let $d=\prod_{i=1}^\ell p_i^{k_i}$ be the prime decomposition of $d$. By \Cref{fact:fundamental_theorem}, $\mathscr{M}\cong \bigoplus_{i=1}^\ell \mathscr{M}_i$ where $\mathscr{M}_i$ has size $|\mathscr{M}_i| = p_i^{nk_i}$ and decomposes as $\mathscr{M}_i \cong \bigoplus_{j=1}^{r_i} \mathbb{Z}/p_i^{\lambda_{ij}}\mathbb{Z}$ with $\lambda_{ij}\geq 1$ and $\sum_{j=1}^{r_i} \lambda_{ij} = nk_i$. Note that $r_i \leq 2n$ for all $i\in[\ell]$ since $\mathscr{M}$ has at most $2n$ generators. We can similarly write $\mathscr{X} \cong \bigoplus_{i=1}^\ell \mathscr{X}_i$ and by the condition $\mathscr{M}/\mathscr{X} \cong \mathbb{Z}/p_a\mathbb{Z}$ for $p_a|d$, then $\mathscr{X}_i = \mathscr{M}_i$ for $i\neq a$ and $\mathscr{M}_a/\mathscr{X}_a \cong \mathbb{Z}/p_a\mathbb{Z}$. We therefore just need to count maximal proper $\mathbb{Z}/p_a^{k_a}\mathbb{Z}$-submodules of $\mathscr{M}_a$.
    
    We claim that there is a one-to-one correspondence between maximal proper submodules $\mathscr{X}_a \subset \mathscr{M}_a$ and kernels of surjective homomorphisms $\phi: \mathscr{M}_a \to \mathbb{Z}/p_a\mathbb{Z}$. Indeed, by isomorphism theorems, $\mathscr{M}_a/\operatorname{ker}(\phi) \cong \mathbb{Z}_p$, meaning that $\operatorname{ker}(\phi)$ must be a maximal proper submodule, while given a maximal proper submodule $\mathscr{X}_a$, the natural projection $\pi: \mathscr{M}_a \to \mathscr{M}_a/\mathscr{X}_a$ defined by $\pi(x) = x + \mathscr{X}_a$ for all $x\in\mathscr{M}_a$ is a surjective homomorphism with $\operatorname{ker}(\pi) = \mathscr{X}_a$. We now count the number of surjective homomorphisms $\phi: \mathscr{X}_a \to \mathbb{Z}/p_a\mathbb{Z}$. A homomorphism is determined by the images of all $r_a$ generators of $\mathscr{M}_a$, which can be mapped to any element in $\mathbb{Z}/p_a\mathbb{Z}$, totaling $p_a^{r_a}- 1$ possible surjective homomorphisms, where we excluded the zero homomorphism since it is not surjective. On the other hand, different homomorphisms can have the same kernel. More specifically, two surjective homomorphisms $\phi$ and $\varphi$ have the same kernel if and only if $\phi = \lambda \varphi$ for $\lambda\in\mathbb{Z}_p\setminus\{0\}$. Since there are $p_a-1$ possible values of $\lambda$, there are $\frac{p_a^{r_a}-1}{p_a-1}$ possible kernels of surjective homomorphisms $\phi: \mathscr{X}_a \to \mathbb{Z}/p_a\mathbb{Z}$, and thus of maximal proper submodules.
\end{proof}

\subsection{Proofs for properties of stabiliser groups}
\label{app:stabiliser_groups}

\begin{lemma}
    Any partial stabiliser group $\mathcal{X}\subset\mathscr{P}^n_d$ can be written as $\mathcal{X} = \{\omega^{s(\mathbf{x})} \mathcal{W}_{\mathbf{x}} :  \mathbf{x}\in \mathscr{X}\}$, where $\mathscr{X}\subset\mathbb{Z}_d^{2n}$ is an isotropic submodule and $s:\mathbb{Z}_d^{2n}\to\mathbb{Z}_d$. If $d$ is odd, then $s$ is linear, i.e., $s(\mathbf{x} + \mathbf{y}) = s(\mathbf{x}) + s(\mathbf{y})$. 
    If $\mathcal{X}$ is a stabiliser group, then $\mathscr{X}$ is also Lagrangian and $|\mathscr{X}| = d^n$.
\end{lemma}
\begin{proof}
    First notice that, for $d$ odd, any operator $\tau^s\mathcal{W}_\mathbf{x}$ can be expressed as $\omega^{s'}\mathcal{W}_\mathbf{x}$ for $s'=2^{-1}s$, where $2^{-1}$ is the multiplicative inverse of $2 \bmod d$. For $d$ even, no operator $\tau^s \mathcal{W}_{\mathbf{x}}$ for $s$ odd can belong to $\mathcal{X}$ otherwise $-\mathbf{I}\in \mathcal{X}$, so all elements of $\mathcal{X}$ are of the form $\omega^s\mathcal{W}_{\mathbf{x}}$ for any $d$. There cannot be two operators $\omega^{s}\mathcal{W}_{\mathbf{x}},\omega^{s'}\mathcal{W}_{\mathbf{x}}\in\mathcal{X}$ with $s\neq s'$ since $\mathcal{X}\cap\mathrm{Z}(\mathscr{P}_d^n) = \{\mathbf{I}\}$. Thus $\mathcal{X} = \{\omega^{s(\mathbf{x})} \mathcal{W}_{\mathbf{x}} : \mathbf{x}\in \mathscr{X}\}$ for some set $\mathscr{X}\subset\mathbb{Z}_D^{2n}$ and $s:\mathscr{X}\to\mathbb{Z}_d$. That $\mathscr{X}$ is a submodule follows from $\mathcal{X}$ being a group. Moreover, given $\omega^{s(\mathbf{x})}\mathcal{W}_{\mathbf{x}}, \omega^{s(\mathbf{y})}\mathcal{W}_{\mathbf{y}}\in\mathcal{X}$, then $(\omega^{s(\mathbf{x})}\mathcal{W}_{\mathbf{x}})^\dagger (\omega^{s(\mathbf{y})}\mathcal{W}_{\mathbf{y}})^\dagger (\omega^{s(\mathbf{x})}\mathcal{W}_{\mathbf{x}})(\omega^{s(\mathbf{y})}\mathcal{W}_{\mathbf{y}}) = \omega^{[\mathbf{x},\mathbf{y}]}\cdot\mathbf{I}$ is an element of $\mathcal{X}$, which implies that $[\mathbf{x},\mathbf{y}]_d = 0$ since $\mathcal{X}\cap \mathrm{Z}(\mathscr{P}_d^n) = \{\mathbf{I}\}$. Now notice that $(\omega^{s(\mathbf{x})}\mathcal{W}_{\mathbf{x}})(\omega^{s(\mathbf{y})}\mathcal{W}_{\mathbf{y}}) = \tau^{[\mathbf{x},\mathbf{y}]}\omega^{s(\mathbf{x}) + s(\mathbf{y})}\mathcal{W}_{\mathbf{x}+\mathbf{y}}$, therefore, by the uniqueness of phase of each Weyl operator, $s(\mathbf{x}) + s(\mathbf{y}) = s(\mathbf{x} + \mathbf{y})$ if $d$ is odd, and $\omega^{s(\mathbf{x} + \mathbf{y})} = \omega^{s(\mathbf{x}) + s(\mathbf{y})}\tau^{[\mathbf{x},\mathbf{y}]}$ if $d$ is even. Moreover, $s(\mathbf{0}) = 0$ again due to $\mathcal{X}\cap\mathrm{Z}(\mathscr{P}_d^n) = \{\mathbf{I}\}$. Therefore, for $d$ odd, $\omega^{s(\cdot)}$ is a character and we can write $s(\mathbf{x}) = [\mathbf{a},\mathbf{x}]$ for some $\mathbf{a}\in\mathbb{Z}_d^{2n}$. Furthermore, notice that we can further restrict $\mathscr{X}$ to be a submodule of $\mathbb{Z}_d^{2n}$ (since $\mathcal{W}_{\mathbf{x}} = \pm \mathcal{W}_{\mathbf{x}\operatorname{mod}d}$) according to the rule $\omega^{s(\mathbf{x})}\mathcal{W}_{\mathbf{x}} = \omega^{s(\mathbf{x}\operatorname{mod}d)}\mathcal{W}_{\mathbf{x}\operatorname{mod}d}$ (otherwise $\mathcal{X}\cap \mathrm{Z}(\mathscr{P}_d^n) \neq \{\mathbf{I}\}$). Together with the fact above that $[\mathbf{x},\mathbf{y}]_d = 0$ for all $\mathbf{x},\mathbf{y}\in\mathscr{X}$, this proves that $\mathscr{X}$ is isotropic. Finally, we prove that $\mathscr{X}$ is Lagrangian if $\mathcal{X}$ is maximal. Assume by contradiction that exists $\mathbf{y}\in\mathscr{X}^{\independent}$ such that $\mathbf{y}\notin\mathscr{X}$. This would mean that $\omega^{s(\mathbf{y})}\mathcal{W}_{\mathbf{y}}\notin\mathcal{X}$ while $\mathcal{W}_{\mathbf{y}}$ commutes with all elements of $\mathcal{X}$, a contradiction since $\mathcal{X}$ is maximal. Therefore  $\mathbf{y}\in\mathscr{X}$ and $\mathscr{X}$ is Lagrangian. As a consequence, $|\mathscr{X}| = d^n$.
\end{proof}

\begin{lemma}
    Let $\mathcal{X} = \{\omega^{s(\mathbf{x})}\mathcal{W}_{\mathbf{x}}:\mathbf{x}\in\mathscr{X}\}$ be a partial stabiliser group and let $V_{\mathcal{X}} \triangleq \{|\Psi\rangle\in(\mathbb{C}^{d})^{\otimes n}: \omega^{s(\mathbf{x})}\mathcal{W}_{\mathbf{x}}|\Psi\rangle = |\Psi\rangle, \forall \mathbf{x}\in\mathscr{X}\}$. Then the projector onto $V_{\mathcal{X}}$ is
    \begin{align*}
        \Pi_{\mathcal{X}} = \frac{1}{|\mathscr{X}|}\sum_{\mathbf{x}\in\mathscr{X}}\omega^{s(\mathbf{x})}\mathcal{W}_{\mathbf{x}}.
    \end{align*}
    Moreover, $\operatorname{dim}(V_{\mathcal{X}}) = d^{n}/|\mathscr{X}|$. It then follows that any stabiliser group $\mathcal{S}$ has a unique stabiliser state, i.e., $\operatorname{dim}(V_{\mathcal{S}}) = 1$.
\end{lemma}
\begin{proof}
    Consider the operators $\Pi_{\mathbf{a}} = |\mathscr{X}|^{-1}\sum_{\mathbf{x}\in\mathscr{X}}\omega^{s(\mathbf{x}) + [\mathbf{a},\mathbf{x}]}\mathcal{W}_{\mathbf{x}}$ for $\mathbf{a}\in\mathbb{Z}_d^{2n}$. That $\Pi_{\mathbf{a}}$ is Hermitian can be seen by the change of index $\mathbf{x} \mapsto -\mathbf{x}$ since $s(-\mathbf{x}) = -s(\mathbf{x})$. It is also idempotent,
    \begin{align*}
        \Pi_{\mathbf{a}}^2 = \frac{1}{|\mathscr{X}|^2}\sum_{\mathbf{x},\mathbf{y}\in\mathscr{X}}\omega^{s(\mathbf{x})+s(\mathbf{y})+[\mathbf{a},\mathbf{x}+\mathbf{y}]}\mathcal{W}_{\mathbf{x}}\mathcal{W}_{\mathbf{y}} = \frac{1}{|\mathscr{X}|^2}\sum_{\mathbf{x},\mathbf{y}\in\mathscr{X}}\omega^{s(\mathbf{x}+\mathbf{y}\operatorname{mod}d)+[\mathbf{a},\mathbf{x}+\mathbf{y}]}\mathcal{W}_{\mathbf{x}+\mathbf{y}\operatorname{mod}d} = \Pi_{\mathbf{a}},
    \end{align*}
    where we used that $\omega^{s(\mathbf{x}) + s(\mathbf{y})}\mathcal{W}_{\mathbf{x}}\mathcal{W}_{\mathbf{y}} = \omega^{s(\mathbf{x} + \mathbf{y} \operatorname{mod}d)}\mathcal{W}_{\mathbf{x}+\mathbf{y}\operatorname{mod}d}$. Therefore, $\Pi_{\mathbf{a}}$ (and so $\Pi_{\mathcal{X}}$) is a projector. Moreover, $\Pi_{\mathcal{X}}|\Psi\rangle = |\Psi\rangle$ for any $|\Psi\rangle\in V_{\mathcal{X}}$. On the other hand, given $\Pi_{\mathcal{X}}|\psi\rangle$ for any $|\psi\rangle$, $\omega^{s(\mathbf{x})}\mathcal{W}_{\mathbf{x}}(\Pi_{\mathcal{X}}|\psi\rangle) = \Pi_{\mathcal{X}}|\psi\rangle$ for all $\mathbf{x}\in\mathscr{X}$, and thus $\Pi_{\mathcal{X}}|\psi\rangle\in V_{\mathcal{X}}$. This proves that $\Pi_{\mathcal{X}}$ is the projector onto $V_{\mathcal{X}}$.
    
    Due to the Pontryagin duality theorem (see~\cite[Theorem~1.7.2]{rudin2017fourier}) and to $\mathscr{X}$ being a finite group, there are $|\mathscr{X}|$ characters of $\mathscr{X}$ (since the set of characters of $\mathscr{X}$ is isomorphic to $\mathscr{X}$). Each character gives rise to a distinct projector $\Pi_{C} = |\mathscr{X}|^{-1}\sum_{\mathbf{x}\in\mathscr{X}}\omega^{s(\mathbf{x}) + [\mathbf{a}_C,\mathbf{x}]}\mathcal{W}_{\mathbf{x}}$ where $C \in \mathbb{Z}_d^{2n}/\mathscr{X}^{\independent}$ is a coset and $\mathbf{a}_C$ is any element from $C$. Two different projectors are orthogonal since they belong to different eigenvalues of at least one Weyl operator or, alternatively,
    \begin{align*}
        \Pi_{C}\Pi_{C'} = \frac{1}{|\mathscr{X}|^2}\sum_{\mathbf{x},\mathbf{y}\in\mathscr{X}}\omega^{s(\mathbf{x})+s(\mathbf{y})+[\mathbf{a}_C,\mathbf{x}]+[\mathbf{a}_{C'},\mathbf{y}]}\mathcal{W}_{\mathbf{x}}\mathcal{W}_{\mathbf{y}} = \frac{1}{|\mathscr{X}|^2}\sum_{\mathbf{x},\mathbf{z}\in\mathscr{X}}\omega^{s(\mathbf{z}) + [\mathbf{a}_C,\mathbf{x}]+[\mathbf{a}_{C'},\mathbf{z}-\mathbf{x}]}\mathcal{W}_{\mathbf{z}} = 0,
    \end{align*}
    using that $\omega^{s(\mathbf{x}) + s(\mathbf{y})}\mathcal{W}_{\mathbf{x}}\mathcal{W}_{\mathbf{y}} = \omega^{s(\mathbf{x} + \mathbf{y} \operatorname{mod}d)}\mathcal{W}_{\mathbf{x}+\mathbf{y}\operatorname{mod}d}$ and $\sum_{\mathbf{x}\in\mathscr{X}}\omega^{[\mathbf{a}_C-\mathbf{a}_{C'},\mathbf{x}]} = 0$ since $\mathbf{a}_C - \mathbf{a}_{C'} \neq \mathscr{X}^{\independent}$ (\cref{lem:sum}). This means that the rank of each projector must be a $|\mathscr{X}|$-fraction of the Hilbert space dimension $d^n$, i.e., $d^{n}/|\mathscr{X}|$. This implies that $\operatorname{dim}(V_{\mathcal{X}}) = \operatorname{rank}(\Pi_{C}) = d^{n}/|\mathscr{X}|$.
\end{proof}

\begin{lemma}
    For any non-zero $|\psi\rangle\in(\mathbb{C}^d)^{\otimes n}$, ${\operatorname{Weyl}}(|\psi\rangle)$ is an isotropic submodule.
\end{lemma}
\begin{proof}
    Suppose there are $\mathbf{x},\mathbf{y}\in{\operatorname{Weyl}}(|\psi\rangle)$ such that $[\mathbf{x},\mathbf{y}] \neq 0~(\operatorname{mod}d)$. On the one hand, $\mathcal{W}_{\mathbf{x}}\mathcal{W}_{\mathbf{y}}|\psi\rangle = \omega^{[\mathbf{x},\mathbf{y}]}\mathcal{W}_{\mathbf{y}}\mathcal{W}_{\mathbf{x}}|\psi\rangle = \omega^{[\mathbf{x},\mathbf{y}] + s(\mathbf{x}) + s(\mathbf{y})}|\psi\rangle$, while on the other hand, $\mathcal{W}_{\mathbf{x}}\mathcal{W}_{\mathbf{y}}|\psi\rangle = \omega^{s(\mathbf{x}) + s(\mathbf{y})}|\psi\rangle$, which is a contradiction. Moreover, given $\mathbf{x},\mathbf{y}\in{\operatorname{Weyl}}(|\psi\rangle)$, then $\mathcal{W}_{\mathbf{x}+\mathbf{y}}|\psi\rangle = \tau^{[\mathbf{y},\mathbf{x}]}\mathcal{W}_{\mathbf{x}}\mathcal{W}_{\mathbf{y}}|\psi\rangle = \tau^{[\mathbf{y},\mathbf{x}]}\omega^{s(\mathbf{x}) + s(\mathbf{y})}|\psi\rangle$, but since $[\mathbf{y},\mathbf{x}] = 0~(\operatorname{mod}d)$, either $\tau^{[\mathbf{y},\mathbf{x}]} = 1$ or $\tau^{[\mathbf{y},\mathbf{x}]} = \omega^{d/2}$, the latter for $d$ even only. Thus $\mathbf{x}+\mathbf{y}\in{\operatorname{Weyl}}(|\psi\rangle)$ and ${\operatorname{Weyl}}(|\psi\rangle)$ is a submodule.
\end{proof}

\begin{corollary}
    For all $\mathcal{C}\in\mathscr{C}_d^n$ and $|\psi\rangle\in(\mathbb{C}^d)^{\otimes n}$, $|{\operatorname{Weyl}}(\mathcal{C}|\psi\rangle)| = |{\operatorname{Weyl}}(|\psi\rangle)|$.
\end{corollary}
\begin{proof}
    Write $\mathcal{C} = \mathcal{W}_{\mathbf{a}}\mu(\Gamma)$ for some $\mathbf{a}\in\mathbb{Z}_d^{2n}$ and $\Gamma\in\operatorname{Sp}(2n,d)$. The result follows because $|{\operatorname{Weyl}}(\mathcal{C}|\psi\rangle)| = |{\operatorname{Weyl}}(\mu(\Gamma)|\psi\rangle)|$ and $\Gamma$ is invertible ($\Gamma^{-1} = \Omega^{-1}\Gamma^\top\Omega$), so that ${\operatorname{Weyl}}(\mu(\Gamma)|\psi\rangle) = \{\Gamma\mathbf{x}:\mathbf{x}\in{\operatorname{Weyl}}(|\psi\rangle)\}$ and ${\operatorname{Weyl}}(|\psi\rangle) = \{\Gamma^{-1} \mathbf{x}:\mathbf{x}\in{\operatorname{Weyl}}(\mu(\Gamma)|\psi\rangle)\}$. Indeed, it is not hard to see that $\{\Gamma\mathbf{x}:\mathbf{x}\in{\operatorname{Weyl}}(|\psi\rangle)\} \subseteq {\operatorname{Weyl}}(\mu(\Gamma)|\psi\rangle)$ and $\{\Gamma^{-1} \mathbf{x}:\mathbf{x}\in{\operatorname{Weyl}}(\mu(\Gamma)|\psi\rangle)\} \subseteq {\operatorname{Weyl}}(|\psi\rangle)$. Assume by contradiction that there is $\mathbf{y}\in {\operatorname{Weyl}}(\mu(\Gamma)|\psi\rangle)$ such that $\mathbf{y} \notin \{\Gamma\mathbf{x}:\mathbf{x}\in{\operatorname{Weyl}}(|\psi\rangle)\}$. Then $\Gamma^{-1}\mathbf{y}\in \{\Gamma^{-1} \mathbf{x}:\mathbf{x}\in{\operatorname{Weyl}}(\mu(\Gamma)|\psi\rangle)\}$ but $\Gamma^{-1}\mathbf{y}\notin {\operatorname{Weyl}}(|\psi\rangle)$, a contradiction. Similarly for ${\operatorname{Weyl}}(|\psi\rangle) = \{\Gamma^{-1} \mathbf{x}:\mathbf{x}\in{\operatorname{Weyl}}(\mu(\Gamma)|\psi\rangle)\}$.
\end{proof}

\end{document}